\documentclass[a4paper,fleqn,usenatbib]{mnras}

\usepackage{newtxtext,newtxmath}
\usepackage[T1]{fontenc}
\usepackage{ae,aecompl}

\usepackage{graphicx}	% Including figure files
\usepackage{amsmath}	% Advanced maths commands
\usepackage{amssymb}	% Extra maths symbols
\usepackage{arydshln}
 
\usepackage{verbatim}
\usepackage{breqn}
\usepackage[export]{adjustbox}

\newcommand{\lya}{Ly$\alpha$}
\newcommand{\lyb}{Ly$\beta$}

\newcommand{\kms}{$km s^{-1}$}

\newcommand{\HI}{\mbox{H\,{\sc i}}}

\newcommand{\HeII}{\mbox{He\,{\sc ii}}}
  
\title[Three-point correlation of IGM]{Three- and two-point spatial correlations of IGM at $z\sim 2$: Cloud based analysis using simulations}

\author[Maitra et al.]{Soumak Maitra$^{1}$\thanks{E-mail: soumak@iucaa.in},
	Raghunathan Srianand$^{1}$, Prakash Gaikwad$^{2,3}$, Tirthankar Roy  \newauthor{Choudhury$^{4}$, Aseem Paranjape$^{1}$ \& Patrick Petitjean$^{5}$}
	\\
	\\
	$^{1}$ IUCAA, Postbag 4, Ganeshkhind, Pune, 411007, India\\
	$^{2}$ Institute of Astronomy, University of Cambridge, Madingley Road, Cambridge, CB3 0HA, UK\\
	$^{3}$ Kavli Institute for Cosmology, University of Cambridge, Madingley Road, Cambridge, CB3 0HA, UK\\
	$^{4}$ National Centre for Radio Astrophysics, Tata Institute of Fundamental Research, Pune 411007, India \\
	$^{5}$ Institut d'Astrophysique de Paris, CNRS-SU, UMR 7095, 98bis bd Arago, 75014, Paris, France\\
}
\date{Accepted XXX. Received YYY; in original form ZZZ}

\pubyear{2020}

\begin{document}
	\label{firstpage}
	\pagerange{\pageref{firstpage}--\pageref{lastpage}}
	\maketitle
	
	\begin{abstract}
		
		\lya\ forest absorption spectra decomposed into multiple Voigt profile components (clouds) allow us to study clustering of intergalactic medium (IGM) as a function of \HI\ column density ($N_{\rm HI}$).
		Here, we explore the transverse three-point correlation ($\zeta$) of these \lya\ clouds using mock triplet spectra obtained from hydrodynamical simulations at $z \sim 2$ on scales of 1-5 $h^{-1}$cMpc.
		We find $\zeta$ to depend strongly on $N_{\rm HI}$ and scale  and weakly on angle ($\theta$) of the triplet configuration. We show that the "hierarchical ansatz" is applicable for scales $\ge~ 3h^{-1}$cMpc, and obtain a median reduced three-point correlation (Q) in the range 0.2-0.7.
		We show, $\zeta$ is influenced strongly by the thermal and ionization state of the gas. As found in the case of galaxies, the influence of physical parameters on Q is weaker compared to that of $\zeta$. We show difference in $\zeta$ and Q between different simulations  are minimized if we use appropriate $N_{\rm HI}$ cut-offs  corresponding to a given baryon over-density ($\Delta$) using the  measured $N_{\rm HI}~vs~\Delta$ relationship obtained from individual simulations. Additionally, we see the effect of pressure broadening on $\zeta$ in a model with artificially boosted heating rates. However, for models with realistic thermal and ionization histories the effect of pressure broadening on $\zeta$ is weak and sub-dominant compared to other local effects.
		We find strong redshift evolution shown by $\zeta$, mainly originating from the redshift evolution of thermal and ionization state of the IGM. We discuss the observational requirements for the detection of three-point correlation, specifically, in small intervals of configuration parameters and redshift.

	\end{abstract}
	
	
	%%%%%%%%%%%%%%%%%%%%%%%%%%%%%%%%%%%%%%%%%%%%%%%%%%
	
	%%%%%%%%%%%%%%%%% BODY OF PAPER %%%%%%%%%%%%%%%%%%

	\section{Introduction}\label{Introduction}
	
	It is now well established that the observed properties of the \HI\
	\lya-forest absorption seen in the spectra of distant quasars are governed by the thermal and ionization history of the intergalactic medium (IGM) together with the underlying dark matter distribution. It is also now well established that apart from very small scales (where pressure smoothing effects are important,see \cite{gnedin1998,peeples2010b,kulkarni2015,rorai2017}) baryons follow the dark matter fluctuations very closely.
	However, the connection between the distribution of baryons and distribution of \HI\ optical depth (or transmitted flux) which one obtains directly from the observations is governed by local temperature and radiation field, thermal history and peculiar velocities \citep[see][for a review on IGM]{meiksin2009}.
	
	It is also now well established that in the post re-ionization era (i.e., $z<6$) the ionization evolution of the low density IGM is governed mainly by the photoionization 
	from the uniform meta-galactic UV background. The processes governing the thermal evolution of the unshocked baryonic gas and the existence of temperature-density (${\rm T}-\delta$) relation in simulations with uniform UVB are also well established \citep[see][]{hui1997}. Therefore, it is now possible to reproduce the basic observables like mean transmitted flux, flux probability distribution function (PDF), flux power-spectrum and their redshift evolution using semi-analytical \citep[]{doroshkevich1977,mcgill1990,bi1997,gnedin1996,choudhury2001}, dark matter only N-body simulations \citep[]{Petitjean1995,muecket1996, Peirani2014,Sorini2016} and hydrodynamical N-body simulations \citep[][]{cen1994,zhang1995,miralda1996,hernquist1996,croft1997,theuns1998b,dave1999,viel2004a,springel2001,springel2005,bolton2006,smith2011,bryan2014}.

	It is also well recognised that clustering studies are important to 
	understand the matter distribution in the IGM. Such studies can be done in two ways; correlation in the \lya\ forest along the line of sight of a single quasar (i.e redshift space correlation or longitudial correlation) or between adjacent sightlines for quasar pairs (i.e angular or transverse correlation). 
	Due to easy availability of larger sample size of single quasar sightlines, they have been used widely to characterize the longitudinal (i.e redshift space) two-point correlation function (or power-spectrum) of the \lya\ forest \citep[see for example,][]{mcdonald2000,mcdonald2006,croft2002,seljak2006}. Power spectrum of transmitted flux (fourier transform of longitudinal correlations) have also been used to constrain neutrino masses, warm dark matter, UV radiation field and thermal state of the IGM in addition to constraining the power spectrum of density fluctuations\citep[]{viel2005,viel2013w,bird2012,palanque2015a,gaikwad2017a,gaikwad2018,gaikwad2019,khaire2019a,walther2019,gaikwad2020}.

	Transverse correlations studies involving small scale \lya\ clustering is primarily dominated by projected quasar pairs or gravitationally lensed quasars \citep[][]{Smette1995, rauch1995, petitjean1998,hennawi2010}. At these scales pressure smoothing plays an important role in the spacial distribution of \lya\ absorption and transverse correlations studies are essential to capture this as in the longitudinal direction thermal broadening tends to weaken the signals introduced by the pressure smoothing \cite[see for example,][]{gnedin1998,peeples2010a,peeples2010b,kulkarni2015,rorai2018}.
	At the scales of few Mpc, positive correlation has been detected in the \lya\ forest clustering \citep{aracil2002,rollinde2003,coppolani2006,dodorico2006, maitra2019} and these usually come from clustering around galaxies or cosmic structures involving filaments and sheets \citep{Petitjean1995,cantalupo2014} in the mildly non-linear regime. The density fields can be probed with more than two closely spaced quasar pairs \citep[see][]{cappetta2010,maitra2019} at $z\sim 2$. Sightlines towards  such closed projected quasar groups can be used to determine higher order satistics of matter clustering in IGM in the tranverse plane. In fact, with dense enough grids of closely spaced quasar sightlines, it becomes possible to extend the clustering studies to perform a full tomographic 3D reconstruction of the underlying matter density fields \citep{Pichon2001,mcdonald2003,caucci2008,lee2014,lee2018,Krolewski2018,horowitz2019}.  Baryonic Oscillation spectroscopic survey \citep[BOSS][]{dawson2013} has allowed the measurement of \lya\ forest transmitted 3D flux power-spectrum at large scale \citep[][]{slosar2011} resulting in the measurement of BAO signals at high-$z$ \citep[see][for the latest results]{ata2018}.

	Studying the higher order statistics is very important to probe the non-Gaussianity in the matter distribution (primordial and those introduced by the non-linear evolution of gravitational clustering)  and to understand the evolution of matter beyond the linear approximation. The first significant order beyond the two point correlation function (or power spectrum in the Fourier space) is the three point correlation function (or bi-spectrum in the Fourier space). It has also been pointed out that one will be able to lift the degeneracy between different cosmological parameters (like bias and $\sigma_8$) by simultaneous usage of two- and three-point correlation functions \citep[see][]{peebles1980,fry1994,Bernardeau2002}. Till date, considerable work has been done on three-point correlation function of galaxies from large surveys \citep{gaztanaga1994,kayo2004,jing2004,gaztanaga2005,nichol2006,sefusatti2006,kulkarni2007,mcbride2011a,mcbride2011b,guo2016,moresco2017}. 
	
	Traditionally, the three point correlation is quantified using the dimension less "Q" parameter called the "reduced three-point correlation". It is defined as the three-point correlation function normalized with the cyclic combination of two-point correlation functions associated  with the three-points in question (for details see Eq.~\ref{Q_eq}). This quantity can be thought of as the skewness of the distribution. The typical Q value is found to be $\sim$1.3 with a moderate dependence on the shape of the matter power spectrum and shape of the triangle (i.e configuration). The dependence of Q on the angle of the triangle is used to quantify the nature of the structure probed at different scales. While compact spherically symmetric structures (probed by equilateral configurations) dominate at small scales, the filamentary structures dominate at large scales (probed by linear configurations). It has been found that more luminous and massive galaxies (having high stellar mass) clustered strongly compared to less luminous and less massive (having low stellar mass) \citep{zehavi2005,zehavi2011}. It was also found that three point correlation function exhibits stronger dependence on both galaxy luminosity and stellar mass than Q \citep{kayo2004,mcbride2011a,guo2016}.  No significant redshift evolution is found for the angular dependence of Q at small scales. The redshift evolution noticed for the larger scales (i.e $>15$ Mpc) are consistent with the expectations of growth structures at low-$z$.

	As far as the higher order correlation studies of IGM is concerned very little work has been done with three-point statistics. Most of the earlier
	studies focus on obtaining 1D bi-spectrum of \lya\ forest \citep{viel2004c,viel2009a,hazra2012} considering three points along a
	single sightline. This statistics with inherently weak signals due to \lya\ being in mildly non-linear regime, will be largely affected by the spectral signal to noise and thermal broadening effects at small scales. So, working with closely spaced quasar triplets to probe the transverse clustering will provide us with a better insight into the non-gaussianity involved with the matter distribution at such scales \citep[see][]{tie2019,maitra2019}. \citet{tie2019}, using statistics of transmitted flux along closely space triplet sightline (probing scales of 10-30 h$^{-1}$ Mpc) in their simulation box, have derived Q = -4.5 for \lya\ forest at $z\sim 2$.  Such unusual value of Q (compared to what one used to see in the case of galaxies) largely reflects the complex relation between the transmitted flux field and the matter density field, and not solely from the density field itself. 
	On the other hand, \citet[]{maitra2019} studied three point correlation of \lya\ absorption towards two quasar triplets observed with X-Shooter using voigt profile components and techniques similar to what has been used in clustering studies of galaxies. This work suggested 
	a monotonically increasing trend in three-point correlation with \HI\ column density ($N_{\rm HI}$) thresholds which can be interpreted as increased non-gaussianity in clustering for higher column density clouds. 
	
	In this paper, we seek to study the transverse three point correlation function of the high-$z$ IGM  at scales up to few Mpc using hydrodynamical simulations. The main motivation of this exploratory work is to investigate the utility of three-point correlations to derive parameters related to the physical state of the IGM and cosmological parameters. We mainly focus on scales where the transition from pressure driven matter distribution to the cosmic structures occurs. In particular we use statistics of \lya\ components obtained using Voigt profile fitting and techniques similar to those used in galaxies. Note most of the work in the literature discussed above use statistics based on transmitted flux. Our main aim
	is to understand the dependence of two- and three point correlation function (and the reduced three point correction, Q) on (i) the configuration of background sources (i.e scale and angle of the projected triangle connecting the three sightlines), (ii) \HI\ column density cut-off (i.e for different baryonic and dark matter over-densities), (iii) thermal history and ionization history (i.e simulations obtained with different ionizing UV background (UVB) radiations) and (iv) redshift evolution of the two- and three-point correlation function. We also check the validity of "hierarchical ansatz" of \citet{peebles1980} in the case of IGM clustering. The details of the simulations used in our study is given in Section 2. Results of our studies are presented in detail in Section 3. Redshift evolution  of clustering is discussed in section 4. In section 5 we disuss the detectability.
	We summarise and discuss our findings in Section 6.

	\section{Simulation} \label{Simulation}
	
	\begin{figure*}
		\centering
		\includegraphics[width=15.5cm]{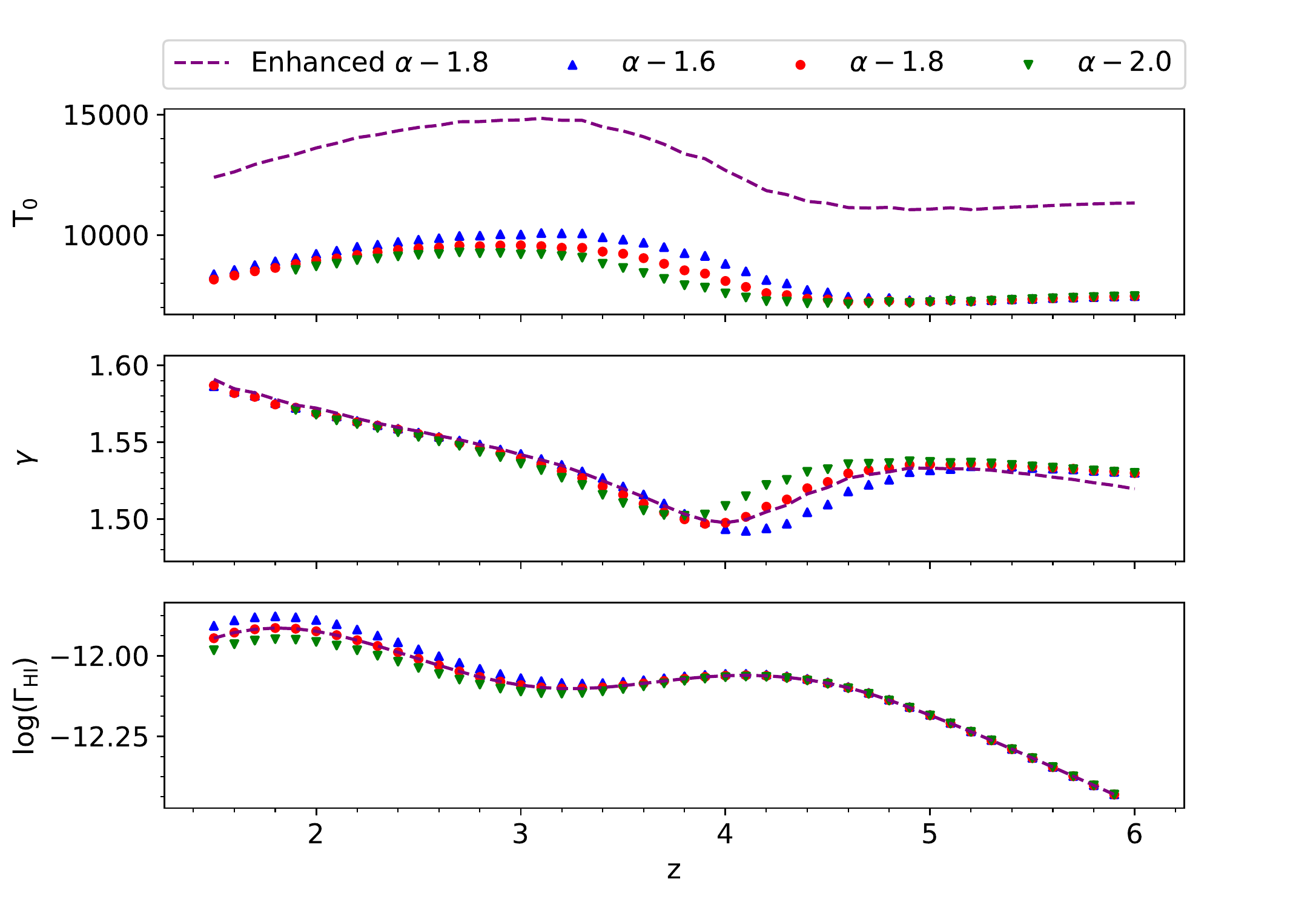}
		\caption{Evolution of the IGM ${\rm T}-\delta$ relation parameters $T_{0}$ (top), $\gamma$ (middle) and \HI\  photoionization rate $\Gamma_{\rm HI}$ (bottom) as a function of redshift for four simulations considered in this work (see Sec.~\ref{Simulation} for details of the simulations). The photoionization rates are given by \citet{khaire2019} for the assumed far-UV quasar spectral index ($f_{\nu}\propto \nu^{-\alpha}$) of $\alpha=1.6,\ 1.8$ and $2.0$. The increase in $T_{0}$ (and slight decrease in $\gamma$) seen around $3 \leq z \leq 4$ is driven by \HeII\ reionization that we model through equilibrium evolution (for details, see \citet{gaikwad2019}). }
		\label{T-g-G}
	\end{figure*}
	
	The simulations that we use in this work have been run using the smoothed particle hydrodynamical code {\sc gadget-3} \citep[a modified version of
	the publicly available {\sc gadget-2}\footnote{\url{http://wwwmpa.mpa-garching.mpg.de/gadget/}} code, see][]{springel2005} which takes care of radiative heating and cooling processes by self-consistently solving the ionization equilibrium and thermal non-equilibrium evolution for a given ionizing metagalactic UV background (UVB). We generate $100 h^{-1} $cMpc simulation box with $2\times 1024^3$ particles using the standard flat $\Lambda$CDM background cosmology ($\Omega_{\Lambda}$, $\Omega_{m}$, $\Omega_{b}$, $h$, $n_s$, $\sigma_8$, $Y$ ) $\equiv$ (0.69, 0.31, 0.0486, 0.674, 0.96, 0.83, 0.24) based on \citet{planck2014}. 
	The initial condition for the simulation is generated at $z=99$ based on second-order lagrangian perturbation theory, using the publicly available {\sc 2lpt}\footnote{\url{https://cosmo.nyu.edu/roman/2LPT/}} \citep{scoccimarro2012} code. The gravitational softening length has been taken as $1/30^{th}$ of the mean inter-particle separation.  In order to run the simulation faster, we convert gas particles with over-density $\Delta>10^3$ and temperature $T<10^5K$ to stars \citep[see][]{viel2004a} by turning on the {\sc quick\_lyalpha} flag in the simulation. Our simulations do not include feedback from Active Galactic Nuclei (AGN) or star formation assisted galactic outflows. We have stored the simulation outputs between $z=6$ and $z=1.5$ with a redshift interval of 0.1.
	
	We ran three such simulations by varying the thermal histories using the ionization and heating rates for H and He  as given by \citet{khaire2019} for the assumed quasar far-UV spectral index ($f_{\nu}\propto \nu^{-\alpha}$) of $\alpha=1.6,\ 1.8$ and $2.0$. Note that this range covers the inferred ionization rates of hydrogen and helium (i.e $\Gamma_{\rm HI}$, $\Gamma_{\rm HeI}$ and $\Gamma_{\rm HeII}$) from observations at high-$z$ \citep{khaire2017,gaikwad2019}. The redshift evolution of the ${\rm T}-\delta$ relation (defined as $\rm  T=T_0$ $(\Delta/\bar{\Delta})^{\gamma-1}$) parameters  $\rm T_0$ (IGM temperature at mean over-density) and $\gamma$ for these simulations have been shown in Fig.\ref{T-g-G}. These were obtained using standard procedure as described in \citet{gaikwad2019}. We consider the simulations run with UV background model with $\alpha=1.8$ as our fiducial model in this work. We also run a simulation with a drastically different thermal history to enhance the effect of thermal history on correlation statistics. In this case, we use the UV background corresponding to $\alpha=1.8$ and then the photoheating rates were artificially doubled while keeping the photoionization rates same (we refer to this as "Enhanced $\alpha=1.8$" simulation). The redshift evolution of thermal parameters for this simulation has also been shown in Fig.\ref{T-g-G} (purple dashed curve). Since the energy injection occurs uniformly for all particles of different densities, the slope of the ${\rm T}-\delta$ relation, $\gamma$, remains same while  $\rm T_0$ is enhanced roughly by a factor of 1.5. Additionally, we also use a  $50 h^{-1} $cMpc simulation box with $2\times 1024^3$ particles using the same cosmology and UVB as our fiducial model for convergence tests.

	\subsection{Generation of transmitted flux skewers}

	For this work, we follow the transmitted flux generation scheme from the simulation box as discussed in \citet{maitra2019} (see Section 3.1 in their paper).
	We shoot lines of sight through the simulation box and generate the 1D neutral hydrogen density ($n_{\rm HI}$), temperature  ($\rm T$) and the peculiar velocity ($v$) fields using SPH smoothing of these quantities from the nearby (within a smoothing scale) particles. We sample each line of sight with 2048 uniformly sampled grids in comoving length of the box. 
	Using the $n_{\rm HI}$, temperature and velocity fields, we obtain the \lya\ optical depth $\tau$ as a function of wavelength along the sightlines \citep[see Eq.30 of][]{choudhury2001}. 
	The optical depth is then negative exponentiated to get the \lya\ transmitted flux spectrum $F$ ($F=e^{-\tau}$).

	We then add the effects of instrumental resolution and noise to our simulated \lya\ transmitted flux to mimic the typical quasar spectrum. The transmitted flux is convolved with the instrumental line spread function (assumed to be a gaussian) with FWHM $\sim 50$ \kms. The data is then rebinned to $\sim 15$ \kms pixels to match the pixel sampling of a typical X-Shooter spectra. This also corresponds to the typical spectral resolution of upcoming multi-object spectrographs in 30m class telescopes.  
	As the final step, we add Gaussian noise to the transmitted flux corresponding to Signal-to-Noise Ratio (SNR) of 20. Note that the effect of SNR in our analysis will decide the $N_{\rm HI}$ completeness. We consider SNR along all the triplet sightlines to be same. This will not be the case in reality as brightness of the background quasars will not be identical. In this case, completeness in $N_{\rm HI}$ will be set by the spectrum with lowest SNR. The flux PDF obtained using 4000 such randomly generated transmitted flux skewers are shown in the top panel of Fig.~\ref{Validation} for our 3 simulation boxes with different thermal histories for $z=2$.

	\subsection{Voigt profile fitting}
	It is a normal procedure to decompose the \lya\ forest into Voigt profile components (or individual absorbers)  parameterised by redshift ($z$), \HI\ column density ($N_{\rm HI}$) and velocity dispersion ($b$).
	Thus, instead of treating IGM as a continuous fluctuating density field, this approach breaks it down into distinct "clouds". The main feature of "cloud" based statistics is that it allows us to study clustering using techniques used for studying galaxy clustering and probe its dependence on \HI\ column density thresholds. The downside is that fitting statistically significant number of sightlines with voigt profiles is a computationally expensive exercise compared to using transmitted flux. However, with the help of high performance computing and the automated parallel Voigt profile fitting code {\sc viper} \citep[see][for details regarding {\sc viper}]{gaikwad2017b}, it is now possible to generate voigt profile fits for a large number of sightlines in a short time. 
	We use {\sc viper} to identify the \lya\ absorption lines and obtain the $N_{\rm HI}$, $b$ and $z$ for individual components. The number of Voigt profile components used to fit an absorption region is objectively decided based on the the Akaike Information Criterion with Correction \citep[AICC;][]{akaike1974,liddle2007,king2011}. The code then assigns a rigorous significance level \citep[RSL, as described in][]{keeney2012} to these fitted Voigt profile components. We consider only components for which the RSL$>3$ in our analysis to avoid false identifications. Note that present versions of {\sc viper} fits only the \lya\ forest lines and does not use additional optical depth constraints coming from \lyb\ and other higher Lyman series lines. This means that fits to the highly saturated lines may not be accurate.
	
\begin{figure}
		\centering
		\includegraphics[viewport=2 0 310 260,width=7cm, clip=true]{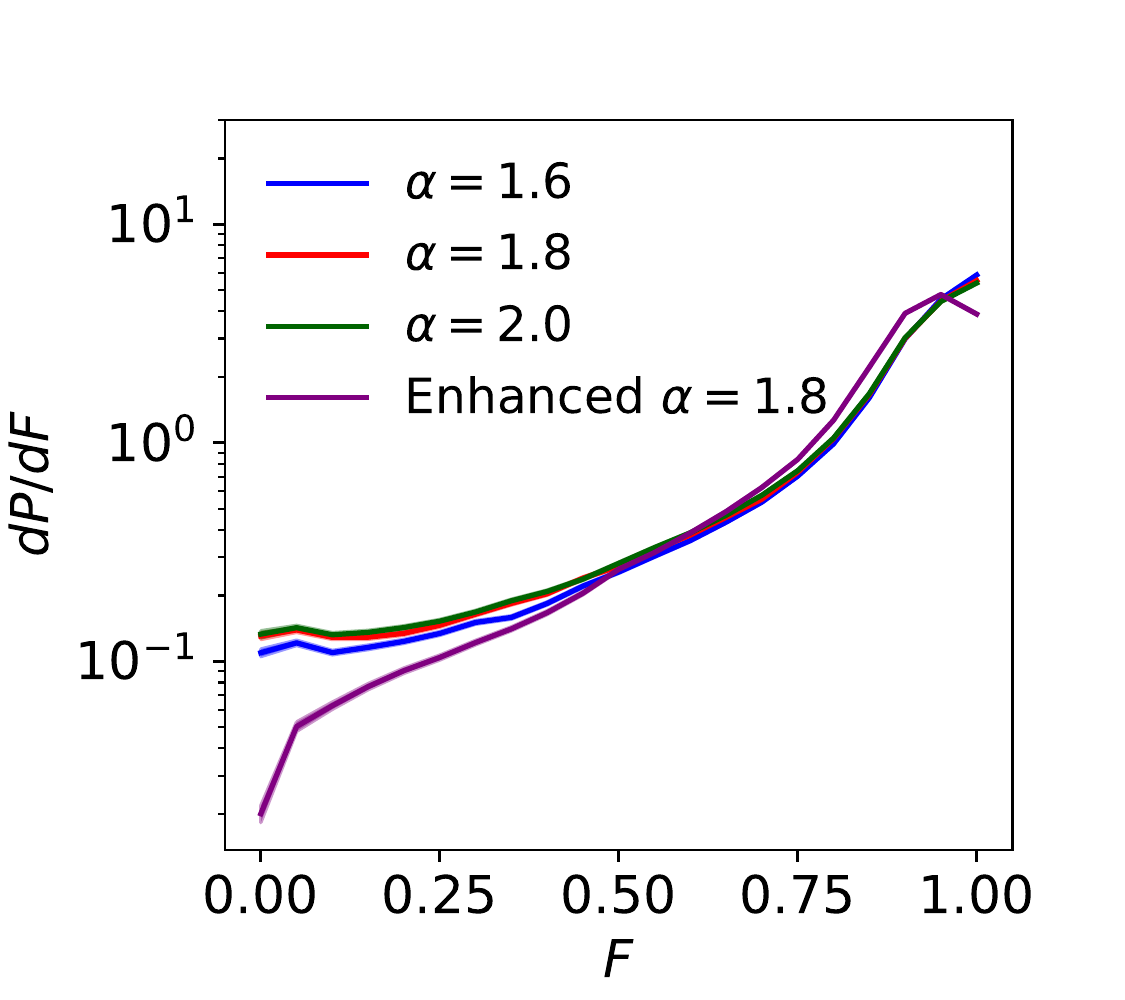}
		\includegraphics[viewport=30 10 350 290,width=7.2cm, clip=true]{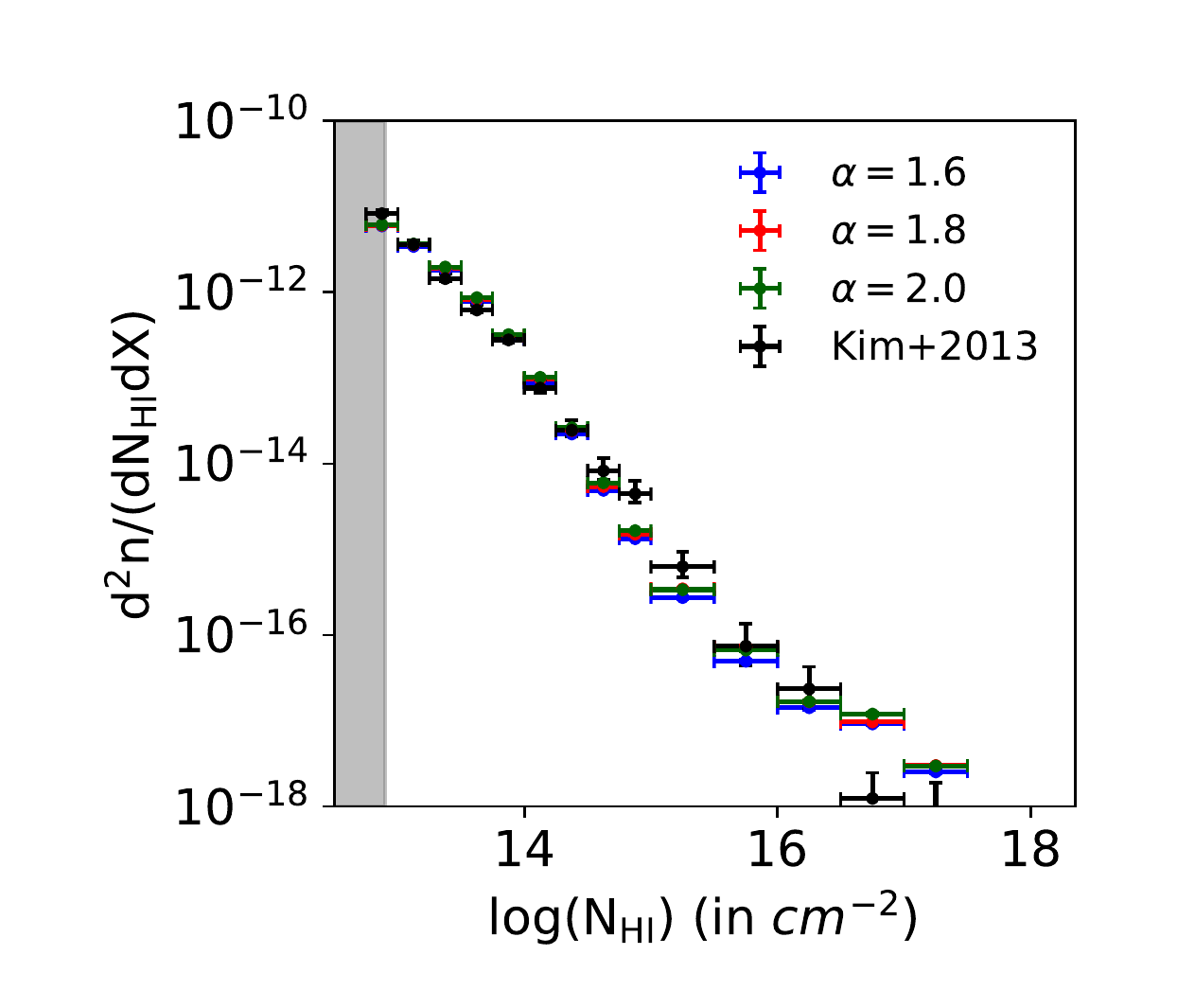}
		\caption{Probability distribution function (PDF) of the transmitted \lya\ flux (\textit{top}) and neutral hydrogen column density distribution per unit absorption distance for different simulations (\textit{bottom}) at $z=2$.  The 1$\sigma$ confidence interval has been calculated by bootstrapping using 4000 simulated sightlines. For the flux PDF, it has been shown as a shaded region while for neutral hydrogen column density distribution, it is too small to be resolved in the figure. The relative differences between neutral hydrogen column density distributions for different simulations are within 10\% for $N_{\rm HI}=10^{13-14.5}$cm$^{-2}$. }
		\label{Validation}
	\end{figure}

	Next, we compare the \HI\ column density distribution function (CDDF) of these absorbers obtained in our simulations with the observed ones \citep{kim2013}. The CDDF, $f(N_{\rm HI},X)$, is defined as the number of \HI\ absorbers within absorption distance interval $X$ and $X+dX$ and within column density interval $N_{\rm HI}$ and $N_{\rm HI}+dN_{\rm HI}$. 
	The absorption distance $X$ is defined as ,
	\begin{equation}
	X(z)=\int_0^z dz \frac{H_0}{H(z)}(1+z)^2  \ ,
	\end{equation}
	by \citet{bahcall1969}. In the bottom panel in Fig.~\ref{Validation}, we compare our CDDF with the one given in \citet{kim2013}. In \citet{kim2013}, the CDDF has been computed in the redshift range of 1.9-2.4 for the \lya\ absorbers and from observed spectra with resolution $\sim$6\kms. The error plotted for \citet{kim2013} corresponds to 1$\sigma$ range. For simulations, we calculate the 1$\sigma$ confidence interval due to bootstrapping over 4000 simulated sightlines, which is too small to be properly resolved in the figure. To keep things same, we use the same binning scheme as in \citet{kim2013}. 
	
	Given the resolution of 50 \kms and the SNR value of 20 per pixel for our simulated skewers, we get an approximate lower completeness limit of $N_{\rm HI}=10^{12.9}$cm$^{-2}$ ($5\sigma$ detection limit). This is given as a grey shaded region in the bottom panel in Fig.~\ref{Validation}. So, while the lowest $N_{\rm HI}$ bin, which is below our completeness limit, gives a slightly lower value of $f(N_{\rm HI},X)$, we obtain good matching with \citet{kim2013} in log($N_{\rm HI}$)=[13.0,14.5]. The differences between our models are also minimum in this range. Disagreement at higher \HI\  column densities can arise from incompleteness in sampling due to finite simulation box size. Also, for saturated absorption profiles, the AICC will favour a fit with minimum number of Voigt profile components of the absorption. The accurate number of components can be determined by the simultaneous fit to the \lya\ and the corresponding \lyb\ profile which has a lower absorption cross-section as compared to \lya. Henceforth, we will fix a lower $N_{\rm HI}$ thresholds of $10^{13}$cm$^{-2}$ for our correlation studies. Additionally, we also investigate the effect of using different $N_{\rm HI}$ thresholds on the clustering properties. 
	In \citet{maitra2019}, we have shown that our simulations also reproduce the observed transverse two-point correlation of transmitted flux as a function of angular separation as measured by \citet{coppolani2006}.

	\subsection{The $N_{\rm HI}$ vs over-density ($\Delta$) relation}
	One of our aims is to study the three-point correlation as a function of $N_{\rm HI}$ at different redshifts. To interpret these results, it will be good to have relation between $N_{\rm HI}$ and the baryonic over-density from our models. Due to peculiar velocities along the line of sight, there is no one to one correspondence between $N_{\rm HI}$ and baryonic over-density. But statistically, it has been found that there exists a power law relationship between the optical depth ($\tau$) weighted over-density $\Delta$ and $N_{\rm HI}$ given by
	
	\begin{equation}\label{delta-N}
	\Delta=\Delta_0 N_{14} ^{\eta}\ .
	\end{equation}
	Here, $N_{14}$ is the $N_{\rm HI}$ given in units of $10^{14}$cm$^{-2}$. Based on our simulated spectra, we can compute the best fit $\Delta_0$ and $\eta$ values. To find the corresponding $\Delta$ values associated with the $N_{\rm HI}$ absorbers, we assign an optical depth ($\tau$) weighted baryonic over-density to these absorbers \citep[see][]{dave1999,schaye1999}. For a given absorber in the velocity space whose pixel value corresponding to the absorption peak identified by index $j$ (in practice this is the centroid of the voigt profile), we associate the over-densities of all the other pixels $i$ in real space that contribute to the optical depth at the position of the absorber. This association comes in the form of a weight factor $\tau_{ij}$ which gives the optical depth contribution of the over-density at pixel $i$ to the optical depth at the position of absorber $j$. The $\tau$ weighted over-density of absorber $j$ is then given as,
	\begin{equation}
	\bar{\Delta}_j=\frac{\sum\limits_i \tau_{ij} \Delta_i}{\sum\limits_i \tau_{ij}}
	\end{equation}
	where the summation is over all the pixels $i$ in the spectra (see \citet{gaikwad2017a}). 
	\begin{figure}
		\centering
		\includegraphics[viewport=7 0 310 300,width=8cm, clip=true]{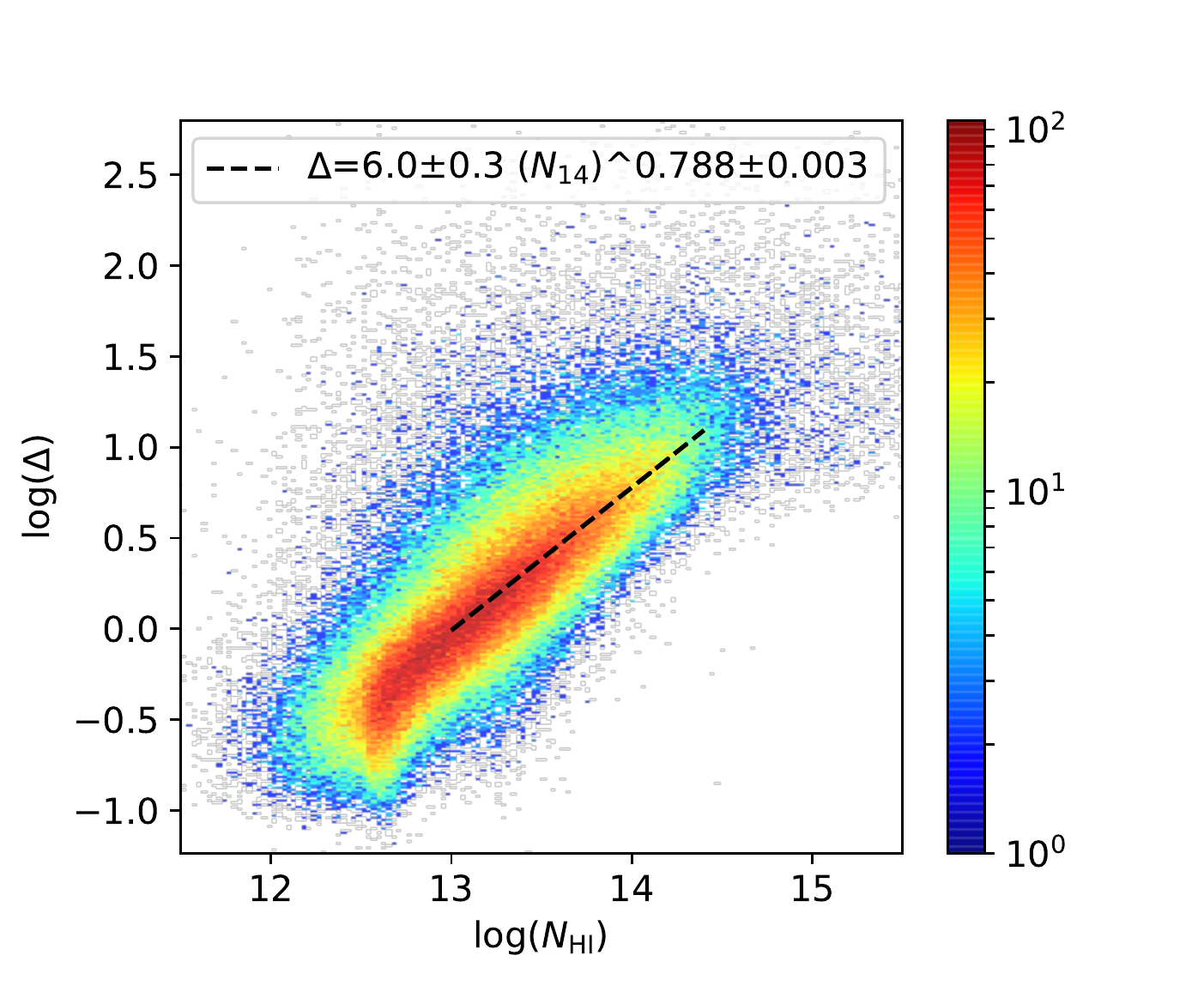}
		\caption{Neutral hydrogen column density ($N_{\rm HI}$) vs $\tau$ weighted baryon over-density ($\Delta$) at $z=2$ for our fiducial model using 4000 simulated spectra with SNR=20 and instrumental FWHM=50 \kms. The color represents the number of points in a certain grid (see the color scale in the right). The black dashed line represents the best fit relationship (also given in the figure) followed by the median $\Delta$ in the range of log($N_{\rm HI}$)=(13,14.5).  }
		\label{N_vs_Delta}
	\end{figure}
	In Fig.~\ref{N_vs_Delta}, we show the log-log density plot of $N_{\rm HI}$ vs the corresponding $\tau$ weighted baryonic over-densities $\Delta$ at $z=2$ from the fiducial model. To fit the power law, we consider $N_{\rm HI}$ over $10^{13}$cm$^{-2}$, to be above the completeness limit set by spectral noise and resolution. We also fix the upper limit to be $10^{14.5}$cm$^{-2}$ as the population of higher column density clouds are limited by the finite box size. We bin the data in equispaced log$_{10}$($N_{\rm HI}$) bins of length 0.1 in the range [13.0,14.5] and find the median $\Delta$ in these bins. We then compute the best fit power law for these points which is shown as a black-dashed line in Fig.\ref{N_vs_Delta}. At $z=2$, the best fit parameter are $\Delta_0=6.0\pm0.3$ and $\eta=0.788\pm 0.003$. This $\Delta_0$ is close to the optimal over-density $\bar{\Delta}=5.69$ at $z=2.05$ as shown by \cite{becker2011}. The $\Delta_0$ value appears to be smaller than $\Delta_0=6.6\pm0.1$ given in \citet{dave2010} at $z=2$ for their no-wind simulation. Also, our $\eta$ value is slightly higher than their value of $0.741\pm 0.003$. This difference may come from the fact that their simulations use \citet{haardt2001} ionizing background radiation while our simulations use that from \citet{khaire2019}. 
	
	The best fit parameters $\Delta_0$ and $\alpha$ for all our simulations are given for $z=2$ in Table.~\ref{N_Delta}. For the fiducial case, we provide these for different redshifts at $1.8\leq z \leq 2.5$. Based on these parameters, we also compute the $\Delta$ corresponding to the fixed $N_{\rm HI}$ thresholds of $10^{13}$ and $10^{13.5}$cm$^{-2}$ at each redshifts that we will use for our correlation studies. This gives us an estimate of what baryonic over-densities one probes for a fixed column density threshold when we study different models and for fixed UVB model. This also allows us to compute $N_{\rm HI}$ thresholds at different redshifts that will correspond to a fixed $\Delta$. We will discuss about this further in the next sections.

	\begin{table*}
		\centering
		\caption{Best fit parameters defining $N_{\rm HI}$ vs $\Delta$ relationship.}
		\begin{tabular}{cccccc}
			\hline
			Redshift & $\Delta_0$ & $\eta$ &\multicolumn{2}{c}{$\Delta$ for} \\
			&     &      & $N_{\rm HI}=10^{13}$cm$^{-2}$ & $N_{\rm HI}=10^{13.5}$cm$^{-2}$\\
			\hline
			\hline
			
			1.8 ($\alpha=1.8$) & $8.5\pm 0.4$ & $0.818\pm 0.003$ & 1.294 & 3.319 \\
			2.0 ($\alpha=1.8$) & $6.0\pm 0.3$ & $0.788\pm 0.003$ & 0.979 & 2.427 \\ 
			2.0 ($\alpha=1.6$) & $6.5\pm 0.3 $ & $0.807\pm 0.004 $ & 1.014 & 2.567 \\ 
			2.0 ($\alpha=2.0$) & $5.6\pm 0.3 $ & $0.793\pm 0.004 $ & 0.902 & 2.247 \\ 
			2.0 (Enhanced $\alpha=1.8$) & $6.9\pm 0.5 $ & $0.801\pm 0.005 $ & 1.091 & 2.744 \\ 
			2.2 ($\alpha=1.8$) & $4.3\pm 0.4$ & $0.776\pm 0.006$ & 0.728 & 1.778 \\
			2.4 ($\alpha=1.8$) & $3.1\pm 0.2$ & $0.749\pm 0.005$ & 0.560 & 1.326 \\
			2.5 ($\alpha=1.8$) & $2.6\pm 0.2$ & $0.721\pm 0.006$ & 0.502 & 1.152 \\
			\hline
		\end{tabular}
		\label{N_Delta}
	\end{table*}

	\subsection{Configuration of triplet skewers}
\begin{figure}
		\hspace{-0.3cm}
		\includegraphics[viewport=130 450 550 710, width=9.6cm,clip=true]{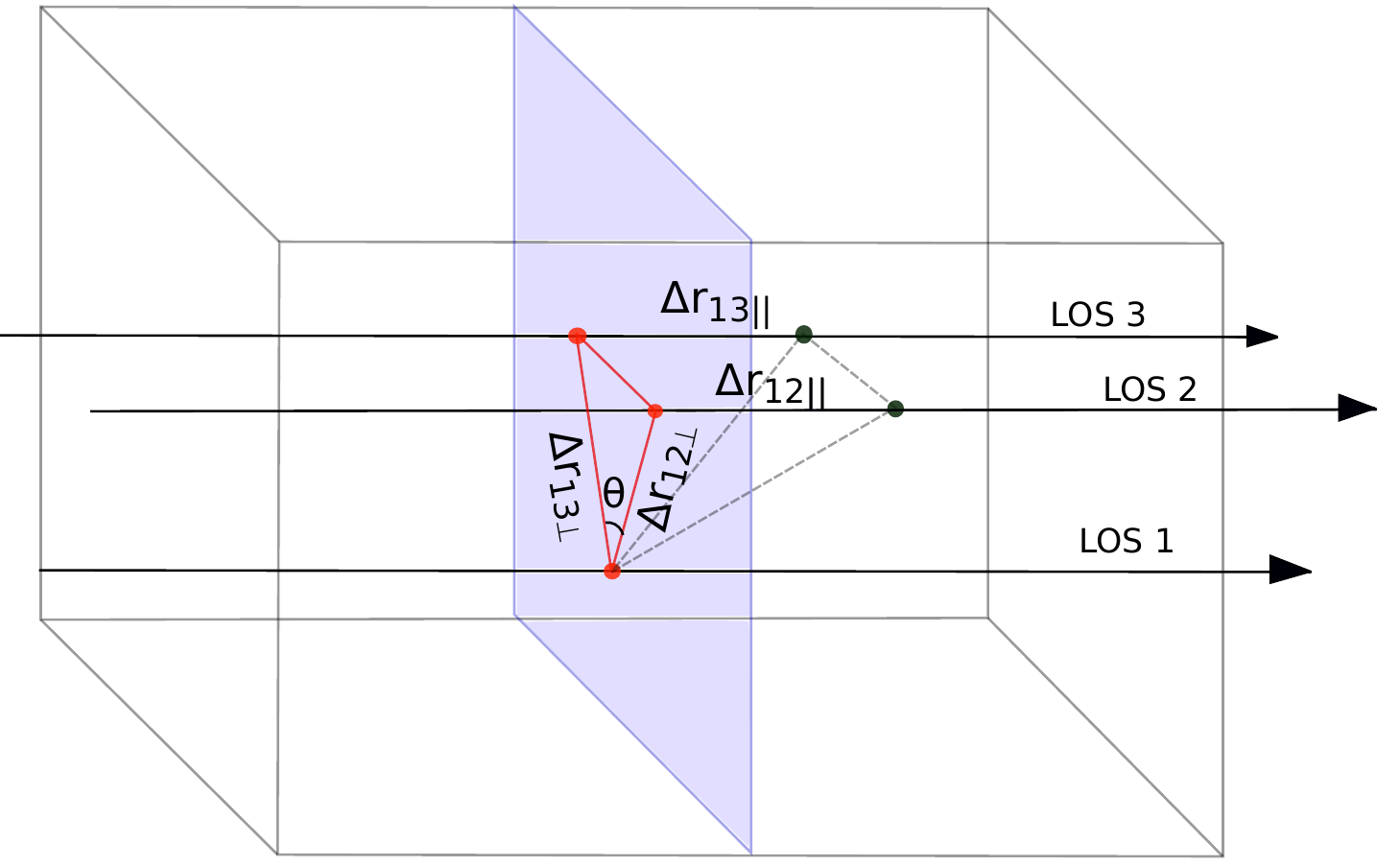}
		\caption{Schematic representation of quasar lines of sight for a triplet configuration.}
		\label{3-point_fig}
	\end{figure}

	The aim of this paper is to study transverse three-point correlations of \lya\ absorbers for various projected configurations of the background sources. For this, we shoot several triplet lines of sight through our simulation box arranged in certain triangular configuration. As shown in Fig.~\ref{3-point_fig}, in general the three-point correlation is a function of 5 variables, $\zeta=\zeta(\Delta \textbf{r}_{12\parallel},\Delta\textbf{r}_{13\parallel},\Delta\textbf{r}_{12\perp},\Delta\textbf{r}_{13\perp},\theta)$. For a fixed point in one of the lines of sight (say LOS 1), $\Delta \textbf{r}_{1i\parallel}$ denotes the longitudinal separation between the points in LOS 1 and $i^{th}$ ($i=2,3$) LOS , $\Delta \textbf{r}_{1i\perp}$ denotes the transverse separation between the points in LOS 1 and $i^{th}$ LOS and $\theta$ denotes the angle subtended by LOS 2 and 3 on LOS 1 in the sky plane. Thus, $(\Delta\textbf{r}_{12\perp},\Delta\textbf{r}_{13\perp},\theta)$ denotes the quasar triplet configuration in the sky-plane while $\Delta \textbf{r}_{1i\parallel}$ denotes the longitudinal space (i.e along the redshift) separation.
	To generate such triplet sight lines, we assume $(\Delta\textbf{r}_{12\perp},\Delta\textbf{r}_{13\perp},\theta)$ to lie on a face of the simulation box and then we shoot lines of sight through the 3 vertices in a direction perpendicular to the plane of this configuration. The position of this configuration is chosen randomly in the xy, yz or xz plane of the simulation box. For a given triplet sightlines, we will be able to construct range of triangular configuration using different combinations of transverse and longitudinal separations. For simplicity, we consider only the configurations with $\Delta\textbf{r}_{12\perp}=\Delta\textbf{r}_{13\perp}=(1,2,3,4,5)\ h^{-1}$cMpc and with $\theta=(10^{\circ},30^{\circ},60^{\circ},90^{\circ},120^{\circ},150^{\circ}$ and $170^{\circ})$. For each of these source configurations, we generate spectra along 4000 triplet sightlines of length 100$h^{-1}$cMpc.

	\section{Three-point correlation} \label{Correlation_Cloud}

	As we discussed before, first we decompose the IGM \lya\ absorption into distinct absorbers or clouds along the line of sight using an automated Voigt profile fitting routine  \citep[{\sc viper},][]{gaikwad2017b}.
	Each of these clouds represent an absorber which can be associated with a neutral hydrogen column density ($N_{\rm HI}$) and line-width parameter ($b$). Next, we construct the three-point correlation of these absorbers based on their positions (i.e, $z$). We choose absorbers having $N_{\rm HI}$ above the chosen threshold. This allows us to probe non-gaussianity in IGM clustering as a function of $N_{\rm HI}$ thresholds using the position of clouds as one does in the case of galaxies.
	
	The clustering in the data skewers generated using Voigt profile fitting of the triplet sightlines is compared with that of random distribution of clouds to quantify the three-point correlation. We define $D_1$, $D_2$ and $D_3$ as three data skewers belonging to the triplet sightlines after Voigt profile fitting and $R_1$, $R_2$ and $R_3$ as the corresponding random skewers. For a sightline, the random distribution (in position) of clouds is generated with number of clouds equal to the mean number of expected clouds having $N_{\rm HI}$ above the assumed threshold based on the known redshift distribution of clouds. These clouds are positioned randomly along the skewer based on a uniform distribution along the sightline. For each of the data skewers, the correlation is constructed by averaging over 100 random skewers. Following \citet{szapudi-szalay1998}, we use the estimator for three-point correlation as
	
	\begin{equation}\label{szapudi-szalay}
	\zeta(\Delta r_{1\parallel},\Delta r_{2\parallel})=\frac{D_1D_2D_3-DDR_{(123)}+DRR_{(123)}-R_1R_2R_3}{R_1R_2R_3} \ ,
	\end{equation}  
	where $DDR_{(123)}=D_1D_2R_3+D_1R_2D_3+R_1D_2D_3$ and $DRR_{(123)}=D_1R_2R_3+R_1D_2R_3+R_1R_2D_3$. $DDD$ is the data-data-data triplet counts measured in the longitudinal bin $(\Delta r_{1\parallel},\Delta r_{2\parallel})$, $DDR$ is the data-data-random triplet counts and likewise. As done in \citet{maitra2019}, the triplet counts are summed over $\pm 2h^{-1}$cMpc ($\sim \pm 200$\kms\ at $z=2$) along each of the longitudinal directions for the calculation of our transverse three-point correlation. This is similar to the redshift space integration done in case of galaxies  \citep[say over 20$h^{-1}$cMpc as in][]{mcbride2011b} to produce the redshift space three-point correlation function. We will discuss how the results change if we change our choice of this binning scale in the Appendix~\ref{Binning}. The triplet counts for $DDD$ are normalized by  dividing with $n(n-1)(n-2)$ where $n$ is the number of clouds along a sightline. Similar process is followed for the other triplet counts. Note that the three-point correlation measured this way is sensitive to the number of clouds along the line of sight and hence, to the line of sight length (100$h^{-1}$cMpc) assumed here. We also discuss this further in Appendix~\ref{Size_test}.
	
	We also compute the individual transverse two-points correlations associated with the 3 arms of the assumed triplet configuration. The transverse two-points correlations is calculated using the $Landy-Szalay$ estimator \citep{landy_szalay1993}. The transverse correlation between two data skewers $D_i$ and $D_j$ (i,j=1,2,3 and i$\neq$j) along two closely spaced sightlines using random skewers $R_i$ and $R_j$ is defined as 
	\begin{equation}
	\xi(\Delta r_{\parallel},\Delta r_{\perp})=\frac{D_iD_j-D_iR_j-R_iD_j+R_iR_j}{R_iR_j} \ .
	\end{equation}
	Here, DD, RR and DR are data-data, random-random and data-random pair counts respectively measured at a separation of $\Delta r_\parallel$. Similar to the three-point correlation, the counts are summed over $\pm 2h^{-1}$cMpc along each of the longitudinal directions and then normalized by dividing with $n(n-1)$.
	
	In the case of galaxies, it has been a common procedure to express the observed three-point correlation in terms of the cyclic combination of associated two-point correlation using a \textit{hierarchical ansatz} \citep{peebles1980} of the form
	
	\begin{dmath}\label{Q_eq}
		\zeta(\textbf{r}_{12\perp},\textbf{r}_{23\perp},\textbf{r}_{13\perp})=Q\ [\xi(\textbf{r}_{12\perp})\xi(\textbf{r}_{23\perp})+\xi(\textbf{r}_{23\perp})\xi(\textbf{r}_{13\perp}) +\xi(\textbf{r}_{12\perp})\xi(\textbf{r}_{13\perp})] = Q [\xi * \xi] \ . 
	\end{dmath}\label{Q_eqn}
	Here, Q, usually referred to as the "reduced three-point correlation", is a scaling quantity denoting the hierarchy existing between the three-point and two-point correlations. For galaxies, Q is found to be constant (Q $= 1.29 \pm 0.21$) and insensitive to the size or shape of the configuration of points \citep{groth1977}. In the perturbative regime, while Q is predicted to be independent of scale it has a strong dependence on the configuration \citep{fry1984}. \citet{fry1994} also showed that the shape dependence can be used to measure the galaxy distribution bias parameter 'b' independent of $\Omega$.
	
	In what follows, we calculate the transverse three-point correlation and investigate its dependence on $N_{\rm HI}$ (and hence baryonic over-density) and source configuration (scale and angle) at $z=2$. We also investigate the hierarchy between the transverse three-point correlation and the cyclic combination of the 3 associated two-point correlations (hereafter to be referred as $\xi*\xi$) of the \lya\ forest. In addition to this, we also investigate its dependence on $N_{\rm HI}$ thresholds and configuration of the triplet source. We check for dependence of the transverse three-point correlation, two-point correlation and Q on thermal history using simulations with different UVBs. As mentioned before, we calculate the three-point and two-point correlation statistics over 4000 triplet sightlines for each of the cases considered. The quoted errors are  68\% confidence interval about the mean value obtained by bootstrapping over these 4000 triplet sightlines. We calculate Q for each of the triplet sightlines and then we consider its median value over 4000 sightlines. We choose to work with median Q since Q value for individual realizations may blow up arbitrarily for a small two-point correlation and artificially enhance the mean-based statistics. We also find that median Q closely follows the "hierarchical ansatz" that one expects from the scaling relation between three-point and associated two-point correlation at large scales, as has been discussed in Sec.~\ref{zeta_vs_xisq}. Further discussions about definition of Q has been provided in Appendix~\ref{Mean_Median_q}.

	\subsection{$N_{\rm HI}$ dependence} \label{N_HI dependence}
	
	\begin{figure*}
		\centering
		
		\includegraphics[viewport=0 30 300 300, width=6cm,clip=true]{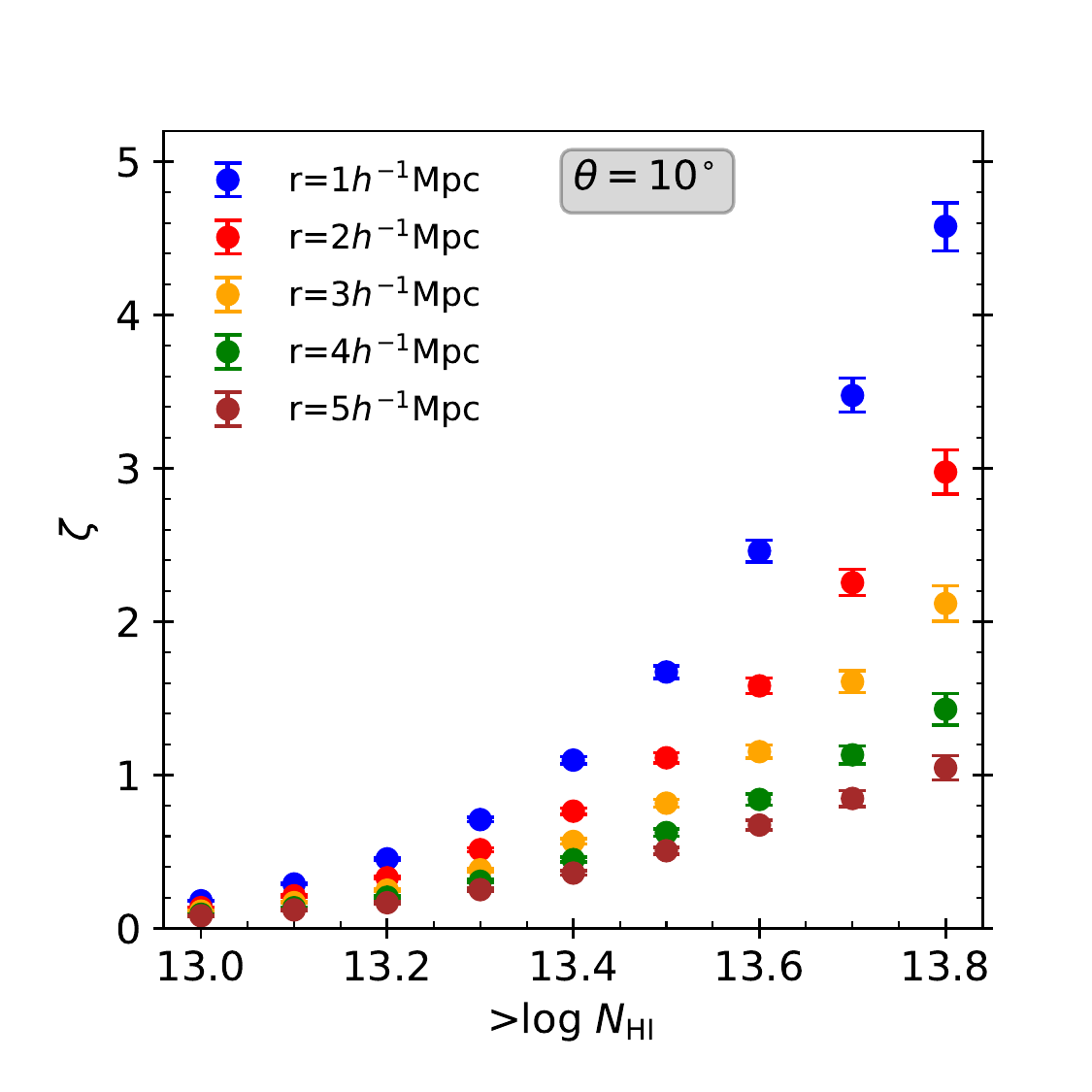}%
		\includegraphics[viewport=0 30 300 300, width=6cm,clip=true]{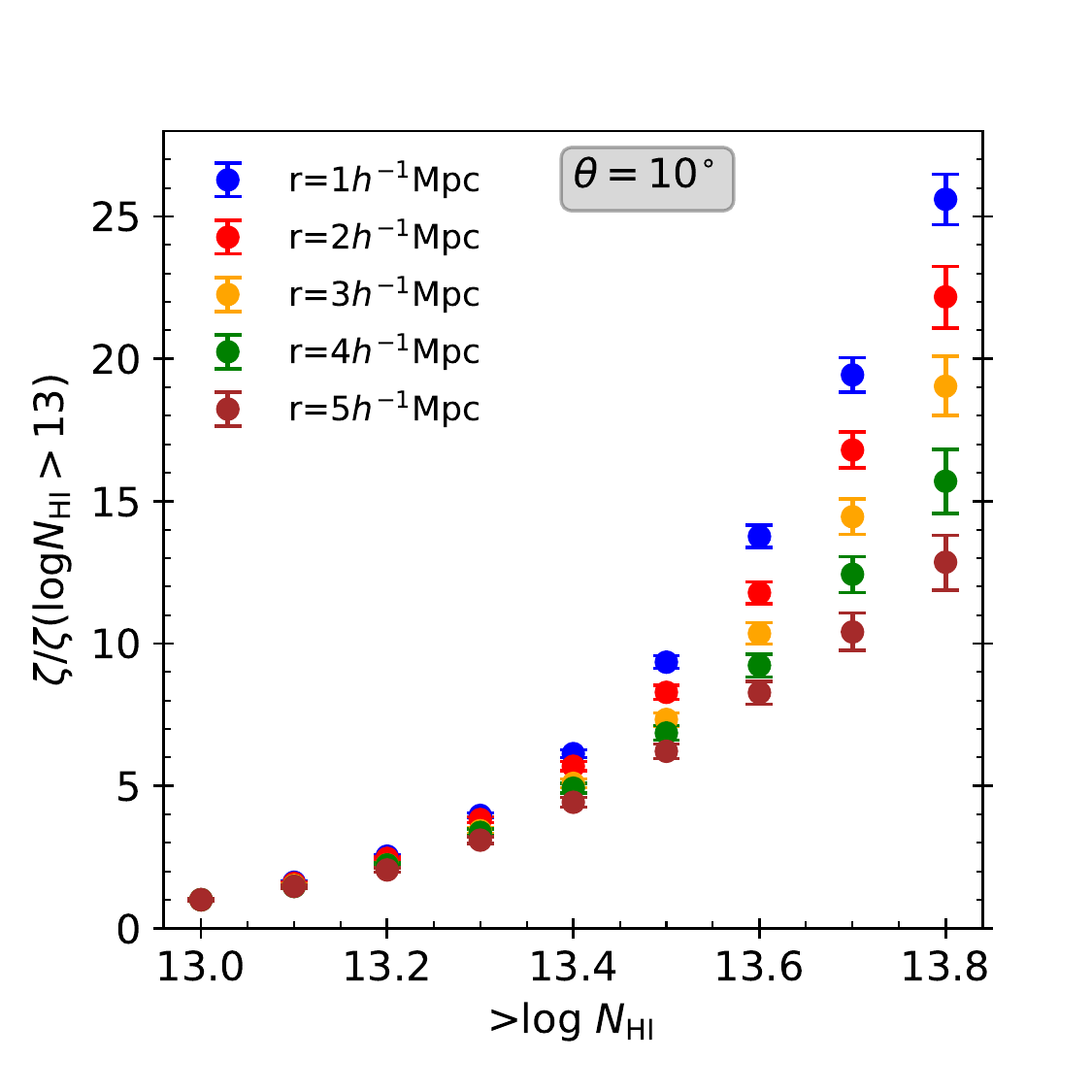}%
		\includegraphics[viewport=0 30 300 300, width=6cm,clip=true]{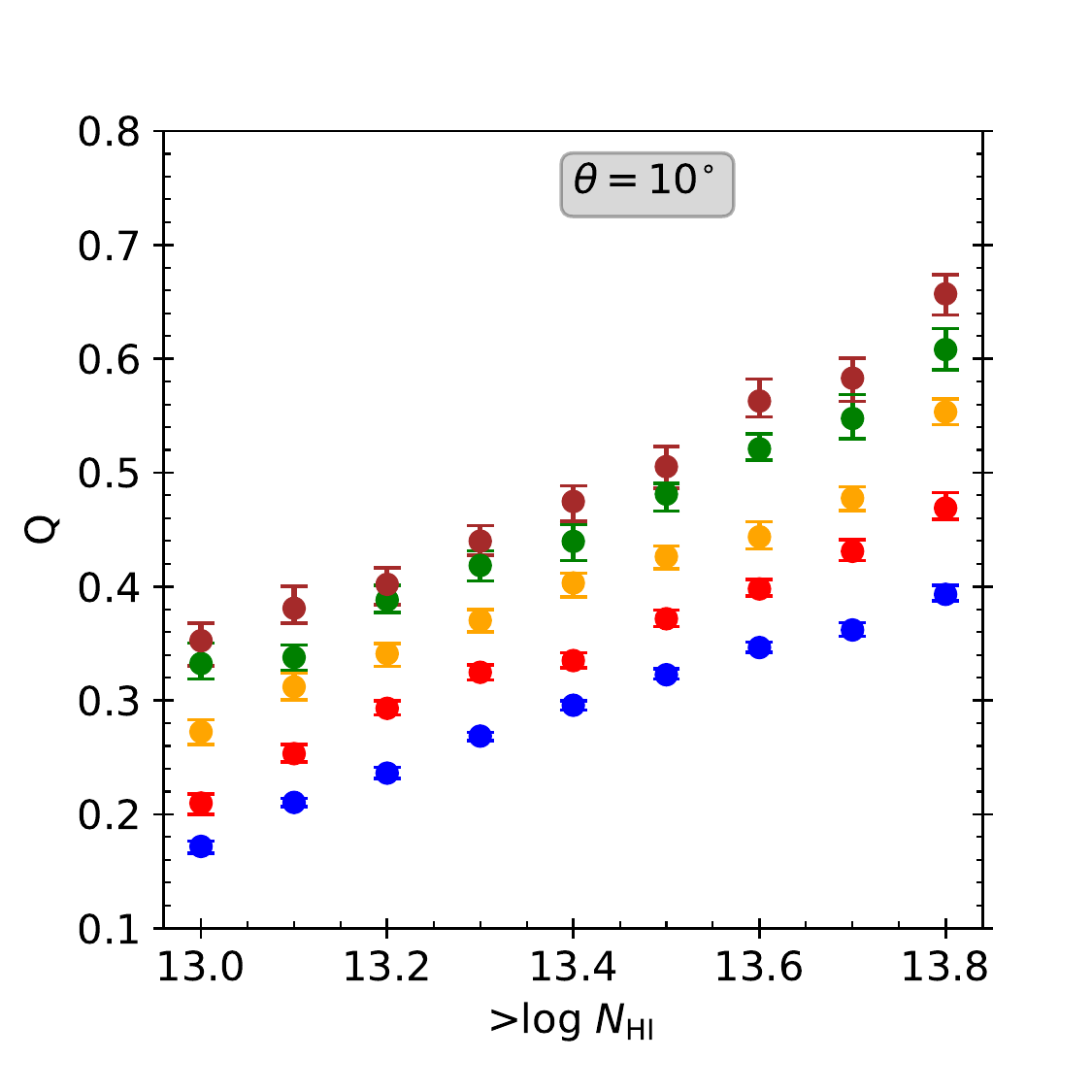}%
		
		\includegraphics[viewport=0 30 300 300, width=6cm,clip=true]{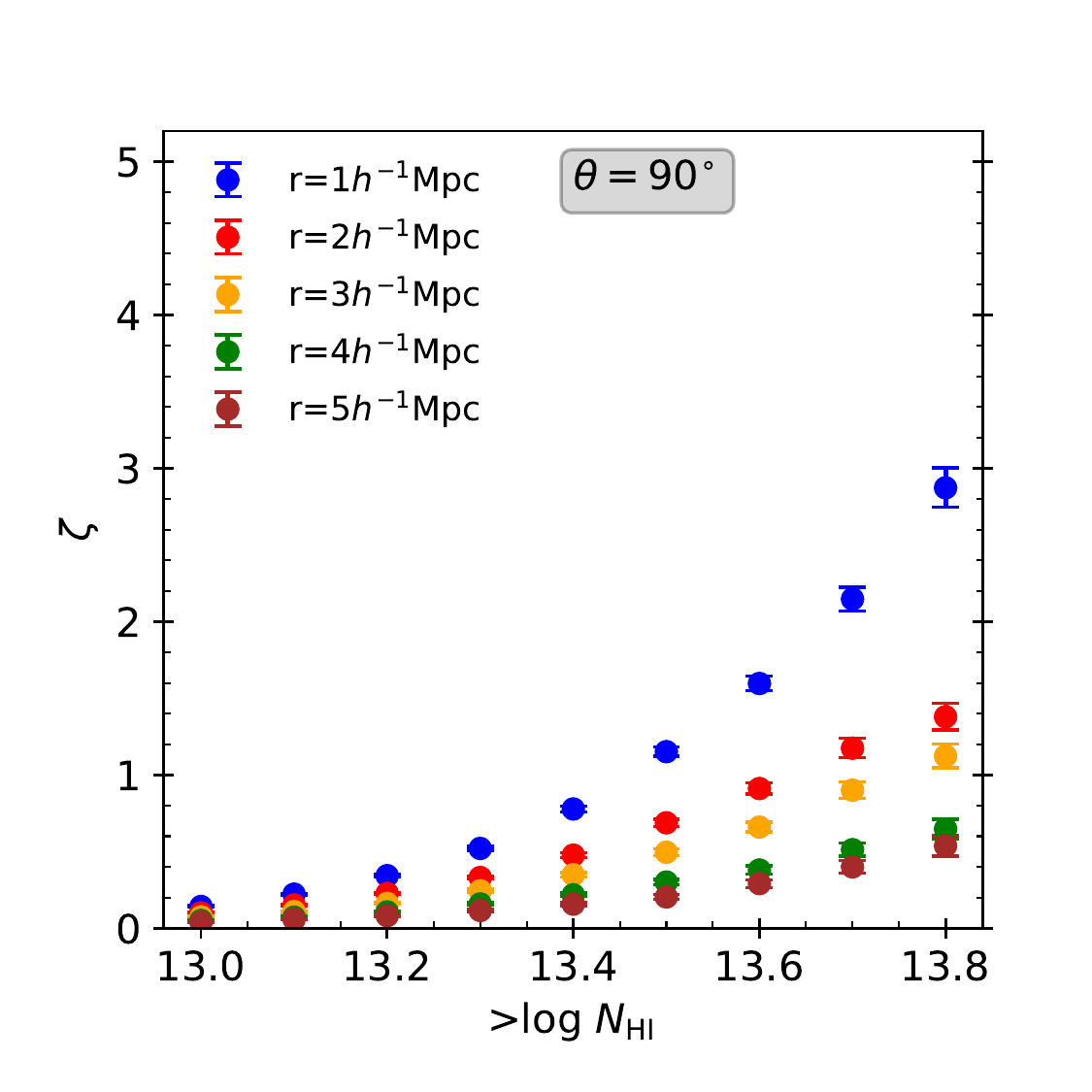}%
		\includegraphics[viewport=0 30 300 300, width=6cm,clip=true]{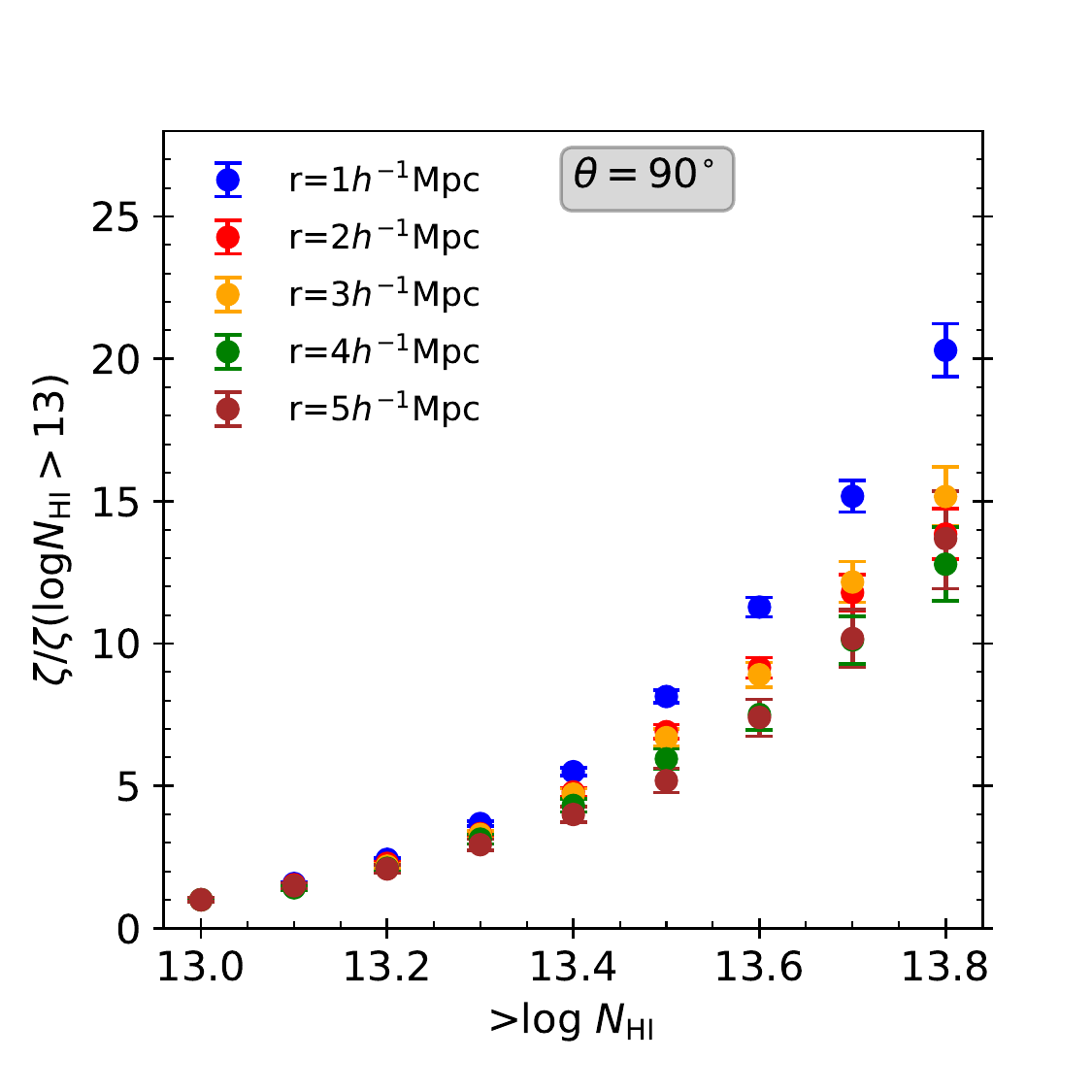}%
		\includegraphics[viewport=0 30 300 300, width=6cm,clip=true]{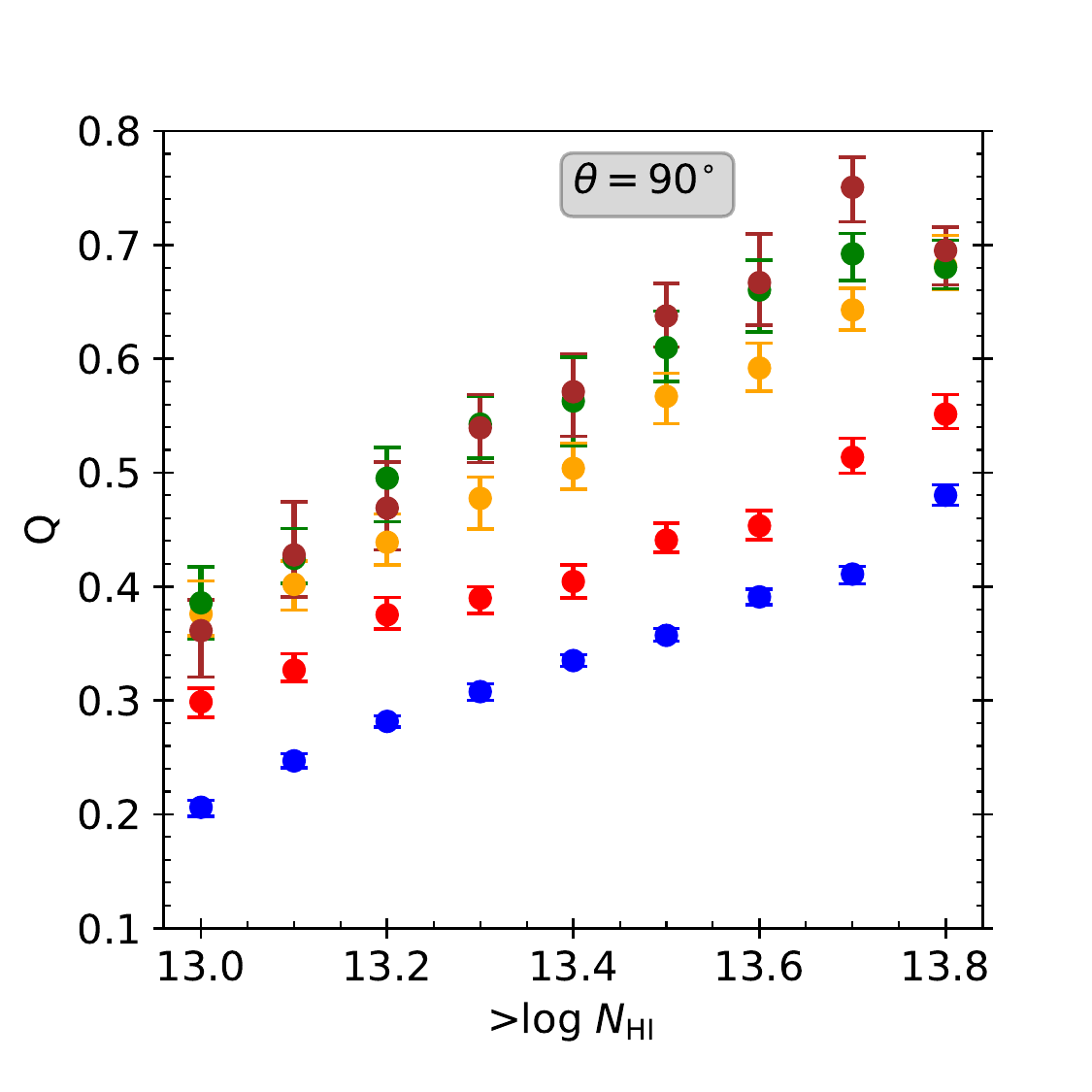}%
		
		\includegraphics[viewport=0 10 300 300, width=6cm,clip=true]{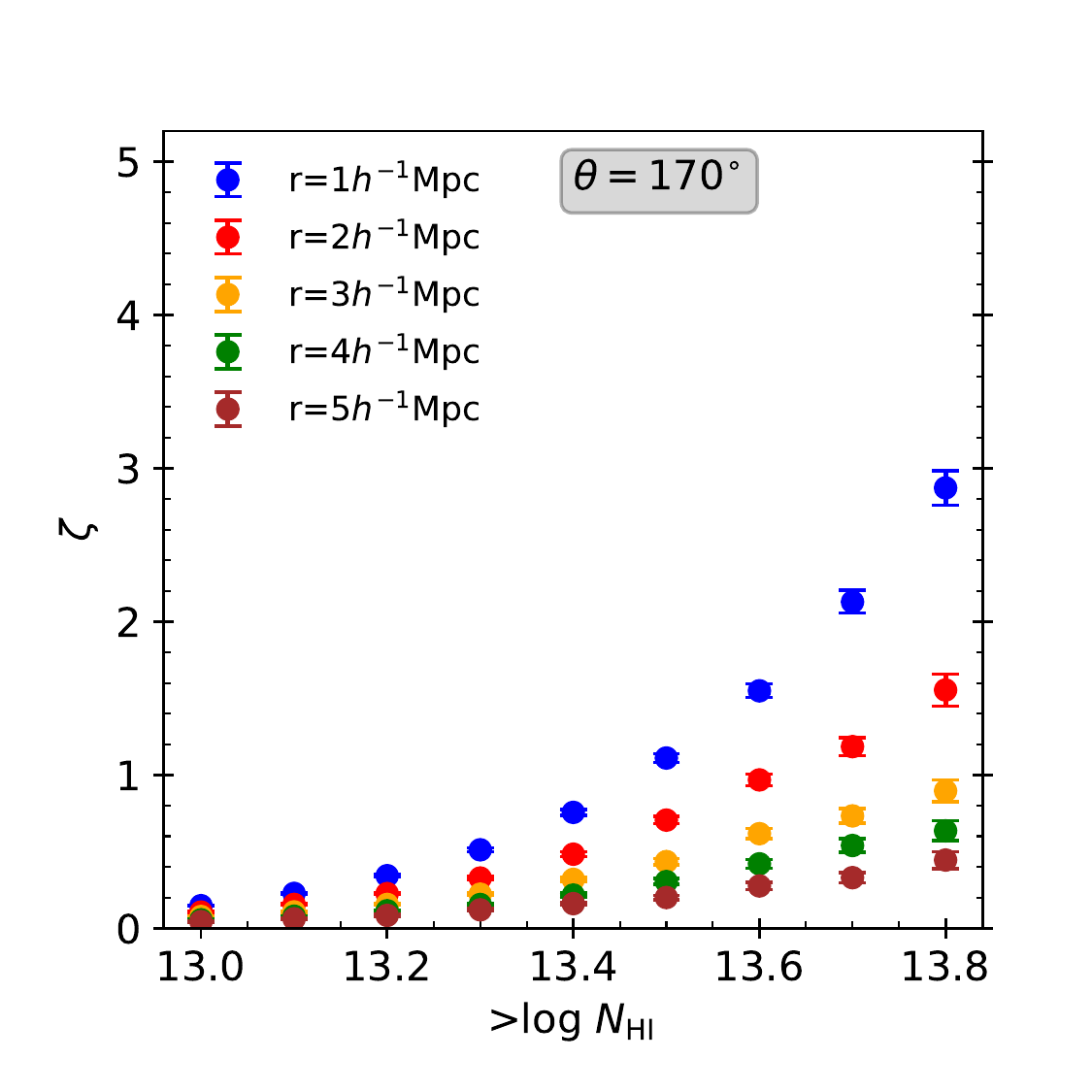}%
		\includegraphics[viewport=0 10 300 300, width=6cm,clip=true]{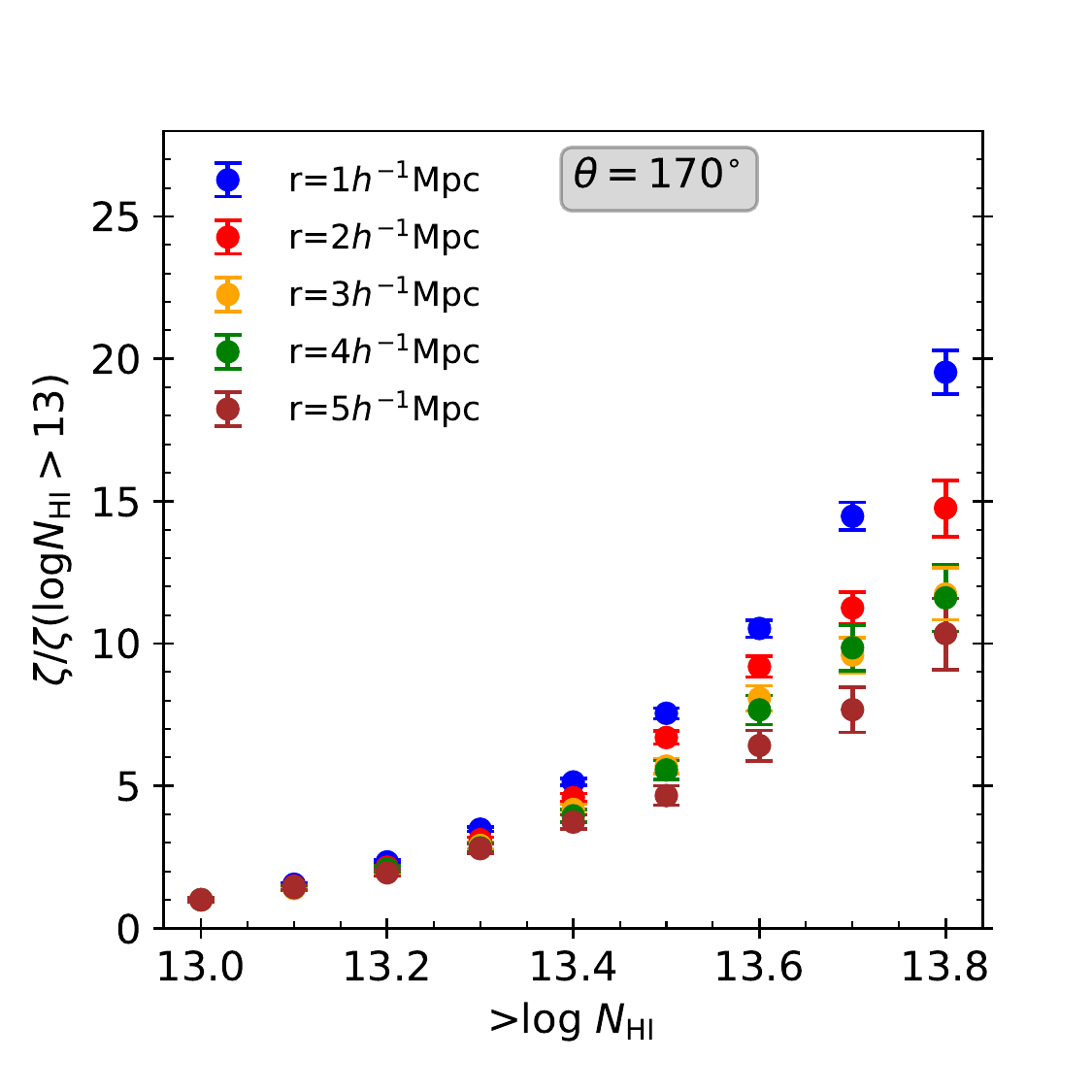}%
		\includegraphics[viewport=0 10 300 300, width=6cm,clip=true]{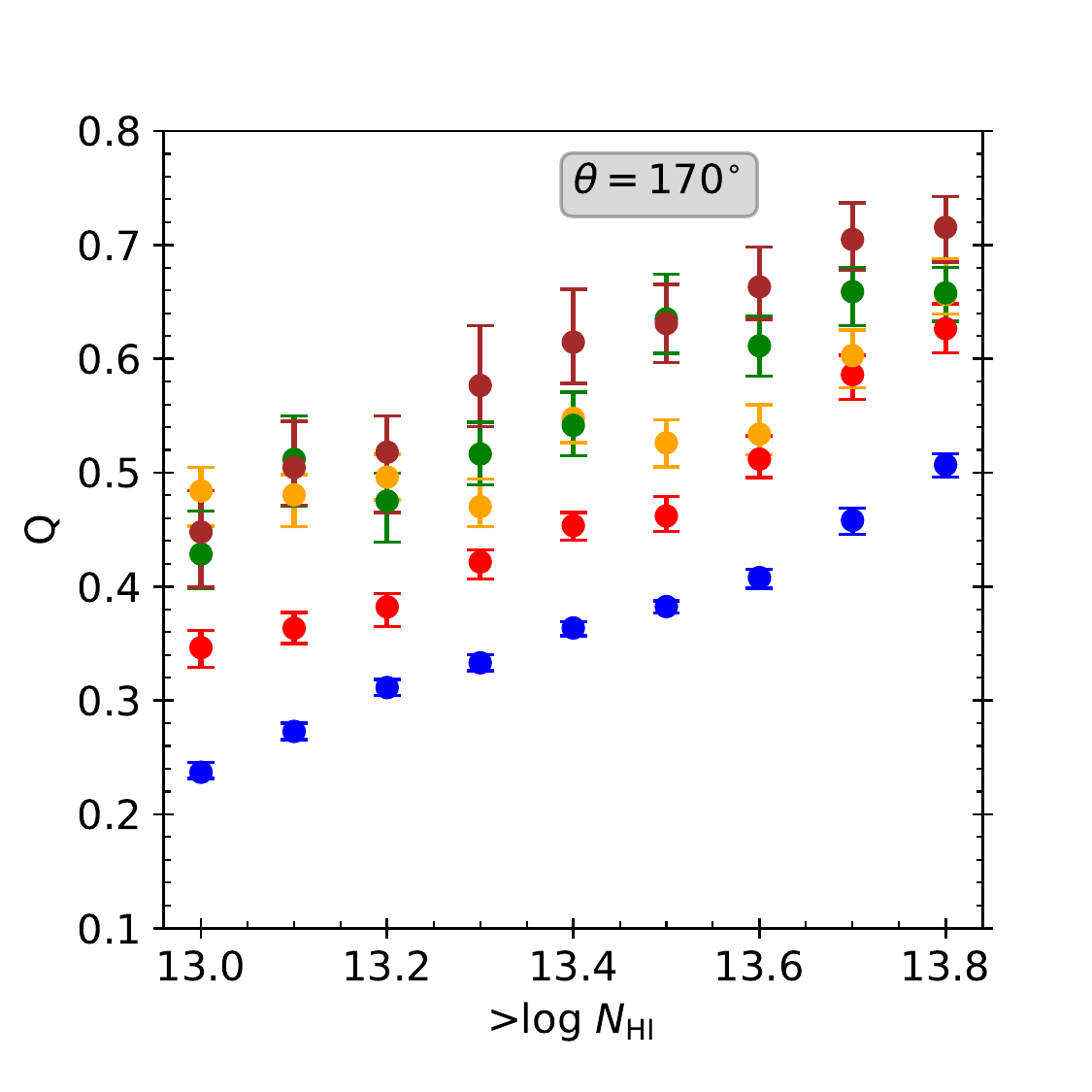}%
		
		\caption{ The left most panels show mean transverse three-point correlation functions and the middle panels show the same normalized to 1 at $N_{\rm HI}>10^{13}$cm$^{-2}$ as a function of $N_{\rm HI}$ thresholds for different scales. The right most panels show median reduced three-point correlation function Q as a function of $N_{\rm HI}$ thresholds for different scales. The correlations shown here are at $z=2$. The angle of the configurations are taken as $\theta=10^{\circ}, 90^{\circ}$ and $170^{\circ}$ for top, middle and bottom rows, respectively. Results are presented for five different scales shown in different colors.}
		\label{Corr_cloud_NHI}
	\end{figure*}
	
\begin{figure*}
		\centering
		
		\includegraphics[viewport=0 10 300 300, width=6cm,clip=true]{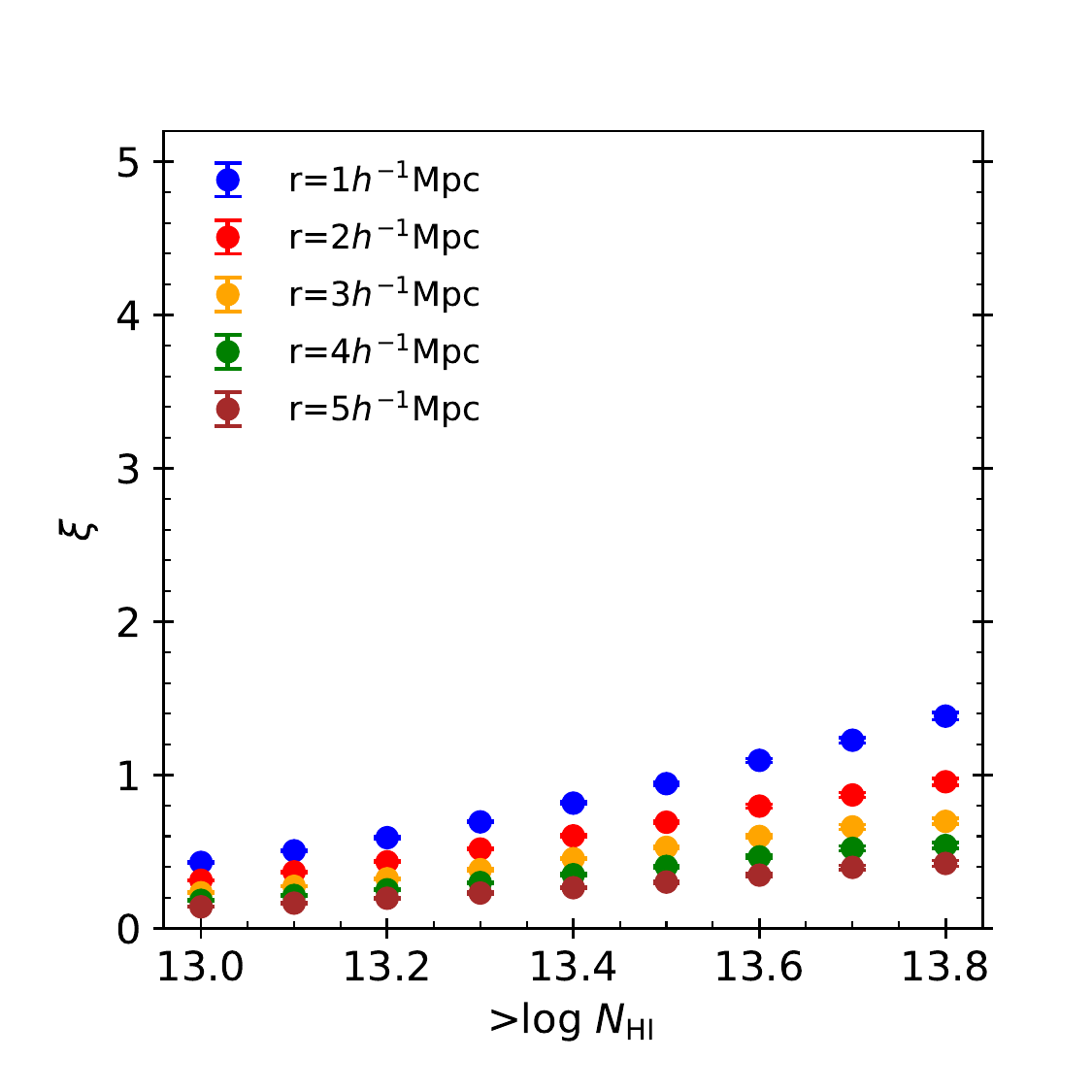}%
		\includegraphics[viewport=0 10 300 300, width=6cm,clip=true]{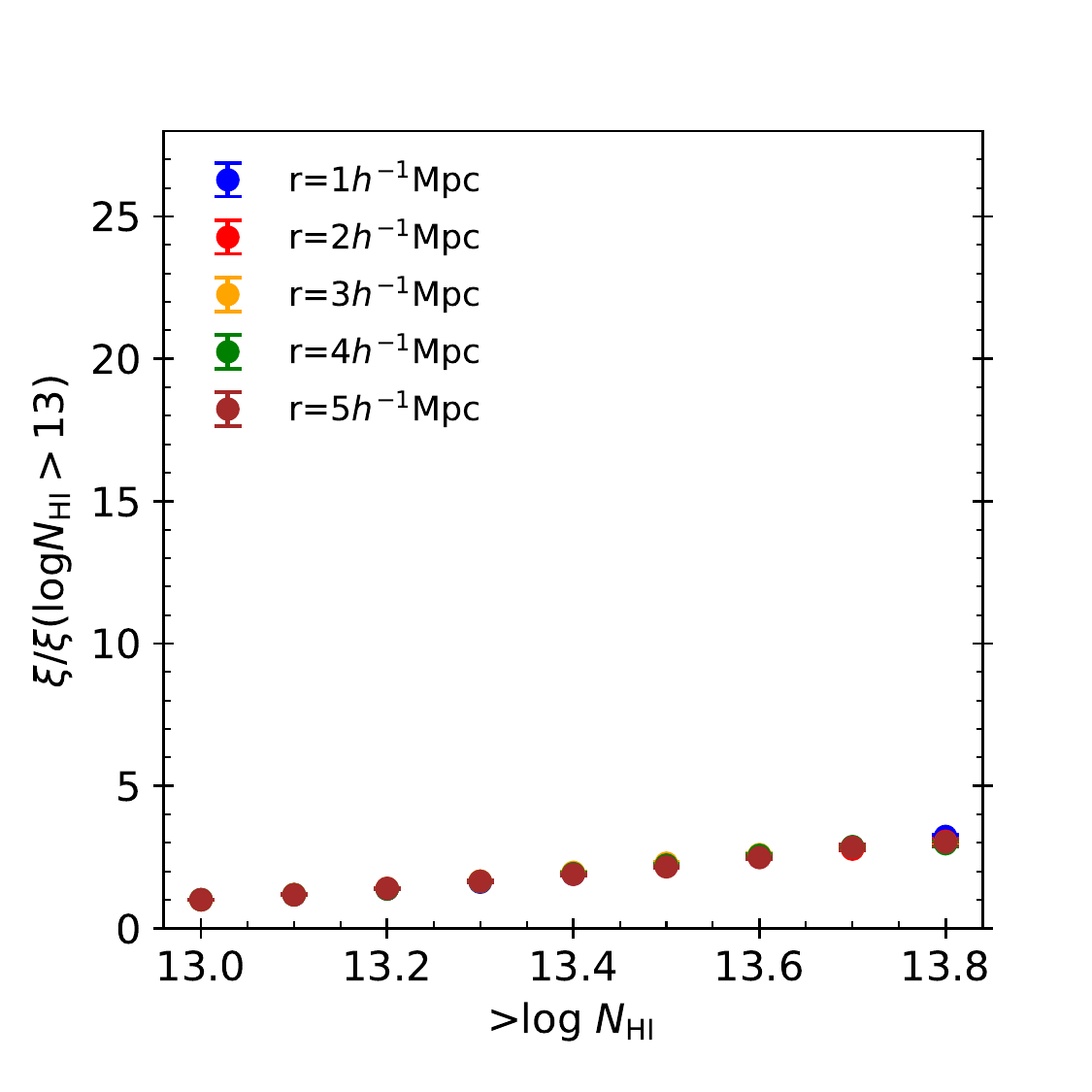}%
		
		\caption{ Plots of transverse two-point correlation (Left)  and the same normalized to 1 at $N_{\rm HI}>10^{13}$cm$^{-2}$ (Right) as a function of $N_{\rm HI}$ thresholds for five different scales.}
		\label{Corr2_cloud_NHI}
	\end{figure*}

	We calculate transverse three-point correlation of clouds having $N_{\rm HI}$ thresholds in the range of $N_{\rm HI}>10^{13}$cm$^{-2}$ to $N_{\rm HI}>10^{13.8}$cm$^{-2}$. We do not take clouds having higher $N_{\rm HI}$ thresholds since this makes the number of clouds along a sightline too less due to finite box size of our simulation. We also compute the transverse two-point correlations for these $N_{\rm HI}$ thresholds and configurations along the equidistant arms ($r=\Delta r_{12\perp}=\Delta r_{13\perp}$). The mean transverse three-point correlation for different $r$ values are plotted as a function of $N_{\rm HI}$ thresholds in the left most panels of Fig.~\ref{Corr_cloud_NHI}. The corresponding plot for the mean two-point correlations measured along the equal arms $r$ are plotted in the left panel of Fig.~\ref{Corr2_cloud_NHI}. Q is computed for each individual line of sight and the obtained median values are plotted in the right panel of Fig.~\ref{Corr_cloud_NHI}. Here we summarize our results for 5 length scales (denoted by different colour symbols in each panel) and 3 angular scales ($10^{\circ},90^{\circ}$ and $170^{\circ}$ from top to bottom). We see a positive three-point and two-point correlation  which increases monotonically with increasing $N_{\rm HI}$ thresholds for all the scales and angles considered. We also find the three-point and two-point correlation to decrease with increasing scale for a given $N_{\rm HI}$ threshold. The transverse three-point correlation becomes less than 0.1 for $\Delta r_{12\perp} \geq 5h^{-1}$cMpc, for the range of $N_{\rm HI}$ thresholds considered here.
	
	Looking specifically at the $N_{\rm HI}$ dependence of transverse three-point and two-point correlation, we see a stronger $N_{\rm HI}$ dependence of three-point correlation compared to two-point correlation at all scales and angles given by a steeper slope. To visualize this better, we have plotted three-point and two-point correlations normalized to 1 for log $N_{\rm HI}$>13 sample as a function of $N_{\rm HI}$ thresholds in the middle column of Fig.~\ref{Corr_cloud_NHI} and right panel of Fig.~\ref{Corr2_cloud_NHI} respectively. \textit{Interestingly, the $N_{\rm HI}$ dependence of the normalized two-point correlation does not change with scale. In case of three-point correlation though, the $N_{\rm HI}$ dependence is stronger at smaller scales compared to larger scales irrespective of the angle.} At the smallest scale of 1$h^{-1}$cMpc probed here, we see a very sharp increase of normalised three-point correlation with increasing $N_{\rm HI}$ thresholds. This dependence weakens as one goes to higher scales. We also notice that this dependence flattens for $\theta\geq 90^{\circ}$ for high $N_{\rm HI}$ thresholds. Thus the three-point correlation gets amplified much sharply with increasing $N_{\rm HI}$ thresholds as one goes to smaller scales.
	
	From the right panels of Fig.~\ref{Corr_cloud_NHI}, we see the inferred Q strongly depends on the $N_{\rm HI}$ threshold used (it increases with increasing $N_{\rm HI}$). The Q values obtained here (in the range of 0.2-0.7) are less than 1.29 found for galaxies and seem to increase with increasing length scales. There are also indications for slight increase in Q values with increasing $\theta$ for a given $N_{\rm HI}$ threshold and scale. We explore this further in Sec.~\ref{Angular_dependence}. For a given $\theta$, we find the $N_{\rm HI}$ threshold dependence of Q is much weaker than that of $\zeta$. These findings are consistent with the finding that dependence of $\zeta$ for galaxies on stellar mass and luminosity is stronger than that of Q. In the case of galaxies, it is understood as the effect of stronger dependence of $\zeta$ on bias parameter compared to Q.
	
\begin{figure}
		\centering
		\includegraphics[viewport=0 0 300 260, width=7.2cm,clip=true]{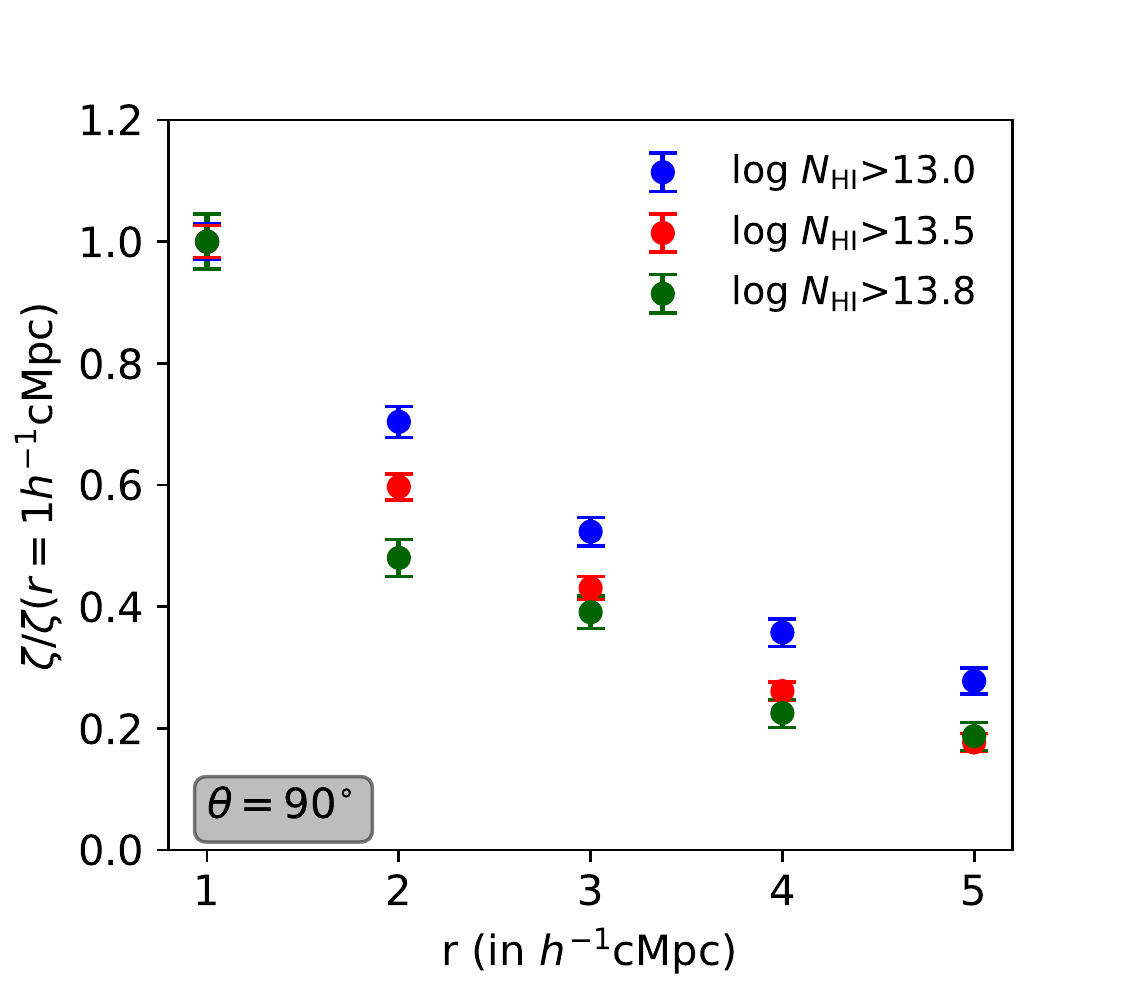}
		\includegraphics[viewport=0 0 300 260, width=7.2cm,clip=true]{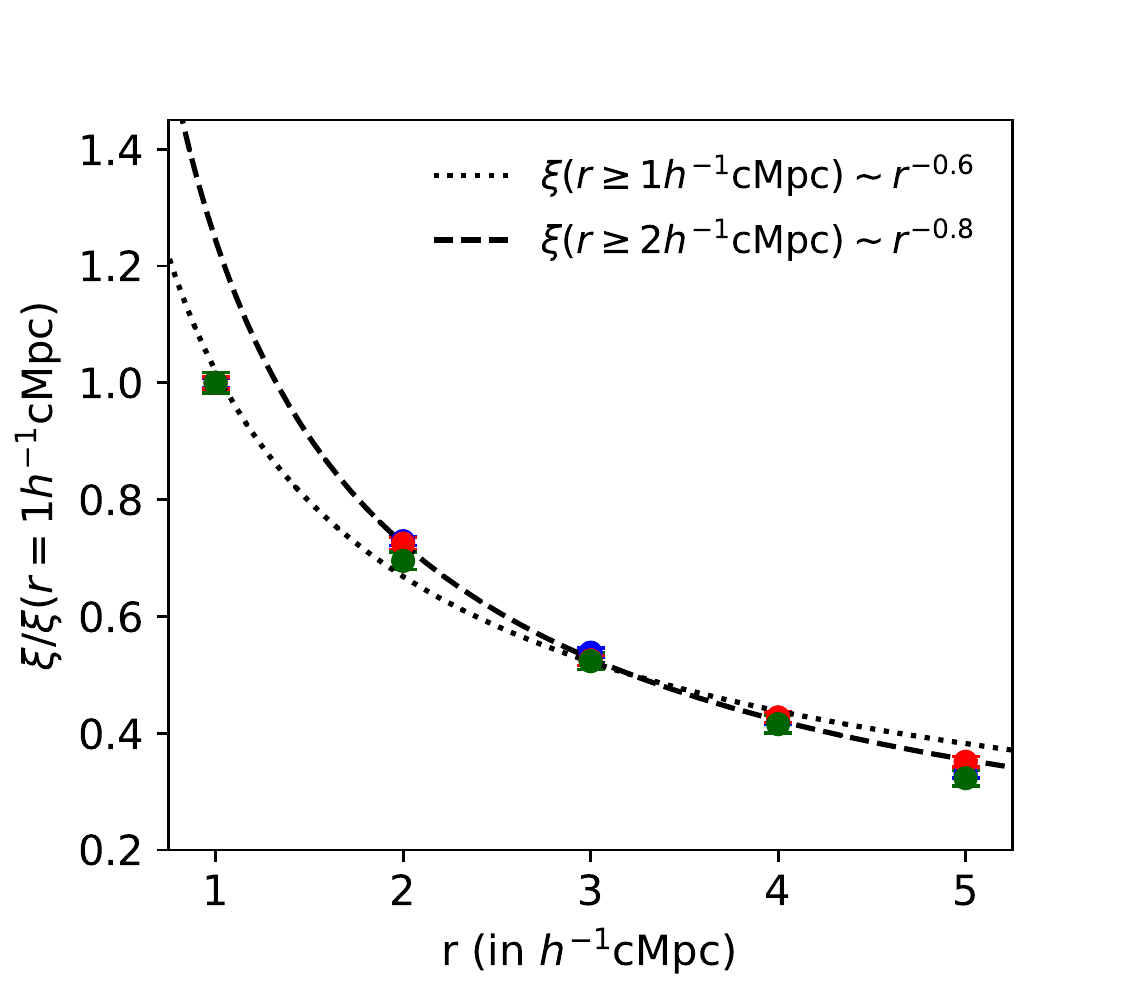}
		
		\caption{Plots of transverse three-point correlation (top panel) and transverse two-point correlation (bottom panel) as a function of scale for the configuration $\theta=90^{\circ}$ at $z=2$. The three-point and two-point correlations have been normalized to 1 at r=$1h^{-1}$cMpc. The plots have been shown for three $N_{\rm HI}$ thresholds.}
		\label{Corr_cloud_scale_NHI}
	\end{figure}

	Physically, two-point correlation probes spatially averaged radial density profile of matter probed by the \lya\ forest. For different $N_{\rm HI}$ thresholds, one can define such a radial profile. The normalized two-point correlation vs $N_{\rm HI}$ threshold (in right panel of Fig.~\ref{Corr2_cloud_NHI}) being independent of scale implies that the shape of average radial density profile does not change as a function of $N_{\rm HI}$ threshold (or baryonic over-density, see Eq.~\ref{delta-N}). Instead, it just gets uniformly amplified at all scales. This is shown in the bottom panel of  Fig.~\ref{Corr_cloud_scale_NHI}. We approximate $\xi(r) \propto r^{-\beta}$ for $r$ in the range 1-5$h^{-1}$cMpc and found $\beta\sim 0.65$. It is clear from Fig.~\ref{Corr_cloud_scale_NHI}, that the single power-law fit may not be the good representation of $\xi(r)$. Much better fit is obtained if we ignore $r=1h^{-1}$cMpc point with $\beta\sim 0.87$. The steep radial profile is what one expects at large scale with smoothing of the density field at smaller scales due to pressure smoothing effects. Note that a steeper profile of $\beta\sim 1.77$ is found for galaxies (\citet{groth1977}). 
	
	\begin{figure}
		\centering
		
		\includegraphics[viewport=5 0 300 290, width=7.3cm,clip=true]{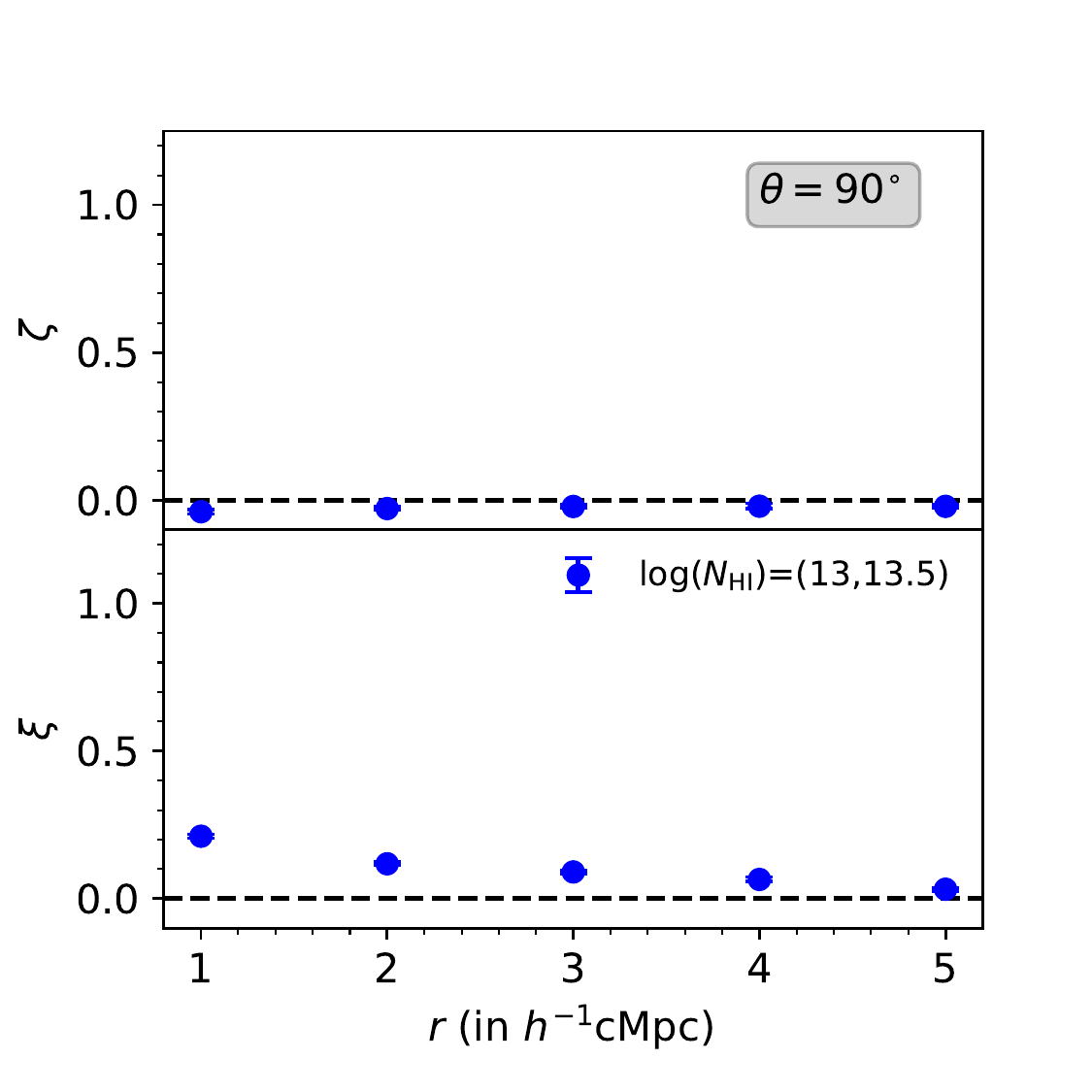}%
		\caption{ Plots of transverse three-point (Top) and two-point (Bottom) correlations for clouds having  $10^{13}$cm$^{-2} < N_{\rm HI}<10^{13.5}$cm$^{-2}$ as a function of scale for $\theta=90^{\circ}$ at $z=2$.}
		\label{Corr_cloud_NHI_13-13.5}
	\end{figure}
	
	One thing to notice in Fig.~\ref{Corr_cloud_NHI} is that while transverse three-point correlation is stronger for clouds with $N_{\rm HI}>10^{13.5}$cm$^{-2}$ than those with $N_{\rm HI}>10^{13}$cm$^{-2}$ by roughly 10 times at 1$h^{-1}$cMpc and $\theta=10^{\circ}$, for the two-point correlation shown in Fig.~\ref{Corr2_cloud_NHI} the difference is much weaker (roughly 2 times). So, the obvious question to ask is what brings the correlation down so strongly in the case of three-point correlation as one lowers the $N_{\rm HI}$ thresholds. To understand this stronger dependence of three-point correlation on $N_{\rm HI}$ thresholds, we investigate the clustering of clouds having column densities in the range $10^{13}$cm$^{-2} < N_{\rm HI}<10^{13.5}$cm$^{-2}$ in Fig.~\ref{Corr_cloud_NHI_13-13.5} for $\theta=90^{\circ}$ configuration. We plot the transverse three-point (top panel) and two-point (bottom panel) correlation for these clouds as a function of scale. It is evident from the plot that clouds in the above mentioned $N_{\rm HI}$ range have a negligible three-point correlation at all scales albeit being close to zero (slightly negative at small scales). This result is found to be valid even for other angles. The negligible and slightly negative three-point correlation in this $N_{\rm HI}$ range is what brings the total three-point correlation down strongly for $N_{\rm HI}>10^{13}$cm$^{-2}$ thresholds. Note when we consider $N_{\rm HI}>10^{13}$cm$^{-2}$ clouds, there are roughly equal number of clouds with $N_{\rm HI}$ above and below $10^{13.5}$cm$^{-2}$. What is making the case interesting is the fact that the two-point correlations for clouds with $10^{13}$cm$^{-2} < N_{\rm HI}<10^{13.5}$cm$^{-2}$ are non-zero and positive. So, the two-point correlation is not as strongly reduced when one goes from $N_{\rm HI}>10^{13.5}$cm$^{-2}$ to $N_{\rm HI}>10^{13}$cm$^{-2}$. One thing to be kept in mind is that in this column density range, one is probing around the mean IGM over-density ($N_{\rm HI}=10^{13}$cm$^{-2}$ corresponds to $\Delta\sim 1$, refer to Table.~\ref{N_Delta}). These mean density clouds are weakly correlated with each other indicated by a small positive two-point correlation while having negligible three-point correlation. Therefore, sharp decrease in $\zeta$ when we lower the $N_{\rm HI}$ threshold can also be attributed to the dilution effect produced by the low $N_{\rm HI}$ clouds.

	\subsection{Scale dependence}\label{Scale_dependence}
	
	\begin{figure*}
		
		\includegraphics[width=15cm]{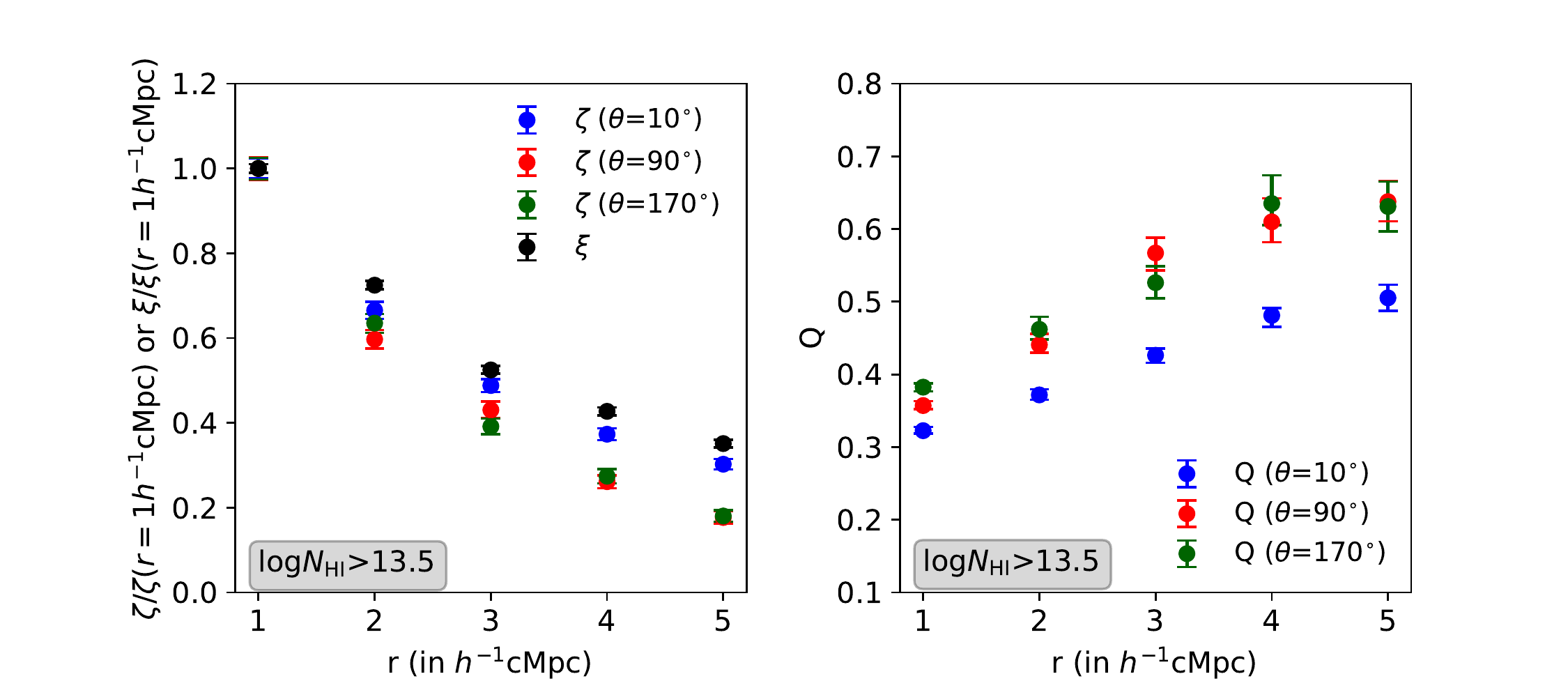}%
		\caption{Left panel: Transverse three-point correlation (or two-point correlation for the equal arm) normalized to 1 at r=1$h^{-1}$cMpc as a function of scale. Right panel: Reduced three-point correlation as a function of scale three different $\theta$ as shown. The correlations have been plotted for configurations having $\theta=10^{\circ},90^{\circ}$ and $170^{\circ}$ at $z=2$.}
		\label{zeta_norm_vs_r}
	\end{figure*}
	
	In Sec.~\ref{N_HI dependence}, we had already seen that transverse three-point correlation decreases monotonically with increasing scale. In Fig.~\ref{zeta_norm_vs_r}, we plot transverse three-point (or two-point) correlation normalized to 1 at r=1$h^{-1}$cMpc as a function of transverse scale for $N_{\rm HI}>10^{13.5}$cm$^{-2}$ (left panel). We consider three configurations with $\theta=10^{\circ},\  90^{\circ}$ and $170^{\circ}$. We notice a steeper profile of three-point correlation with scale, as compared to the two-point correlation. Also, the $\zeta$ profile is similar for the cases with $\theta=90^{\circ}$ and $170^{\circ}$, but these are shallower compared to $\theta=10^{\circ}$. For $\theta=90^{\circ},170^{\circ}$, the three-point correlation at 5$h^{-1}$cMpc falls to 10\% of its value at 1$h^{-1}$cMpc while for $\theta=10^{\circ}$, it goes to 25\%. This suggests the possible importance of configuration for the three-point correlation (see Sec.~\ref{Angular_dependence}). Also we need to remember for our choice of $\Delta r_{12\perp}=\Delta r_{13\perp}$ small angles will mean $\Delta r_{23\perp}$ probing stronger two-point correlation.
	
	We also plot the scale dependence of median reduced three-point correlation Q for $N_{\rm HI}>10^{13.5}$cm$^{-2}$ (right panel in Fig.~\ref{zeta_norm_vs_r}) with $\theta=10^{\circ}, 90^{\circ}$ and $170^{\circ}$ configurations. It is evident that for a given $\theta$ and $N_{\rm HI}$ threshold, Q increases slowly with increasing scale. It is also interesting to note Q as a function of $r$ is nearly identical for $\theta=90^{\circ}$ and $\theta=170^{\circ}$ at large $r$ values. We see Q to be smallest for $\theta=10^{\circ}$ and nearly same for $\theta=90^{\circ}$ and $\theta=180^{\circ}$ for any given scale. We explore this angular dependence further in Sec.~\ref{Angular_dependence}.
	
	\subsection{Angular dependence}\label{Angular_dependence}
	
	\begin{figure*}
		\centering
		\includegraphics[viewport=40 30 1200 320,width=20cm, clip=true]{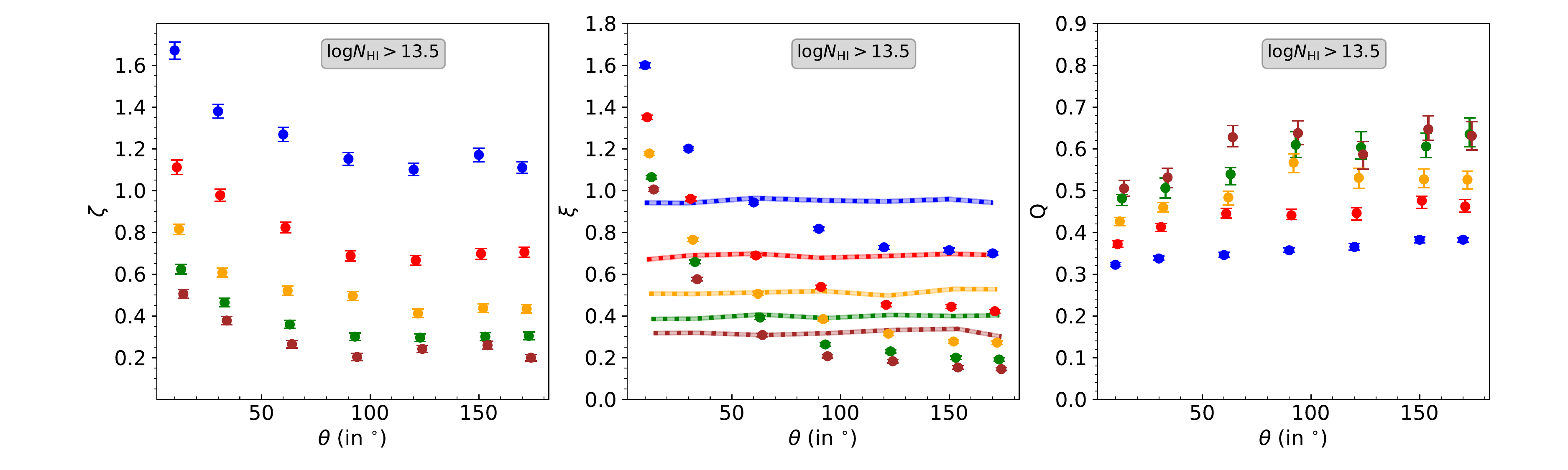}
		
		\includegraphics[viewport=40 10 1200 320,width=20cm, clip=true]{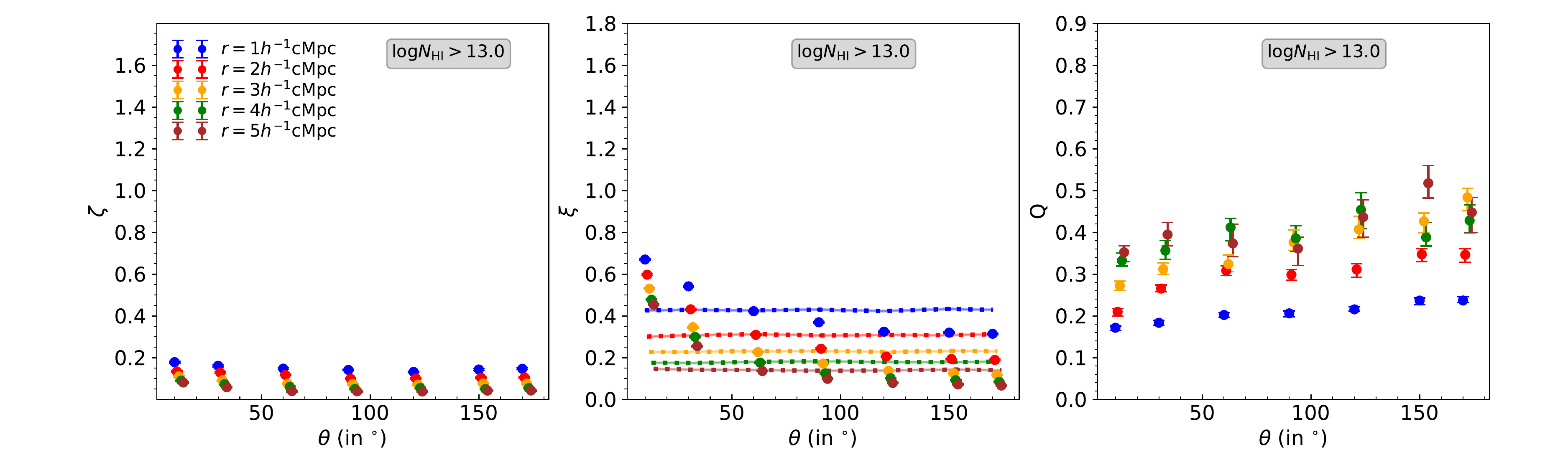}
		\caption{Angular dependence of transverse three-point correlation (left panels), the associated two-point correlation (middle panels) and the reduced three-point correlation (right panels) functions for neutral hydrogen column density threshold $N_{\rm HI}>10^{13.5}$cm$^{-2}$ (top panels) and $N_{\rm HI}>10^{13}$cm$^{-2}$ (bottom panels) at scales 1-5$h^{-1}$cMpc and at $z=2$.}
		\label{Corr_cloud_angle}
	\end{figure*}
	
	The left column panels in Fig.~\ref{Corr_cloud_angle} show the transverse three-point correlation of clouds for different triplet source configurations and with different column density thresholds [$N_{\rm HI}>10^{13.5}$cm$^{-2}$ (\textit{Top panel}) and $N_{\rm HI}>10^{13}$cm$^{-2}$ (\textit{Bottom panel})] as a function of the angle of the configuration at $z=2$. For $N_{\rm HI}>10^{13.5}$cm$^{-2}$, we see clear angular dependence of three-point correlation, where it decreases up to $\theta=90^{\circ}$ and then flattens out (i.e it becomes weakly dependent on $\theta$). As discussed before, the signals are much weaker in case of $N_{\rm HI}>10^{13}$cm$^{-2}$.
	
	In the middle panels of Fig.~\ref{Corr_cloud_angle}, we plot the corresponding transverse two-point correlation functions for the triplet source (one corresponding to the equal arms which are $\theta$ independent and the other corresponding to the third arm which is $\theta$ dependent). It is seen that below $60^{\circ}$, as expected, the two-point correlation in the third arm of the triplet is stronger compared to the other two equal sized arms. Purely based on this, we expect a stronger three-point correlation at small angles (i.e. $\theta<60^{\circ}$), if we assume a constant Q. Beyond $\theta=60^{\circ}$, the two-point correlation in the third arm begins to get weaker compared to the other two arms. So, the angular dependence of three-point correlation is expected to get weaker beyond $\theta=60^{\circ}$ if Q is $\theta$ independent.
	However, this expectations for the transverse three-point correlation are contrary to our finding that there is a flattening in $\zeta$ beyond certain angle. Upon careful observation, we can even see a very small increase in three-point correlation in going to $170^{\circ}$ at $r=1$ and $2h^{-1}$cMpc. 
	
	We also plot the corresponding reduced three-point correlation Q in the right panels of Fig.~\ref{Corr_cloud_angle}. Compared to $N_{\rm HI}$ thresholds and scale, we find that Q has a much weaker dependence on $\theta$. However, we do see a steady rise in Q with increasing angular scale. In the galaxy literature, Q at  $\theta\sim 0^{\circ}$ and $\theta\sim 180^{\circ}$ are thought to be influenced by the linear structures and $\theta=90^{\circ}$ (right angled configurations) are thought to probe more spherical distribution. Thus, the angular dependence of Q is found to be either U-shaped (typically at small scales) or V-shaped curve (typically at large scales). Recently, \citet{moresco2017} have found in their {\sc VIPERS} survey of galaxies at $0.5<z<1.1$ that $\theta$ dependence of Q evolves with redshift for largest scale considered in their study. In their highest redshift bin, the Q vs $\theta$ does not show a U or V shape, rather a slowly increasing function with $\theta$ even at large scales. They interpreted this as an indication of buildup of filaments with cosmic time, that enhances the $\zeta$ in the elongated configuration ($\theta\sim 0^{\circ}$ or $180^{\circ}$) while reducing it in the equitorial configuration ($\theta=90^{\circ}$) as one goes towards lower $z$. Lack of strong dependence of Q on $\theta$ could mean equal distribution of elongated and spherical configurations in the IGM at $z\sim 2$ over the scales we have probed here. In addition, the reduction in Q at small $\theta$, in particular for small $r$ may also come from pressure smoothing effects, that are important in the case of IGM.

	\subsection{Exploring the hierarchy between three-point and two-point correlation}\label{zeta_vs_xisq}

	\begin{figure*}
		\centering
		\includegraphics[viewport=50 30 790 620,width=16cm, clip=true]{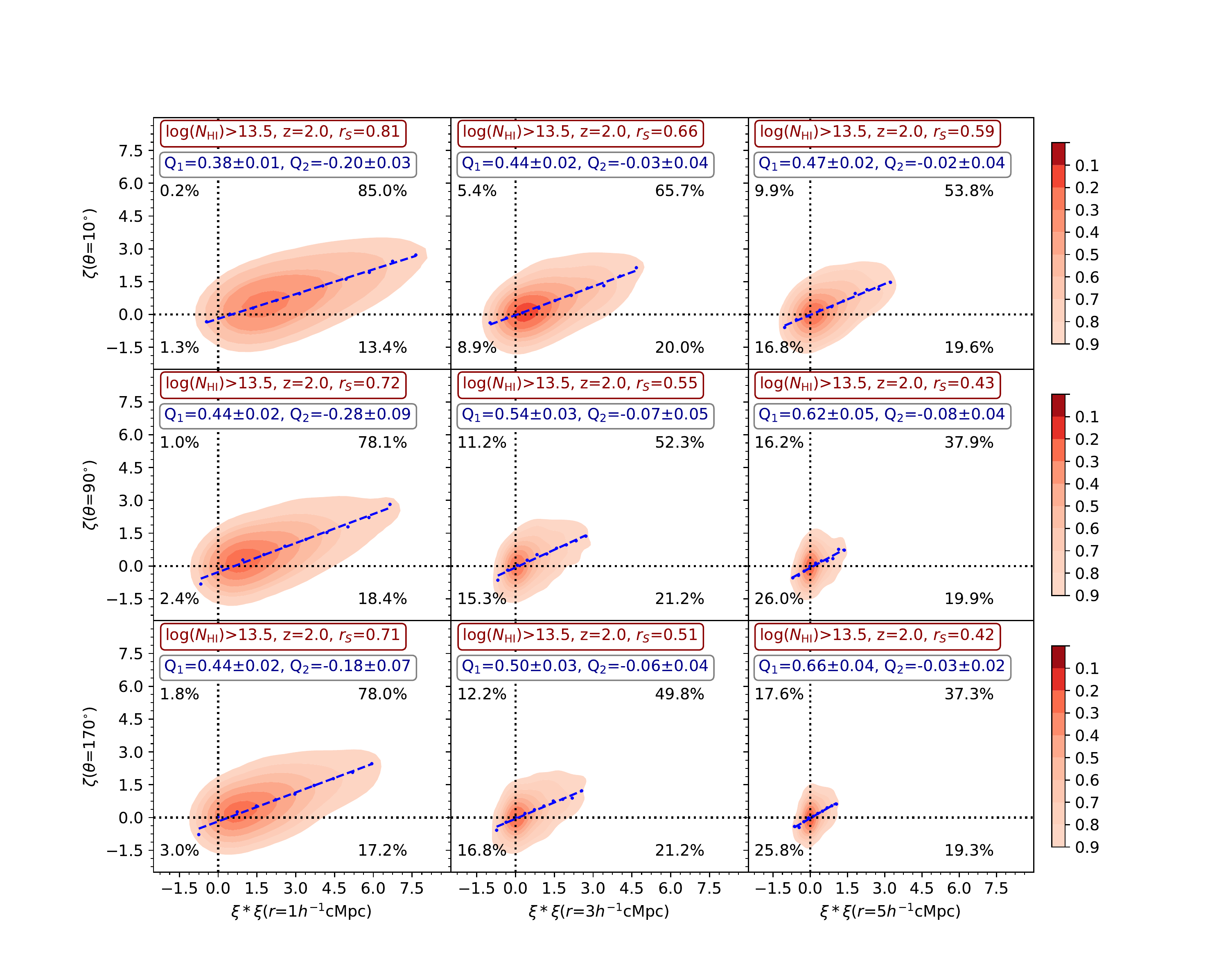}
		\caption{$\zeta$ vs $\xi * \xi$ plots for different configurations of clouds having $N_{\rm HI}>10^{13.5}$cm$^{-2}$ at $z=2$. The plots are made in a $3\times 3$ grid. We vary the scale along the plots in horizontal direction from 1 to 5$h^{-1}$cMpc and angle along the plots in vertical direction from $10^{\circ}$ to $170^{\circ}$. $r_S$ is the Pearson correlation coefficient between $\zeta$ and $\xi * \xi$ for each configuration. The blue dashed line is the linear fit to $\zeta$ vs $\xi * \xi$ and $\rm Q_1$ and $\rm Q_2$ are the slope and y-intercept of the linear fit. We also show the percentage out of the total sightlines in each quadrant of $\zeta$ vs $\xi * \xi$ plot.}
		\label{zeta_vs_xisq_scale}
	\end{figure*}

	Before going any further, it will be important to check the validity of the "hierarchical ansatz" for the \lya\ forest. In Fig.~\ref{zeta_vs_xisq_scale}, we make contour plots of $\xi*\xi$ (see Eq.~\ref{Q_eq}) vs $\zeta$ at different source configurations for $N_{\rm HI}>10^{13.5}$cm$^{-2}$ considering the measurement around individual sightlines.  We make shaded contours comprising of the fractions of total population of ($\xi*\xi$, $\zeta$) points (corresponding to individual correlation values ($\xi*\xi$, $\zeta$) derived from 4000 simulated realizations of triplet skewers for each configuration) ranging from 0.1 to 0.9 in steps of 0.1. The plots in Fig.~\ref{zeta_vs_xisq_scale} are arranged in a $3\times 3$ grid. Plots showing the effect of varying scales are arranged in horizontal direction from 1 to 5$h^{-1}$cMpc and  the same for the angles are arranged along the vertical direction from $10^{\circ}$ to $170^{\circ}$. According to the "hierarchical ansatz" (Eq.~\ref{Q_eq}), we expect a strong correlation between $\zeta$ and $\xi*\xi$. The slope of this correlation should give the Q value.
	
	We assign a Pearson's correlation coefficient $r_S$ between $\zeta$ and $\xi*\xi$, to quantify how tightly the three-point and two-point correlations are correlated for all these configurations. For $N_{\rm HI}>10^{13.5}$cm$^{-2}$, we see a very tight correlation at r=$1h^{-1}$cMpc with $r_S\sim 0.8$ for $\theta=10^{\circ}$ which decreases to $r_S\sim 0.7$ for $\theta=170^{\circ}$. It is also evident that for a given $\theta$, the correlation is found to decrease with increasing scales. In general, it is seen that there is a stronger decrease in the correlation coefficient in going from $\theta=10^{\circ}$ to $\theta=90^{\circ}$ than in going from $\theta=90^{\circ}$ to $\theta=170^{\circ}$. This is consistent with the trend we found for Q in Fig.~\ref{zeta_norm_vs_r}.
	
	The $\xi*\xi$ vs $\zeta$ contour plots also provides the percentage of line of sights having positive two-point as well as three-point correlations corresponding to coherent non-linear structures in the IGM probed by the triplet sightlines. To visualize this, we provide the percentages of lines of sight in each quadrant of $\xi*\xi$ vs $\zeta$ contour plot in  Fig.~\ref{zeta_vs_xisq_scale}. It is seen that at small scales, one obtains a lot of non-linear connected structures indicated by positive and larger values of $\xi*\xi$ and $\zeta$. Also, points (line of sights) with negative $\xi*\xi$ are negligible in number. One sees some negative $\zeta$ points which correspond to structures which are not present coherently in all the three sightlines. With increasing scale, the number of lines of sight with non-linear connected structures indicated by positive $\xi*\xi$ and $\zeta$ decreases and one begins to see more and more sightlines with negative  $\xi*\xi$ and $\zeta$. At scales beyond $3h^{-1}$cMpc, we have significant number of sightlines with negative  $\xi*\xi$ as well as $\zeta$ which correspond to anti-correlated regions or coherent gaps present in the triplet sightlines. The configurations with $\theta=90^{\circ}$ and $\theta=170^{\circ}$ in general sample similar kind of regions indicated by similar values of perecentages in each of the quadrants. It is also seen that for $\theta=10^{\circ}$ configurations one samples regions with positive $\zeta$ and $\xi * \xi$ more frequently. 
	
	Additionally, since we find that a definite hierarchy exists between three-point and two-point correlations, we assign a slope to determine how they are connected. We then compare this slope with the reduced three-point correlation Q. To find the slope, we use a linear fitting function of the form
	\begin{dmath}\label{Q1_Q2_def}
		\zeta=\mathrm Q_1 (\xi*\xi) + \mathrm Q_2  \ , 
	\end{dmath}
	with $\rm Q_1$ and $\rm Q_2$ being the slope and y-intercept of the linear fit. We bin the points in $\xi*\xi$ in the range of the 0.9th contour's x-projection into 10 equi-spaced bins. We find the median $\zeta$ values for each $(\xi*\xi)$ bin, given by blue dots in the figure, and obtain a linear fit for this given by blue dashed line. 
	The $\rm Q_1$ and $\rm Q_2$ values for each fits are given in each panel of Fig.~\ref{zeta_vs_xisq_scale}. Note that $\rm Q_2$ should be zero for "hierarchical ansatz" to be applicable perfectly. This seems to be the case for $r\geq 3h^{-1}$cMpc where $\rm Q_2$ is consistent with zero within 2$\sigma$ level. However, for $r\sim 1h^{-1}$cMpc, $\rm Q_2 \neq 0$. This simply means at small scales, the median $\zeta\sim 0$ (or mildly negative) even for sightlines showing non-negligible two-point correlation (i.e $\xi * \xi \sim 0.5$). 
	It is also evident that $\rm Q_1$ increases with increasing scale. The evolution is stronger between $\theta=10^{\circ}$ to $\theta=90^{\circ}$ but flattens for $\theta>90^{\circ}$. This is roughly consistent with the behaviour seen in Fig.~\ref{Corr_cloud_angle} for Q. We obtain $\mathrm Q_1\sim 0.4$ at smallest scales of $1h^{-1}$cMpc which increases upto 0.5-0.6 at $5h^{-1}$cMpc. Interestingly, we also obtain a negative y-intercept for these fits ranging from $\sim -0.2$ at 1$h^{-1}$cMpc to 0 at 5$h^{-1}$cMpc.
	\begin{figure*}
		\centering
		
		\includegraphics[viewport=5 5 1150 270,width=18cm, clip=true]{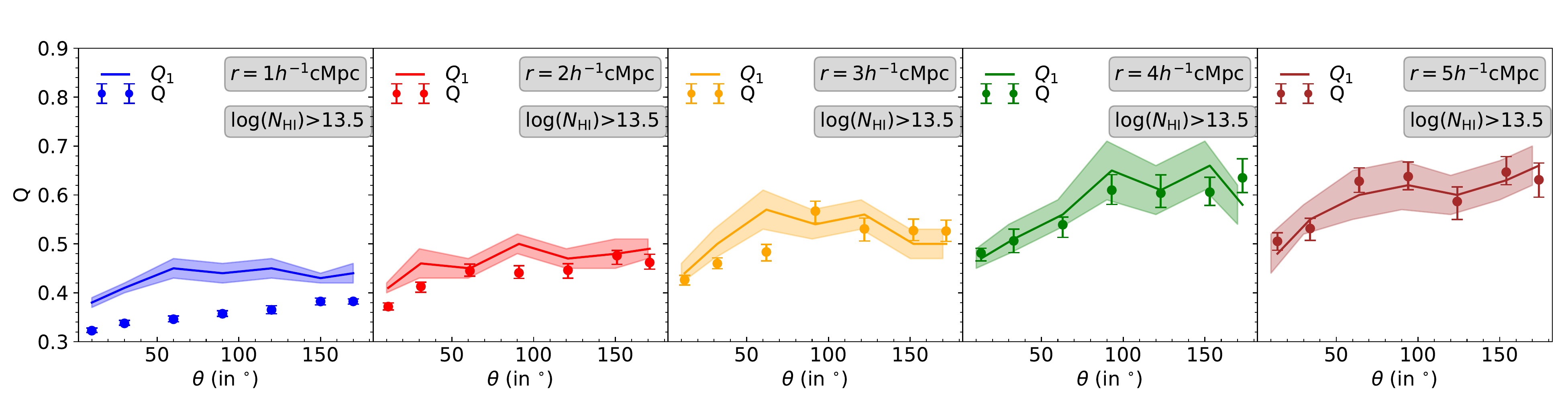}
		\caption{Plot of median Q and $\rm Q_1$ as a function of angle at different scales (ranging from 1 to 5$h^{-1}$cMpc) for clouds having  $N_{\rm HI}>10^{13.5}$cm$^{-2}$ at $z=2$. While there exists an offset at small scales, the median Q follows $\rm Q_1$ at large scales.}
		\label{Q_vs_Q1}
	\end{figure*}
	
	In Fig.~\ref{Q_vs_Q1}, we compare the $\rm Q_1$ obtained for different configurations with the median reduced three-point correlation Q for $N_{\rm HI}>10^{13.5}$cm$^{-2}$ as a function of angle and at different scales. We find a definite offset between Q and $\rm Q_1$ at 1$h^{-1}$cMpc, i.e., Q is always smaller than $\rm Q_1$. With increasing scale, this offset goes away and $\rm Q_1$ follows the median reduced three-point correlation at scales beyond 2$h^{-1}$cMpc. The Q defined as mean three-point correlation normalised to cyclic combination of mean two-point correlations (which we will further refer to as $\rm \bar{Q}$, see Eq.~\ref{Mean_Q} in Appendix~\ref{Mean_Median_q}) has a larger magnitude than median Q and hence would be larger than $\rm Q_1$. Thus, we find median Q to be a better representation of the "hierarchical ansatz" that one expects from the scaling relation between three-point and associated two-point correlation at large scales.
	
	In Sec.~\ref{Scale_dependence}, we suggested that scale dependence of Q is weaker than that of $\zeta$. It is evident from Fig.~\ref{Q_vs_Q1} that the dependence will be further weakened if we consider $\rm Q_1$.  This offset that is seen at smaller scales is commensurate with the non-zero $\rm Q_2$ values we obtained at these scales. The suppression of Q values in comparison to $\rm Q_1$ at smaller scales indicates a small scale suppression of three-point correlation. It will also be interesting to ask what physical process decides the scales below which $\rm Q_2\neq 0$.

	\subsection{UVB dependence}
	
	\begin{figure*}
		\centering
		\includegraphics[width=17.5cm]{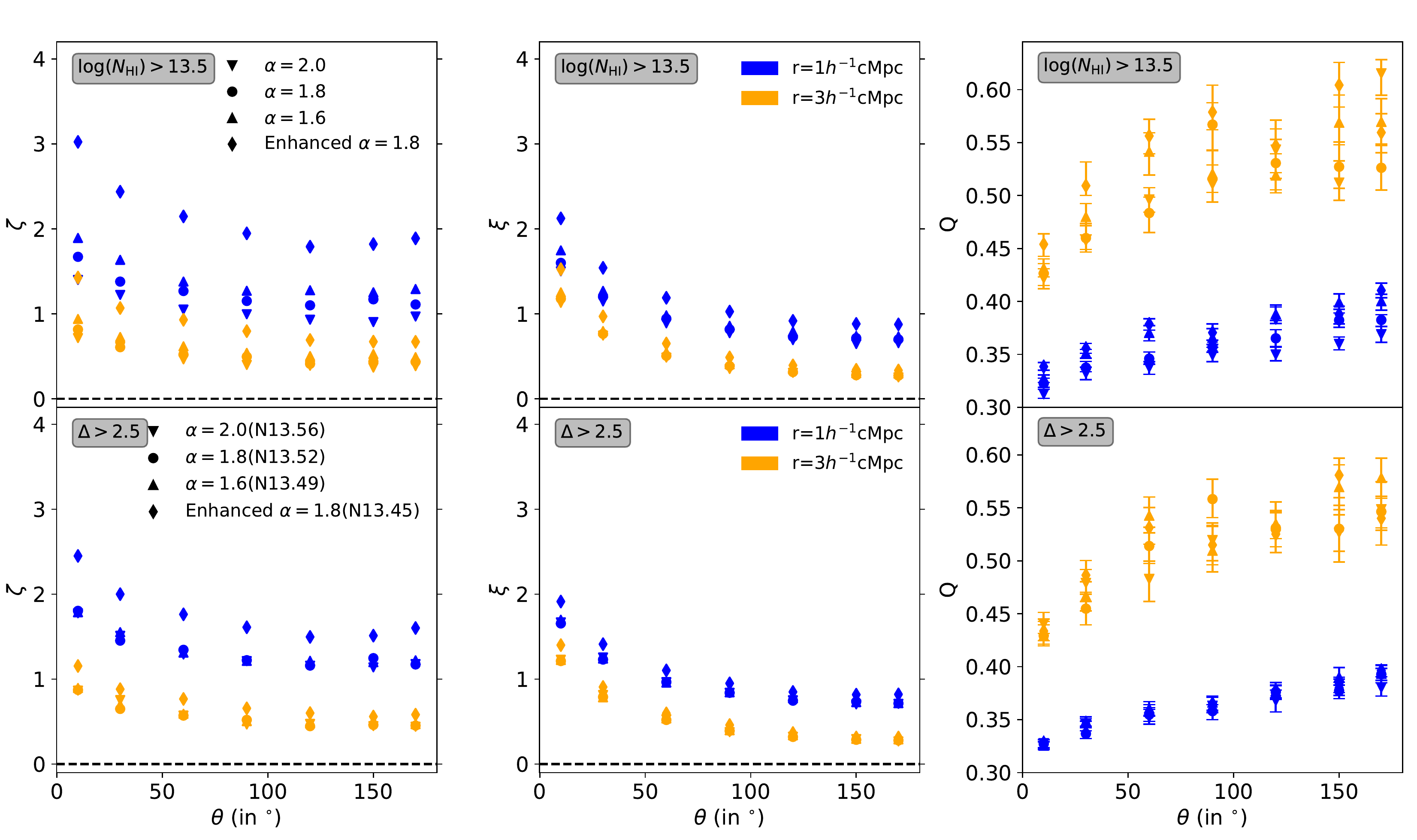}%
		
		\caption{Transverse three-point (\textit{Left column panels}), the associated two-point (\textit{Middle column panels}) and reduced three-point correlation (\textit{Right column panels}) functions for UVB with $\alpha=1.6,1.8$ and $2.0$ and an Enhanced UVB with $\alpha=1.8$ as a function of angle at $z=2.0$. We have plotted the transverse two-point correlation corresponding to the unequal arm of the triplet configuration ($\Delta r_{23\perp}$ which is $\theta$ dependent). We do not plot the errors for transverse three-point and two-point correlation as they are roughly the size of the markers. The correlation statistics has been calculated for a fixed neutral hydrogen column density threshold $N_{\rm HI}>10^{13.5}$cm$^{-2}$ (\textit{Top panels}) and a fixed $\tau$-weighted over-density threshold $\Delta>2.5$ (\textit{bottom panels}) at scales of 1 and 3$h^{-1}$cMpc. The log$N_{\rm HI}$ threshold corresponding to the fixed $\Delta$ threshold has been put in the legend in the form of (N log$N_{\rm HI}$ threshold) for each UVB.}
		\label{Corr_SED}
	\end{figure*}
	
	In this section, we will study the impact of local physical conditions and thermal history on the three-point correlation statistics of IGM at $z\sim 2$ using four different simulation boxes discussed in Sec.~\ref{Simulation}. These simulations essentially have similar dark matter density distributions as they were ran with the same cosmological parameters and slightly different $\Gamma_{\rm HI}$ evolution as a function of $z$ (see Fig.~\ref{T-g-G}). However, these simulations differ in their neutral hydrogen density distribution due to differences in the redshift evolutions for thermal parameters as shown in Fig.~\ref{T-g-G}. The \lya\ absorbers are known to be affected by different thermal histories in three ways: 
	\begin{itemize}
		\item Thermal broadening of \lya\ absorption depends on the local temperature. Density dependence of thermal broadening will be governed by the local ${\rm T}-\delta$ relation.
		\item Ionization fraction of hydrogen depends on photoionization rates and temperature dependent recombination rates, and hence depends on local thermal parameters.
		\item Physical size of baryons for a given dark matter potential depends on pressure broadening scales associated with the thermal history. This manifests as an integrated effect of the thermal history and does not simply depend on local thermal parameters.
	\end{itemize}
	Our aim is to study their effects on the transverse three-point and two-point correlation.
	
	We calculate the transverse three-point correlations for all the four simulations at $z=2$ for $N_{\rm HI}>10^{13.5}$cm$^{-2}$ clouds which are plotted in the top left column panel in Fig.~\ref{Corr_SED} as a function of angle for two different scales ($r=1$ and $3h^{-1}$cMpc). First we consider three models excluding the "Enhanced $\alpha=1.8$" model. A positive increase in three-point correlation is seen with decrease in $\alpha$ from 2.0 to 1.6 for r=1$h^{-1}$cMpc and 3$h^{-1}$cMpc. At $z\sim 2$, all these models have nearly same '$\gamma$', slightly higher $\rm T_0$ and $\Gamma_{\rm HI}$ for $\alpha=1.6$ compared to $\alpha=2.0$. The difference between the $\zeta$ measured for the three models is higher for $r=1h^{-1}$cMpc case. In the case of $r=3h^{-1}$cMpc, the difference in $\zeta$ between these models are not very significant. Maximum three-point correlation is obtained for "Enhanced $\alpha=1.8$" model at both the scales. Note that the baryons in this model have higher temperature (by a factor of $\sim 1.5$) compared to our fiducial model. 
	
	We repeat the entire exercise for two-point correlation and the results are plotted in the top-middle column panel of Fig.~\ref{Corr_SED}. We have plotted the transverse two-point correlation corresponding to the unequal arm of the triplet configuration ($\Delta r_{23\perp}$ which is $\theta$ dependent). For two-point correlation too, maximum amplitude is obtained for "Enhanced $\alpha=1.8$" model followed by decreasing amplitude in going from $\alpha=1.6$ to 2.0. This effect is found to be more pronounced at smaller scales in two-point too (i.e the difference is negligible for $r\geq 3h^{-1}$cMpc). Also, for $\theta=10^{\circ}$ which corresponds to a very small $\Delta r_{23\perp}$ in transverse two-point correlation, the differences between models using different UVBs is maximum. Overall, while difference between different UVB models is present in transverse two-point correlation, these differences are smaller in comparison to transverse three-point correlation.
	
	It is now important to ask whether the differences in correlation amplitudes for different thermal histories is caused by difference in local thermal parameters at $z=2$ (thermal broadening and ionization state), or is it a signature of different clustering properties originating from an integrated effect of different thermal histories (pressure broadening scales). What one needs to keep in mind is that we are fixing $N_{\rm HI}$ thresholds while comparing the correlation statistics between these simulation boxes having different thermal histories. However, due to different thermal properties, same $N_{\rm HI}$ values will not correspond to the same baryonic over-densities (or dark matter over-densities). This is important as we already noticed that high $N_{\rm HI}$ (i.e high $\Delta$) regions tend to show higher values of $\zeta$ and $\xi$.
	
	To understand this further we use the $N_{\rm HI}$ vs $\Delta$ power-law fit values given in Table.~\ref{N_Delta} to fix an over-density threshold of $\Delta>2.5$ (roughly corresponding to $N_{\rm HI}>10^{13.5}$cm$^{-2}$ for our fiducial model) and scale the $N_{\rm HI}$ thresholds accordingly for the four UVBs. Doing so fixes the effect from local thermal parameters, that is, thermal broadening and different ionization states, by probing clustering for a fixed baryonic over-density threshold. However, the impact of pressure broadening may still not be accounted for which affects the relationship between the baryonic density field and the dark matter field.

	Plots of three-point and two-point correlations with $N_{\rm HI}$ scaled for a fixed $\Delta$ is shown in the bottom left and middle column panels respectively in Fig.~\ref{Corr_SED}.
	While the differences in three-point and two-point correlations obtained for simulations having $\alpha=1.6,1.8$ and $2.0$ UVB are very much reduced, we still see that the "Enhanced $\alpha=1.8$" simulations has higher three-point and two-point correlations. Since fixing the baryonic over-density should have fixed the effects from the difference in the thermal broadening and ionization states, this difference must be coming from differences in pressure broadening scales of the baryonic density fields as dark matter field is statistically same for these models. 
	
	\begin{figure}
		\centering
		\includegraphics[width=8cm]{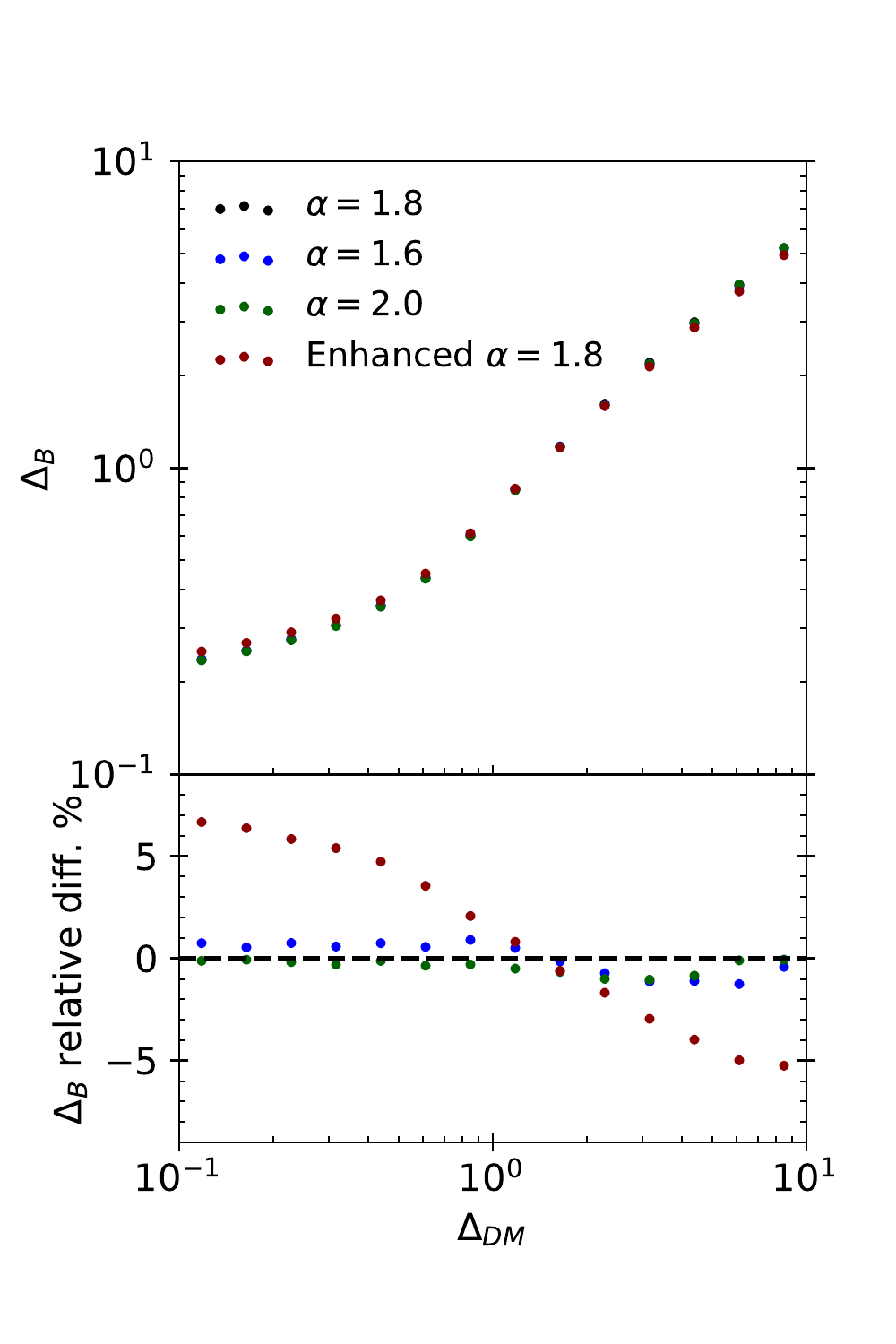}%
		
		\caption{Plots of dark matter over-density ($\Delta_{DM}$) vs baryonic over-density ($\Delta_B$) for different UVB models with $\alpha=1.6,1.8$ and $2.0$ and an Enhanced UVB with $\alpha=1.8$ (\textit{Top})  and their relative differences in percentage with respect to the fiducial $\alpha=1.8$ model (\textit{Bottom}) at $z=2$.}
		\label{Delta_B_vs_Delta_DM}
	\end{figure}
	
	To further investigate this effect, we calculate the 1D dark matter over-densities $\Delta_{DM}$ from our simulation along the lines of sight using a standard Cloud-in-Cell algorithm for each of the grid-points. We calculate the $\Delta_{DM}$ corresponding to 4000 lines of sight for which we already have the baryonic over-densities $\Delta_B$. We then logarithmically bin $\Delta_{DM}$ values in the range of 0.1 to 10 and find the median $\Delta_B$ corresponding to each $\Delta_{DM}$ bin. We consider $\Delta_{DM}$ and $\Delta_B$ values for all the grid points along 4000 lines of sight. $\Delta_{DM}$ vs $\Delta_B$ has been plotted in the top panel of Fig.~\ref{Delta_B_vs_Delta_DM} for the four UVB models. In the bottom panel, we plot the relative percentage differences in $\Delta_B$ for each $\Delta_{DM}$ bin with respect to our fiducial model. It is seen that while $\Delta_{DM}$ vs $\Delta_B$ is consistent with each other for $\alpha=1.6,1.8$ and $2.0$ models within 1\% accuracy, the difference between our fiducial model and "Enhanced $\alpha=1.8$" model is considerably more. For underdense regions having $\Delta_{DM}=0.1$, the $\Delta_B$ corresponding to "Enhanced $\alpha=1.8$" is higher than that in our fiducial model by 7\%.  For overdense regions having $\Delta_{DM}=10$, the $\Delta_B$ corresponding to "Enhanced $\alpha=1.8$" is less by 5\%. This clearly shows the effect of pressure smoothing. The box having higher temperature, i.e, "Enhanced $\alpha=1.8$" has higher pressure. This pressure smoothens the baryonic density field such that overdense region becomes less overdense and underdense region becomes less underdense. Since the box having Enhanced $\alpha=1.8$ has a considerably larger temperature field (look Fig.~\ref{T-g-G}), the pressure effect is seen more appreciably. So, the differences observed in transverse two-point and three-point correlation at fixed baryonic over-densities corresponding to different pressure broadening scales ascertains both these statistics as a sensitive probe of the thermal history of IGM. 
	
	Since the effect of using different UVB models changes both three-point and two-point correlation in the same direction, next we examine whether Q also has a UVB dependence. We plot Q for $N_{\rm HI}>10^{13.5}$cm$^{-2}$ clouds in the top right panel of Fig.~\ref{Corr_SED}. The dependence on different UVB models is found to be very small in Q compared to what we have found for $\zeta$. So, while transverse three-point and two-point correlations are found to be sensitive to thermal history especially at smaller scales, Q does not change appreciably with the thermal history,
	In the bottom right column panel of Fig.~\ref{Corr_SED}, we plot Q with $N_{\rm HI}$ threshold scaled for a fixed $\Delta$. We find Q to be independent of the pressure broadening effects that we had seen for three-point and two-point correlation between our fiducial and "Enhanced $\alpha=1.8$" model. The slight differences seen for a fixed $N_{\rm HI}$ threshold at 1$h^{-1}$cMpc is also gone for a fixed $\Delta$.
	
\begin{figure}
		\centering
		\includegraphics[viewport=37 20 580 450,width=10cm, clip=true]{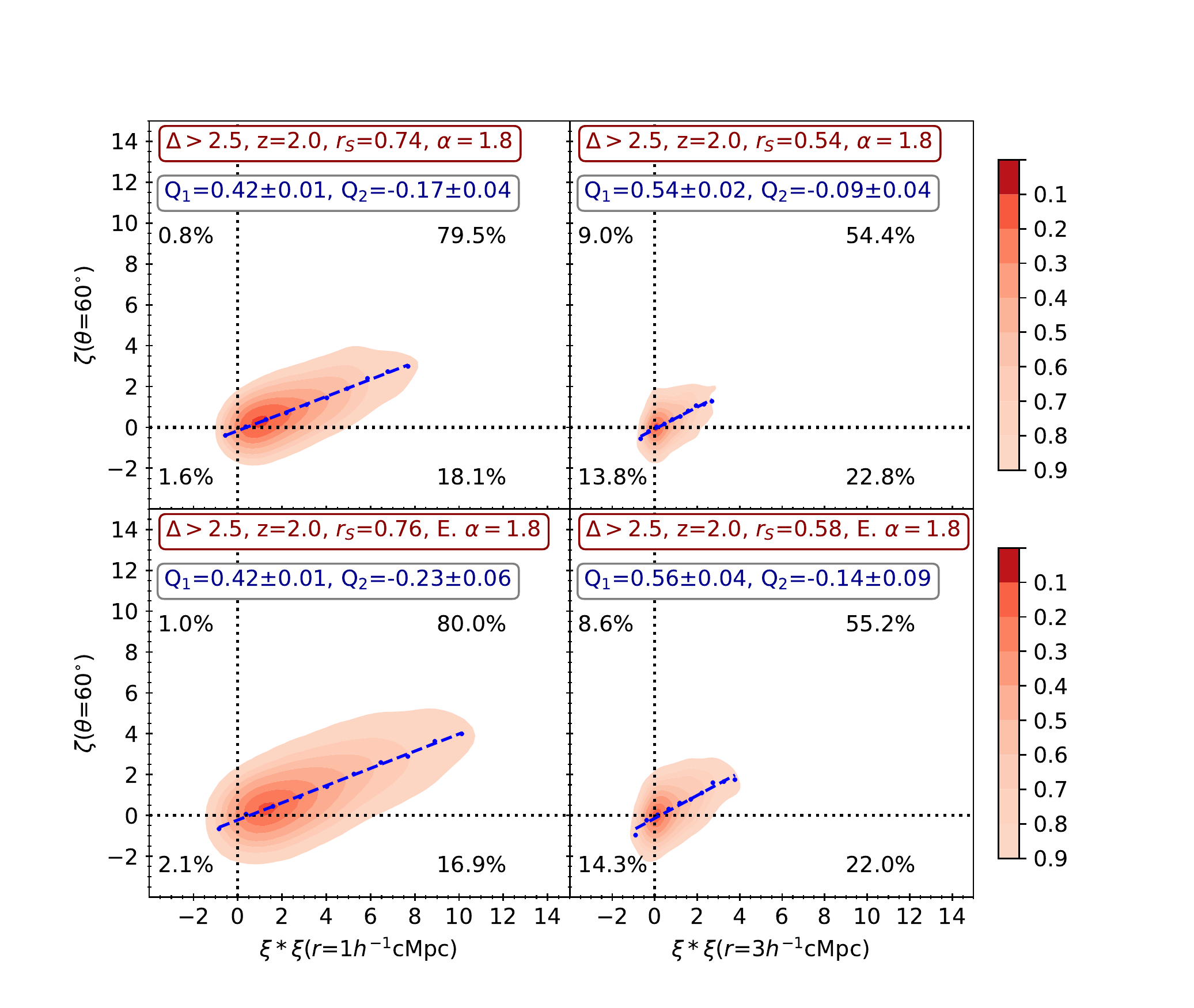}
		\caption{$\zeta$ vs $\xi * \xi$ plots for clouds having $\Delta>2.5$ for simulations with $\alpha=1.8$ (\textit{top row}) and Enhanced $\alpha=1.8$ UVB (\textit{bottom row}) at $z=2$.  We vary the scale along the plots in horizontal direction from 1 to 3$h^{-1}$cMpc and fix the angle to $\theta=60^{\circ}$.}
		\label{zeta_vs_xisq_Hot_vs_cold}
	\end{figure}

	To explore this further, we also examine whether $\rm Q_1$ and $\rm Q_2$ changes with thermal history. In Fig.~\ref{zeta_vs_xisq_Hot_vs_cold}, we plot the ($\xi*\xi$, $\zeta$) contour plots for the fiducial simulation with $\alpha=1.8$ UVB (top row) and for the simulation with Enhanced $\alpha=1.8$ UVB at scales of 1 and 3$h^{-1}$cMpc (Left and right column respectively) and with $\theta=60^{\circ}$ for a fixed baryonic over-density threshold ($\Delta>2.5$). The simulation having the enhanced background has larger contours at both the scales. However, $\rm Q_1$ does not change between the two models. Also, despite having large errors, there is a slight indication of $\rm Q_2$ being larger for the simulation having enhanced background for both the scales considered here. So the x-intercept (i.e $\rm -\ Q_2/Q_1$, which denotes $\xi * \xi$ values beyond which median $\zeta$ becomes non-zero) has increased in the case of "Enhanced $\alpha=1.8$" model and thus may depend on the thermal evolution of the IGM.
	
	\section{Redshift evolution of the \lya\ clustering}
	
	\begin{figure*}
		
		\includegraphics[width=18cm]{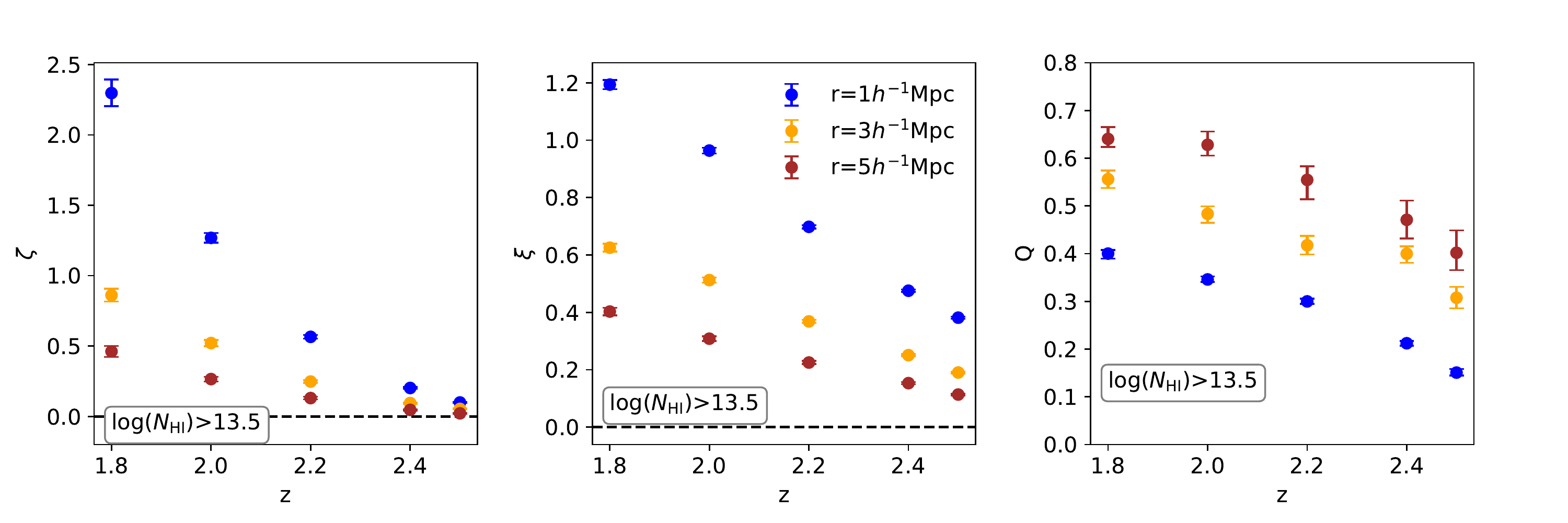} 
		\includegraphics[width=18cm]{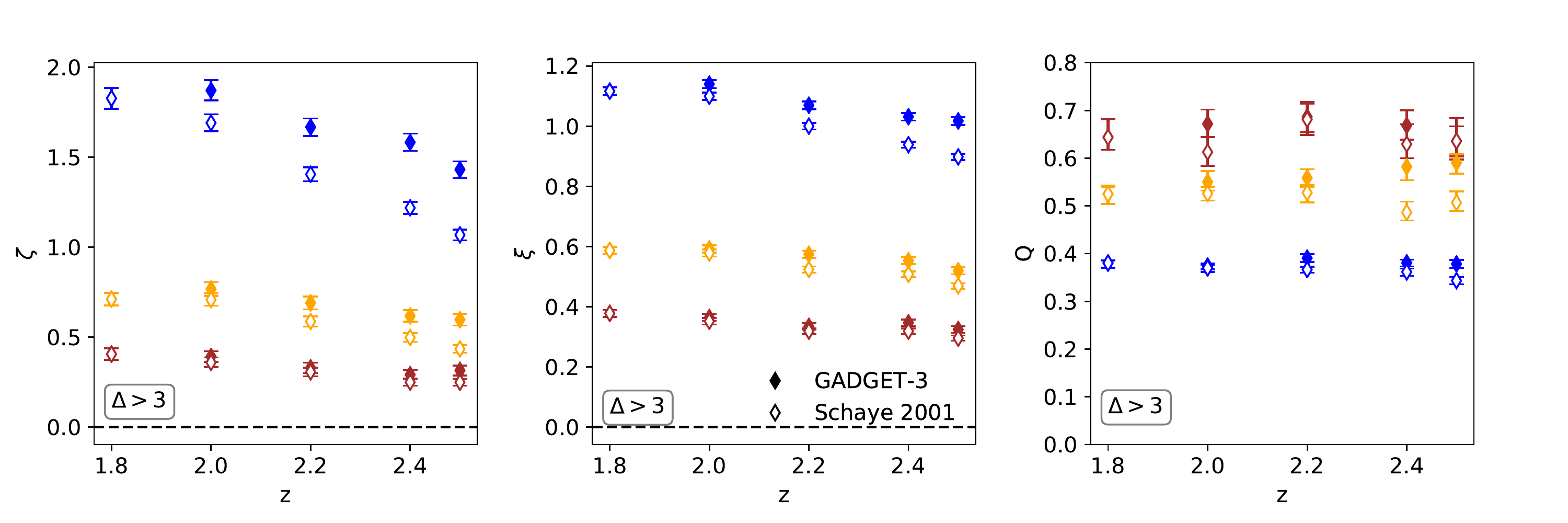}

		\caption{Transverse three-point correlation (\textit{left column}), two-point correlation (\textit{middle column}) and reduced three-point correlation (\textit{right column}) as a function of redshift for scales of $r=1,3$ and $5h^{-1}$cMpc. For transverse three-point correlation and reduced three-point correlation, we consider equilateral configuration (i.e $\theta=60^{\circ}$). The top row shows correlations for a fixed $N_{\rm HI}$ threshold of $N_{\rm HI}>10^{13.5}$cm$^{-2}$. The bottom row shows correlations for a fixed $\Delta$ threshold of $\Delta>3$. The filled markers show fixed $\Delta$ threshold based on $N_{\rm HI} - \Delta$ relation from the simulation. The hollow markers show fixed $\Delta$ threshold obtained from the analytical expression of $N_{\rm HI} - \Delta$ relationship in \citet{schaye2001} under the assumption of hydrostatic equilibrium of Jeans like cloud using physical parameters from the simulation.  }
		\label{Corr_redshift}
	\end{figure*}
	
	Till now we have been focusing our investigations on the 
	IGM probed by \lya\ absorbers at $z=2$. In this section, we study the redshift evolution of the transverse three-point, two-point and reduced three-point correlation of the IGM over $1.8\leq z \leq 2.5$. For this purpose, we consider simulation snapshots from our fiducial model at
	$z$=[1.8, 2.0, 2.2, 2.4 and 2.5]. To start with, we consider the equilateral configuration (i.e $\theta =60^\circ$) and three different length scales $r$ = 1, 3 and 5 $h^{-1}$cMpc. 

	In the top left panel of Fig.~\ref{Corr_redshift}, we show transverse three-point correlation as a function of redshift for $N_{\rm HI}>10^{13.5}$cm$^{-2}$. We plot the same for two-point correlation 
	in the top middle panel. It is clear from these plots that for a given scale three-point correlation function increases more rapidly compared to the two-point correlation function with decreasing redshift. It is evident that $\zeta$ becomes less than 0.1 for $z\sim2.5$ even for $r=1$ $h^{-1}$cMpc.  Interestingly we find  that 
	the redshift evolution of both $\xi$ and $\zeta$ has weak dependence on $r$ for a fixed 
	$N_{\rm HI}$ threshold (i.e shape of $\xi(z)$ and $\zeta(z)$ are weakly dependent on $r$).
	In the the top right panel of Fig.~\ref{Corr_redshift}, we plot $Q$ as a function of $z$. For the all the three length scales considered Q decreases with increasing $z$. We can clearly see that the redshift depepndence of Q is shallower than what we see for $\zeta$.

	Next, we study the effect of clustering of the baryonic density field by using different $N_{\rm HI}$ thresholds at different redshifts corresponding to a given $\Delta$ threshold instead of using one $N_{\rm HI}$ threshold for all redshifts. We plot the redshift evolution of transverse three-point, two-point and reduced three-point correlation
	for $\Delta>3$ in the bottom left, middle and right panels, respectively in Fig.~\ref{Corr_redshift} using filled markers. For this we have used the relationship between $\Delta$ and
	$N_{\rm HI}$ given in Table~\ref{N_Delta} for different redshifts for our fiducial model.
	We see that for a fixed $\Delta$ threshold, the redshift evolution of these quantities are very weak compared to what we found for a fixed $N_{\rm HI}$ threshold. In particular, we see a very slight decrease in the three-point and two-point correlation with increasing redshift for $r = 1h^{-1}$cMpc. For $r= 3$ and 5$h^{-1}$cMpc, we do not see any appreciable change in $\zeta$ and $\xi$ as a function of redshift.  In the case of Q, whatever redshift evolution seen in the case of fixed $N_{\rm HI}$ threshold disappears when we use fixed $\Delta$.
	{\it Therefore, the redshift evolution of $\zeta$ and $\xi$ we see for a fixed $N_{\rm HI}$ threshold (top panels in Fig.~\ref{Corr_redshift}) comes primarily from the evolution of thermal parameters and does not reflect the evolution of clustering properties in the baryonic field itself, or underlying dark matter probed by the \lya\ absorption.}

	It will be possible to measure the corresponding $N_{\rm HI}$ cuts at different redshifts probed in this work, from the observational data by demanding a constant values of $\xi$, $\zeta$ and notably Q. This can then be used to constrain the redshift evolution of physical conditions in IGM.
	Unlike what we could do with simulations, while dealing with the real data we may not have the correct relationship between $\Delta$ and $N_{\rm HI}$. In that case, we check the utility of the analytic expression (see the equation 11) derived by \citet{schaye2001}. We fixed the value of $f_g$ (the baryon fraction) by comparing our fitting function at $z\sim1.8$ with this analytic expression. Then we predicted $N_{\rm HI}$ cut at different redshifts for a given $\Delta$ using the thermal parameters and $\Gamma_{\rm HI}$ for a fixed $f_g$. First we notice that the value of $N_{\rm HI}$ cut predicted by the analytic expression is slightly lower than the value we get at different $z$. The difference is progressively increasing as we go to the higher redshifts. For example, the difference is 0.1 dex at $z\sim2.5$.  The correlation functions obtained for these $N_{\rm HI} $ cuts are also shown in Fig.~\ref{Corr_redshift} using hollow markers. It is clear that to get accurate constraints on physical conditions (i.e better than 10\% accuracy) one needs more accurate relationship between $\Delta$ and $N_{\rm HI}$ as a function of $T_0$, $\gamma$ and $\Gamma_{\rm HI}$. Then one will be able to use the values of $\xi$, $\zeta$ and Q as a function of $\Delta$ to probe the underlying cosmological parameters if thermal parameters are constrained by other observables.
	
	As we discussed in the introduction, redshift evolution of the angular (i.e $\theta$) dependence of Q for galaxies can be used to infer the redshift evolution of the proportion of linear to spherical structures. we notice that for $r=1$ and $3h^{-1}$cMpc, the angular dependence of Q shows very little change in the shape over $1.8\leq z\leq 2.5$. This may suggest minor evolution in the cosmic web over $1.8\leq z\leq 2.5$. This could be due to small redshift range probed and the fact that \lya\ forest studied here probes relatively milder over-densities.

	\section{Observability of three-point correlation function:}
	
	In this section we estimate the minimum number of triplets sightlines (or the redshift path length) required to detect three-point correlation at more than 5$\sigma$ level for a range of scales and $N_{\rm HI}$ thresholds probed here. To start with we consider the configurations with $\theta = 90^{\circ}$. As we saw before $\theta$ dependence of $\zeta$ is very weak beyond $\theta =90^\circ$. Therefore the results we obtain for $\theta=90^{\circ}$ is applicable to $\theta>90^{\circ}$ also. We define the $\sigma$ as,
	\begin{equation}\label{Det}
	\sigma = \sqrt{\sigma_{D}^2+\sigma_{R}^2}.
	\end{equation}
	Here, $\sigma_D$, is the error associated with the data which will depend on the number of sightlines and number of clouds per unit redshift interval (i.e $N_{\rm HI}$ threshold). $\sigma_{R}$ is the zero error associated with the randomly distributed clouds obtained with all three triplet skewers being generated randomly. In general $\sigma_R$ is expected to be much smaller than $\sigma_D$.

\begin{figure}
		\centering
		\includegraphics[viewport=0 5 300 260,width=8cm, clip=true]{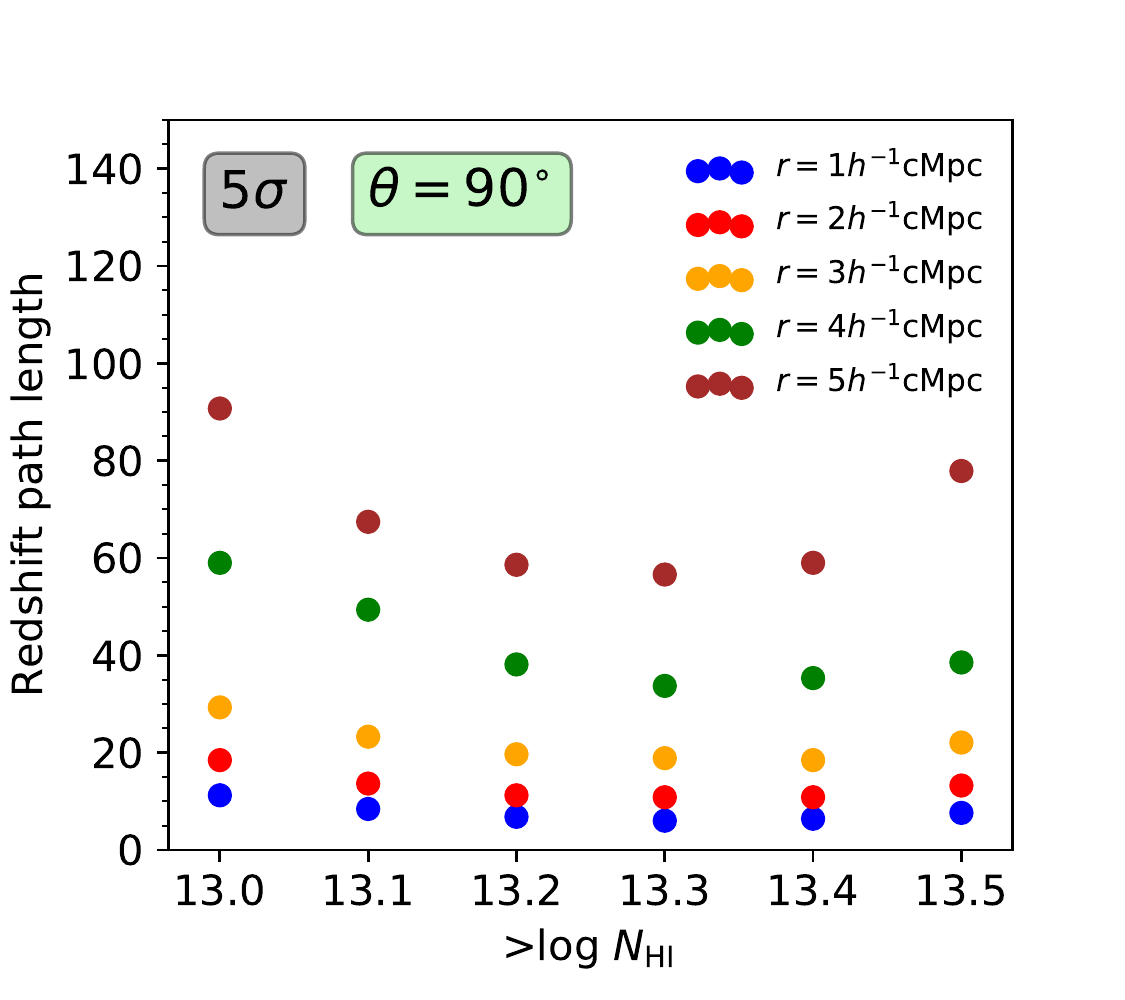}%
		\caption{The minimum redshift path length required for detecting transverse three-point correlation at $z=2$ at $>5\sigma$ level. For sub-samples defined for different $N_{\rm HI}$ thresholds, we have used spectra with FWHM resolution of 50\kms\ and SNR=20. The detectability is plotted for $r=1-5h^{-1}$cMpc and $\theta=90^{\circ}$.}
		\label{Detectibility}
	\end{figure}

	In Fig.~\ref{Detectibility}, we plot the minimum redshift path length (common along the triple sightlines) required for the $5\sigma$ detectability of three-point correlation as a function of $N_{\rm HI}$ threshold for different scales at $z=2$. For these calculations,  we consider spectral resolution $\sim 50$~\kms\ and SNR$\sim$ 20 and individual segments of 100h$^{-1}$ cMpc length. As we discussed before these gives the $N_{\rm HI}$ completeness of $\sim 10^{13}$ cm$^{-2}$. The chosen spectral resolution is typical of what can be achieved with present day spectrographs like X-Shooter and VLT, or upcoming surveys like low resolution mode WEAVE-QSO \citep{pieri2016} or MaunaKea Spectroscopic Explorer (MSE, \cite{mse2019}) and optical spectrographs like Wide Field Optical Spectrograph (WFOS) in the Thirty Metre Telescope (TMT). Spectra having low SNR (along one sight line) will correspond to having higher completeness limit for $N_{\rm HI}$ and one will consider samples with $N_{\rm HI}$ thresholds above this limit in such cases.

	It is clear from the figure that the required redshift path length decreases with
	increasing column density threshold before becoming flat (for lower r values) or
	showing increase for higher $N_{\rm HI}$ thresholds.
	The reason behind this behavior is that at lower $N_{\rm HI}$ threshold, the signal is weaker and the relative $\sigma_{D}$ is larger and hence one needs large number of sightlines. At larger $N_{\rm HI}$ thresholds, while the signal is strong the number of absorbers per unit redshift interval decreases thereby increasing the $\sigma$. For a quasar at $z>2$ we will be able to probe the \lya\ forest over a redshift path length of  $\Delta z \sim 0.4$ (considering region between \lya\ and \lyb). Considering the fact that the redshifts of the background quasars need not be identical one will typically cover $\Delta z \sim 0.2$ along each triplet (or equivalently probing the clustering over a redshift interval of 0.2). From the figure we can conclude that we need about 100  triplet sightlines for $r\sim 3 h^{-1}$cMpc for the $N_{\rm HI}$ thresholds considered here. One may need half these number of triplets to probe the clustering at smaller length scales. As seen in Fig.~\ref{Corr_cloud_angle}, $\zeta$ is usually stronger for $\theta<90^{\circ}$. Therefore, one will need lesser number of triplets to get $5\sigma$ detections.
	
\begin{figure}
		\centering
		\includegraphics[viewport=5 2 310 227,width=9.1cm, clip=true]{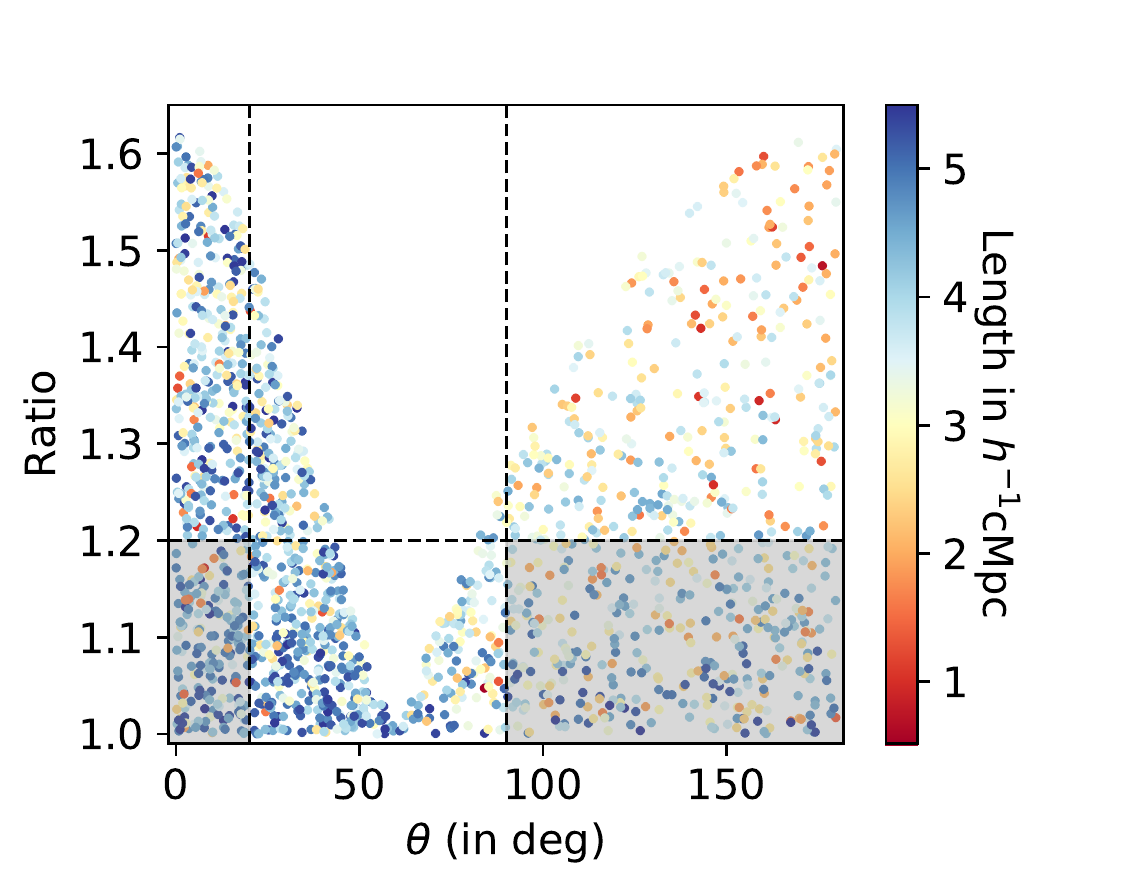}%
		\caption{$\theta$ vs arm length ratio of SDSS quasar triplets for probing in $z=2.0\pm 0.1$ upto a length scale of $r=5.0+0.5 h^{-1}$cMpc. The colorbar indicates the length scale $r$ of each of these triplets. Total number of quasar triplets identified is 1874. Out of this, 358 satisfy the condition of $\theta\geq 90^{\circ}$ and arm length ratio $\leq 1.2$, while, 194 satisfy the condition of $\theta\leq 20^{\circ}$ and arm length ratio $\leq 1.2$. This are indicated with shaded region.}
		\label{Detectibility_theta_vs_ratio}
	\end{figure}

	Next we consider $\theta$ and $r$ dependence for detectability and look at the availability of quasar triplets in the SDSS catalog with an aim to probe IGM in redshift range of $z=2.0\pm 0.1$. We search in SDSS DR14 catalog for projected quasar triplets having at least $\Delta z=0.1$ coincident \lya\ forest (considering \lya\ forest between the \lyb\ and \lya\ emission and excluding the proximity regions within 5000~\kms\ of the quasar emission redshift) within $z=2.0\pm 0.1$.
	For all the resulting quasar triplets, we consider the two arms (denoted as $r_1$ and $r_2$) whose ratio is closest to 1. We take the shorter of these two arms ($r_2$) as the denominator  to define the arm ratio (i.e ratio is $r_1/r_2$), such that the ratio is always greater than 1. We also take $r_2$ as the arm length (or scale) of the triplet.  We only select triplets having arm length less than $5.5 h^{-1}$cMpc (i.e the angular separation less than $\sim 5.5'$). The angle between these two arms is considered as the angle $\theta$ for the configuration. There are 1874 quasar triplets identified in this fashion in the SDSS catalog. In Fig.~\ref{Detectibility_theta_vs_ratio}, we plot the angle vs. arm length ratio for these quasar triplets. The colorbar gives the arm lengths of each of these triplets in units of arcmin.
	
	We then take a sub sample out of the selected triplets. We select triplets with angles greater than $90^{\circ}$ where the angular dependence of three-point correlation is negligible. We also consider only those triplets whose arm length ratio is less than 1.2. There are 358 quasar triplets which satisfy this condition.  Simultaneously, we also consider another sub-sample having $\theta\leq 20^{\circ}$ and having arm ratio less than 1.2. There are 194 quasar triplets which satisfy this condition.
	These two subsamples are represented by grey shaded region in Fig.~\ref{Detectibility_theta_vs_ratio}.
	Next, we distribute the selected triplet sub-samples based on their arm length into equispaced arm length bins centered around $r=[1,2,3,4,5]h^{-1}$cMpc. The resulting number of triplets for the two sub samples are given in the 6th and 7th column of Table.~\ref{Detectibility_SDSS}, while the redshift path length covered by these triplets are given in the 4th and 5th column.
	
	In the 2nd and 3rd columns of Table.~\ref{Detectibility_SDSS}, we give the redshift path length required for $5\sigma$ detectability of three-point correlation for each arm length and for $\theta=10^{\circ}$ and $90^{\circ}$. 
	In the table for each length scale, we consider $N_{\rm HI}\ge10^{13.3}$ cm$^{-2}$ threshold that gives the best detectability as shown in Fig.~\ref{Detectibility}. We compare these redshift path lengths obtained from the simulations and compare it with what one gets from the sub-samples in SDSS. It is clear that the number of known QSO triplets from the SDSS (even if we were to get high SNR spectroscopy) will not give adequate triplets to get 5$\sigma$ detection in a small redshift range for the length and angular scale probed here. However, we expect this situation to dramatically improve with Dark Energy Spectroscopic Instrument DESI \citep{desi2016} and Legacy Survey of Space and Time LSST.
	We then predict the expected significance of detection of three-point correlation at each length scales using the already known quasar triplets in SDSS. It is seen that for $\theta\leq 20^{\circ}$, the three-point correlation can be observed with highest significance at the largest scales of 4 and 5$h^{-1}$cMpc (4.8$\sigma$ and $4.5\sigma$ respectively). For that, we need to observe 70 quasar triplets (210 spectra) having $r=4h^{-1}$cMpc and 86 quasar triplets (i.e 258 spectra) having $r=5h^{-1}$cMpc. For $\theta= 90^{\circ}$, the most significant detection can be achieved with known quasars for intermediate scales of 2 and 3$h^{-1}$cMpc (4.4$\sigma$ and $4.7\sigma$ respectively). We need to observe 42 quasar triplets having $r=2h^{-1}$cMpc and 96 quasar triplets having $r=3h^{-1}$cMpc for this. Thus the discussion presented here suggests that one needs about 100 triplets to probe a given configuration (i.e small range in $r_1/r_2$, $r_2$ and $\theta$). Note while surveys like CLAMATO (\citet{lee2018}) have spectra taken at much smaller separations, the spectral SNR achieved are not sufficient for Voigt profile analysis discussed here.
	
	Note in these discussions we consider the three point correlation computed at the transverse separations with similar redshifts. In the real data one will be able to consider more triangular configurations by including different redshift ranges along different sightlines (i.e length scale defined as $r=\sqrt{r_{\perp}^2+r_{\parallel}^2}$). However, as they will probe larger scales compared to the transverse correlations we expect the significance of detections in such configurations to be less than 5$\sigma$ level. As demonstrated in \cite{maitra2019}, when good quality triplet spectra are available, in addition to probing $\zeta$ and Q, we will be able to study i) Void distribution in coherent gaps, ii) three-point correlation of metals and iii) Cross-correlation (both transverse and longitudinal) between \HI\ and metals, \HI\ and quasars, etc. One can in principle compute the three point correlation in the redshift space (i.e $\theta = 180^\circ$) using single lines of sight. We are investigating this using the high resolution quasar spectra available from VLT and KECK. These results will be discussed in our upcoming paper.

	\begin{table*}
		\centering
		\caption{SDSS sample at $z=2.0 \pm 0.1$.}
		\begin{tabular}{ccccccc}
			\hline
			Scale & \multicolumn{2}{c}{Redshift path length for 5$\sigma$ detection} & \multicolumn{2}{c}{Redshift path length (SDSS)} & \multicolumn{2}{c}{Number of triplets (SDSS)} \\
			($h^{-1}$cMpc)        &   $\theta=10^{\circ}$ &  $\theta=90^{\circ}$   &  $\theta\leq 20^{\circ}$ &  $\theta\geq 90^{\circ}$ & $\theta\leq 20^{\circ}$ & $\theta\geq 90^{\circ}$ \\
			\hline
			\hline
			$1.0\pm 0.5$ & 4.8 & 6.4 & 0.5 ($1.6\sigma$) & 0.1 ($0.9\sigma$) & 3 & 1\\
			$2.0\pm 0.5$ & 7.2 & 10.5 & 1.9 ($2.6\sigma$) & 8.2 ($4.4\sigma$) & 12 & 48  \\
			$3.0\pm 0.5$ & 9.6 & 18.6 & 3.6 ($3.2\sigma$) & 16.4 ($4.7\sigma$) & 23 & 98 \\
			$4.0\pm 0.5$ & 12.8 & 33.3 & 11.8 ($4.8\sigma$) & 19.1 ($3.7\sigma$) & 70 & 115 \\
			$5.0\pm 0.5$ & 17.2 & 56.0 & 14.1 ($4.5\sigma$) & 15.4 ($3.4\sigma$) & 86 & 96 \\
			\hline
		\end{tabular}
		\label{Detectibility_SDSS}
	\end{table*}

	%{\bf Notes: We need to discuss this in a skype session...}
	
	\section{Summary and Discussions}
	
	In this work, we have explored three-point correlation function of the IGM probed by the \lya\ absorption lines at $z\sim 2$ using cosmological simulations and absorption components (called "clouds" in this work) obtained using Voigt profile decomposition. Below we summarize and discuss important findings from this work.
	\begin{enumerate}
		\item[1.] \noindent{\bf Clouds vs flux based statistics:} It is very common to use statistics of transmitted flux for measuring the astrophysical and cosmological parameters using the \lya\ forest data (see references in Sec.~\ref{Introduction}). As pointed out in \citet{maitra2019}, when we use the transmitted flux a particular value of negative three-point correlation in flux can arise either from coherent absorption along all three sightlines or from absorption in one sight line and gaps in the other two sightlines. Such a degeneracy, which is inherent in the flux statistics, is not present in the analysis based on cloud distributions. While early studies of \lya\ forest clustering used Voigt profile components to study two-point correlation along the line of sight \citep[][]{chernomordik1995,cristiani1995,khare1997} not much progress is made in studying the higher order statistics. This is mainly due to difficulty in decomposing the \lya\ forest with Voigt profiles for large number of simulated data. Such studies have now been made possible with high performance computing and with the help of automated Voigt profile fitting code {\sc viper} \citep{gaikwad2017b}.
		
		Here we show that, using cloud based approach one will be able to do clustering analysis of the IGM using the standard techniques routinely used for clustering studies of galaxies. In particular, we show that the clustering depends strongly on the $N_{\rm HI}$ and the dependence is different for two- and three-point correlation functions.  This in principle allows us to probe the linear and non-linear bias parameters as a function of $N_{\rm HI}$ and the scales and angles probed by the triplet configurations.
		In this work we use simulations that do not include feedback processes, however if we use simulations that includes various feedback processes then we will be able to extend our analysis to sub-samples of metallicity (or metal to HI column density ratios). Such a study is important to extract more constraints on the parameters governing the galaxy feedback processes.
		\vskip 0.1in
		
		\item[2.] \noindent{\bf Applicability of "hierarchical ansatz" for the \lya\ forest:}
		We investigate the validity of the "hierarchical ansatz" using $\zeta$ and $\xi*\xi$ obtained for individual triplet sightlines. We fit this distribution using a linear fitting function $\zeta = \mathrm Q_1(\xi*\xi) + \mathrm Q_2$  where $\mathrm Q_2\sim 0$ when the "hierarchical ansatz" is valid.
		For models considered here $\mathrm Q_2\sim 0$ (within $2\sigma$ level) for $r\ge3$ h$^{-1}$ cMpc and $\rm Q_1$ closely follows the median reduced three-point correlation function (${\rm Q}=\zeta/(\xi*\xi)$). 
		However at smaller length scales we find non-zero $\rm Q_2$ with indications of this being dependent on the thermal parameters and thermal history. Thus  for extracting cosmological information like bias one has to use data from configuration having $r\ge3$ h$^{-1}$ cMpc.
		\vskip 0.1in
		\item[3.] \noindent{\bf Dependence on the configuration:} In this study we mainly concentrated on the triangular configurations with equal arms for a full 
		range of angles. As expected from the evolution of gravitational instabilities, 
		three-point correlation function calculated for a given $N_{\rm HI}$ threshold decreases with the increasing length scale. Interestingly the scale dependence of three-point correlation is found to be steeper than that of the two-point correlation for the sub-samples defined with a same $N_{\rm HI}$ threshold.
		We also find the three-point correlation to show mild dependence on $\theta$.  
		The $\zeta$ decreases with increasing $\theta$ for $\theta \le 60^\circ$ and remains nearly constant (or shows very mild increase) with increasing $\theta$. We do not find U or V shape in Q vs $\theta$ distribution as found in the case of low-$z$ galaxies. 
		
		\vskip 0.1in
		\item[4.] \noindent{\bf The reduced three point function and bias:} We also study the median reduced three point correlations (i.e Q) and its dependence on $N_{\rm HI}$ and on the angle and scale of the triplet configurations. While Q is found to increase with increasing $N_{\rm HI}$ (as $\zeta$) the rate of increase is weaker than that seen for $\zeta$. Similarly the dependence on the scale and $\theta$ are found to be weaker for Q. We notice that for higher scales (i.e $r>4$ h$^{-1}$ cMpc) Q values converge.  The median Q values found are typically in the range 0.2 to 0.7. This is lower than Q$\sim$1.3 found for low-$z$ galaxies.
		In the perturbative region the observed Q from a tracer can be related to the Q of the dark matter ($\mathrm Q_m$) as $\bar{\mathrm Q} = ({\mathrm Q_m}/b_1) + (b_2/b_1^2)$ (up to first non-linear term) where $b_1$ is the linear bias and $b_2$ is the non-linear bias \citep{fry1993}. The Q value which is used in the bias equation ($\bar{\mathrm Q}$) is calculated using mean three-point and associated two-point correlation (see Eq.~\ref{Mean_Q}) and different than the median Q that we use in this study. Both Q values have been plotted and compared in Appendix~\ref{Mean_Median_q}.
		
		For an equilateral configuration ($\theta=60^\circ$), $ r = 5$ h$^{-1}$ cMpc and $N_{\rm HI}>10^{13.5}$cm$^{-2}$, we measure $\xi=0.31$ and this corresponds to $b_1 = 1.76$, if we use two-point function of dark matter obtained using linear theory. As for a given scale $\xi$ depends on $N_{\rm HI}$ cut off, we find that $b_1$ to be a strong function of $N_{\rm HI}$. The value of $b_1$ increases from 1.52 to 2.05 when we change the 
		$N_{\rm HI}$ cut off from $10^{13.3}$ to $10^{13.8}$ cm$^{-2}$.
		To obtain $b_2$ we need $\mathrm Q_m$. We also have to consider $\bar{Q}$ instead of median Q. If we take a typical value of $\mathrm Q_m=1.3$ and use $b_1=1.76$ and $\bar{\mathrm Q}=0.91$ obtained from our simulations for $N_{\rm HI}>10^{13.5}$ cm$^{-2}$ we get $b_2=0.53$. For $b_1=2.05$ and $\bar{\mathrm Q}=1.1$ obtained from our simulations for $N_{\rm HI}>10^{13.8}$ cm$^{-2}$ we get $b_2=1.96$. For a given choice of $\mathrm Q_m$, $b_2$ value increases with increasing $N_{\rm HI}$.
		We defer a detailed self-consistent discussion on the bias to our upcoming work. 
		\vskip 0.1in
		\item[5.] \noindent{\bf Dependence on thermal and ionization history:} 
		We consider simulations obtained with four different ionization and thermal history. We show for a given $N_{\rm HI}$ and configuration $\zeta$ is more sensitive compared to $\xi$ or $\mathrm Q$. In particular, our model with enhanced heating (i.e Enhanced $\alpha=1.8$ model)  shows larger values for $\zeta$ and $\xi$ compared to the value obtained from simulations using three self-consistent background models. The difference in $\zeta$ 
		measured between these models decreases with increasing length scale. We found the relationship between $N_{\rm HI}$ and baryonic over-density $\Delta$ for these models at $z\sim 2$.
		When we use different $N_{\rm HI}$ that corresponds to a single $\Delta$ we find $\xi$ and $\zeta$ from all these models match very well. Interestingly, our study also suggests that the scatter in Q due to varying physical conditions is minimum. Thus it could provide a better probe of the underlying matter distribution even in the absence of good understanding of the thermal evolution of the IGM.
		
		\vskip 0.1in
		\item[6.] \noindent{\bf Redshift evolution:} Our study also shows that the three-point correlation function evolves strongly over the redshift range $1.8\le z\le 2.5$ compared to the two-point correlation function for a fixed $N_{\rm HI}$ and configuration. Interestingly the rate of decrease with $z$ seems similar for different length scales for both $\xi$ and $\zeta$.
		We also notice that the evolution of $\mathrm Q$ is much slower than what we see for $\zeta$. We find that if we use different $N_{\rm HI}$ cut offs corresponding to single $\Delta$ at different $z$ then $\zeta$ and $\xi$ are nearly constant over the redshift range considered. This once again suggests that the strong redshift evolution seen over a small redshift range is dominated by the evolution of physical conditions in the IGM. Any physical model that connects between $\Delta$ and $N_{\rm HI}$ will depend on $T_0$, $\gamma$, $\Gamma_{\rm HI}$ and a physical scale that connects between density and column density of HI. We find that even after using correct physical parameters at each redshift we may need a slightly faster redshift evolution if we use the analytical equation given by \citet{schaye2001}. This could either mean redshift evolution in the gas fraction $f_g$ or the length (local Jeans length as assumed by \citet{schaye2001}) connecting density and $N_{\rm HI}$. One possible way out is to get the fitting function connecting $\Delta$ and $N_{\rm HI}$ as a function of redshift 
		considering models with wide range of above mentioned parameters. Having such a fitting function will allow us to probe clustering as a function of $\Delta$ and its redshift evolution. This will then allow us to constrain the cosmological parameters from the observational data.
		\vskip 0.1in
		
		\item[7.]\noindent{\bf Future perspective:}
		We also investigate the spectroscopic requirements for measuring $\zeta$, $\xi$ and $Q$ over a small redshift range $\delta z = 2.0\pm0.1$ for a given length scale (i.e r$\pm$0.5$h^{-1}$cMpc) and smaller angular intervals (i.e $\theta\pm 10^\circ$).  We compute the redshift path length (or number of triplets) required for detecting three point function at $5\sigma$ level.
		For the column density range and length scale probed we need about 70 to 100 triplets for probing the length scale of 4 to 5 cMpc. This roughly corresponds to 200 to 300 quasar spectra with spectral resolution of $50$ kms$^{-1}$ and SNR$\sim$ 20. Such a spectra will also allow us to probe (i) the clustering properties of the \lya\ with metal line detections; (ii) cross-correlation between the \lya\ and metal lines (both in transverse and longitudinal directions) and (iii) connection between quasars and \lya\ absorption in the transverse direction as demonstrated in \citet{maitra2019}. We investigate the availability of triplets for such a study in SDSS quasar catalog. 
		
	\end{enumerate}

	\section*{Acknowledgement}
	We acknowledge the use of High performance computing facilities PERSEUS and PEGASUS at IUCAA. We thank Kandaswamy Subramanian and Nishikanta Khandai for useful discussions.

	
	\bibliographystyle{mnras}
	\bibliography{main} % if your bibtex file is called example.bib
	%    \end{document}	
	
	%%%%%%%%%%%%%%%%%%%%%%%%%%%%%%%%%%%%%%%%%%%%%%%%%%
	
	%%%%%%%%%%%%%%%%% APPENDICES %%%%%%%%%%%%%%%%%%%%%
	\clearpage
	\newpage
	\appendix
	
	\section{Definition of Reduced three-point correlation function}\label{Mean_Median_q}
	
	Based on Eq.~\ref{Q_eqn}, Q is defined as the three-point correlation function normalized to the cyclic combination of the associated two-point correlation functions. In galaxies, it is a common practice to normalize the three-point correlation function averaged over the entire volume with the cyclic combination of averaged two-point functions. In this work, we define a unique Q value for each of the triplet sightlines. We can then obtain a mean or median Q from the distribution of Q values for 4000 triplet sightlines. With mean Q, any arbitrarily small cyclic combination of two-point correlation makes the Q value blow up, thereby artificially enhancing the mean Q value. To get around this issue, we choose to work with median Q value.
	
	\begin{figure}
		\centering
		\includegraphics[width=9cm]{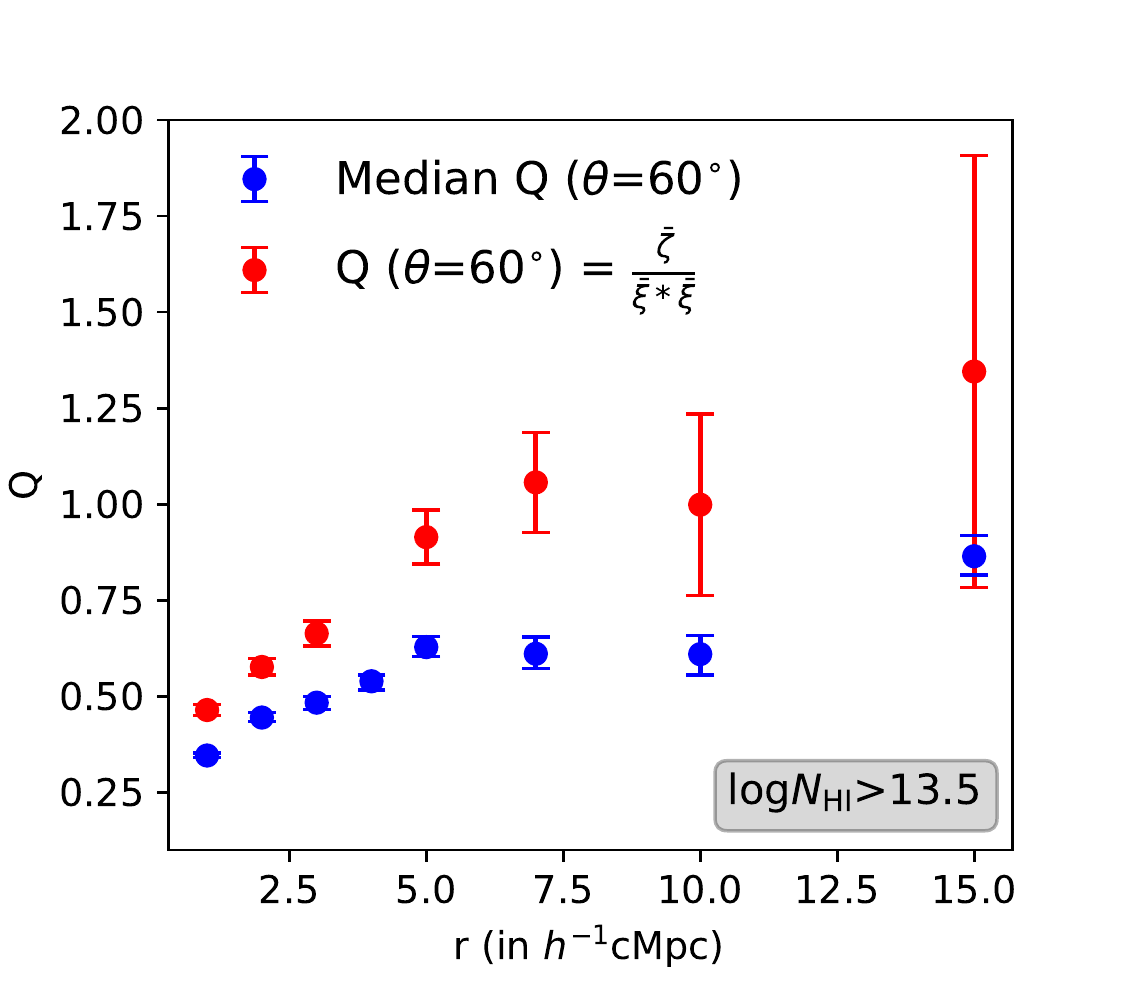}
		\caption{Reduced three-point correlation function defined in two ways (Median Q and Q obtained from Eq.~\ref{Mean_Q}) as a function of scale for $\theta=60^{\circ}$ and $N_{\rm HI}>10^{13.5}$cm$^{-2}$.}
		\label{Mean_vs_Median_Q}
	\end{figure}
	
	One can also choose to define Q using three-point and two-point correlation functions which are averaged over 4000 triplet sightlines as
	\begin{equation}
	{\rm \bar{Q}}=\frac{\bar{\zeta}}{\bar{\xi}*\bar{\xi}}
	\end{equation}\label{Mean_Q}
	We compare the two definitions of Q in Fig.~\ref{Mean_vs_Median_Q} as a function of scale for $\theta=60^{\circ}$ and $N_{\rm HI}>10^{13.5}$cm$^{-2}$. For $\bar{\mathrm Q}$ defined using Eq.~\ref{Mean_Q}, we assign errors by propagating errors associated with mean three-point and two-point correlation functions. It is seen that $\rm \bar{\mathrm Q}$ has larger magnitude compared to median Q, but the associated errors are large. Also, as shown in Fig.~\ref{Q_vs_Q1}, at large scales median Q closely follows the "hierarchical ansatz" that one expects from the distribution of three-point and associated two-point correlation functions over 4000 triplet sightlines.

	\section{Convergence Tests}
	
	We perform convergence tests involving the impact of size of the simulation box on the transverse three-point correlation. Additionally, we examine the effect of taking a smaller longitudinal slice of the triplet sightlines for a fixed simulation box size. We also discuss about our choice of taking a $\pm 2h^{-1}$cMpc longitudinal bin size while calculating the transverse three-point correlation.
	
	\subsection{Simulation box-size and slice-size}\label{Size_test}
\begin{figure*}
		\centering
		\includegraphics[viewport=0 -5 335 290,width=6cm, clip=true]{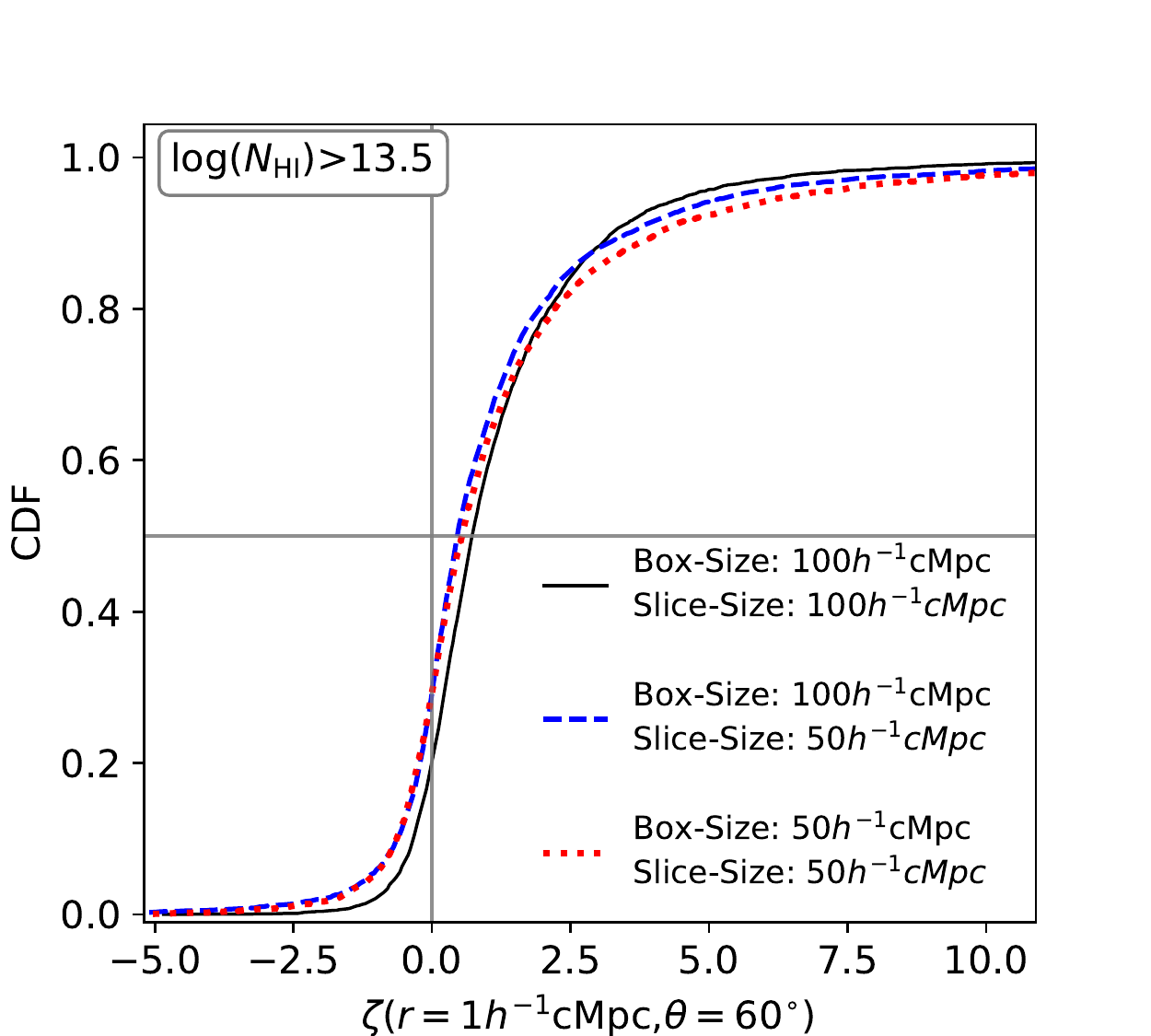}%
		\includegraphics[viewport=10 12 325 290,width=6cm, clip=true]{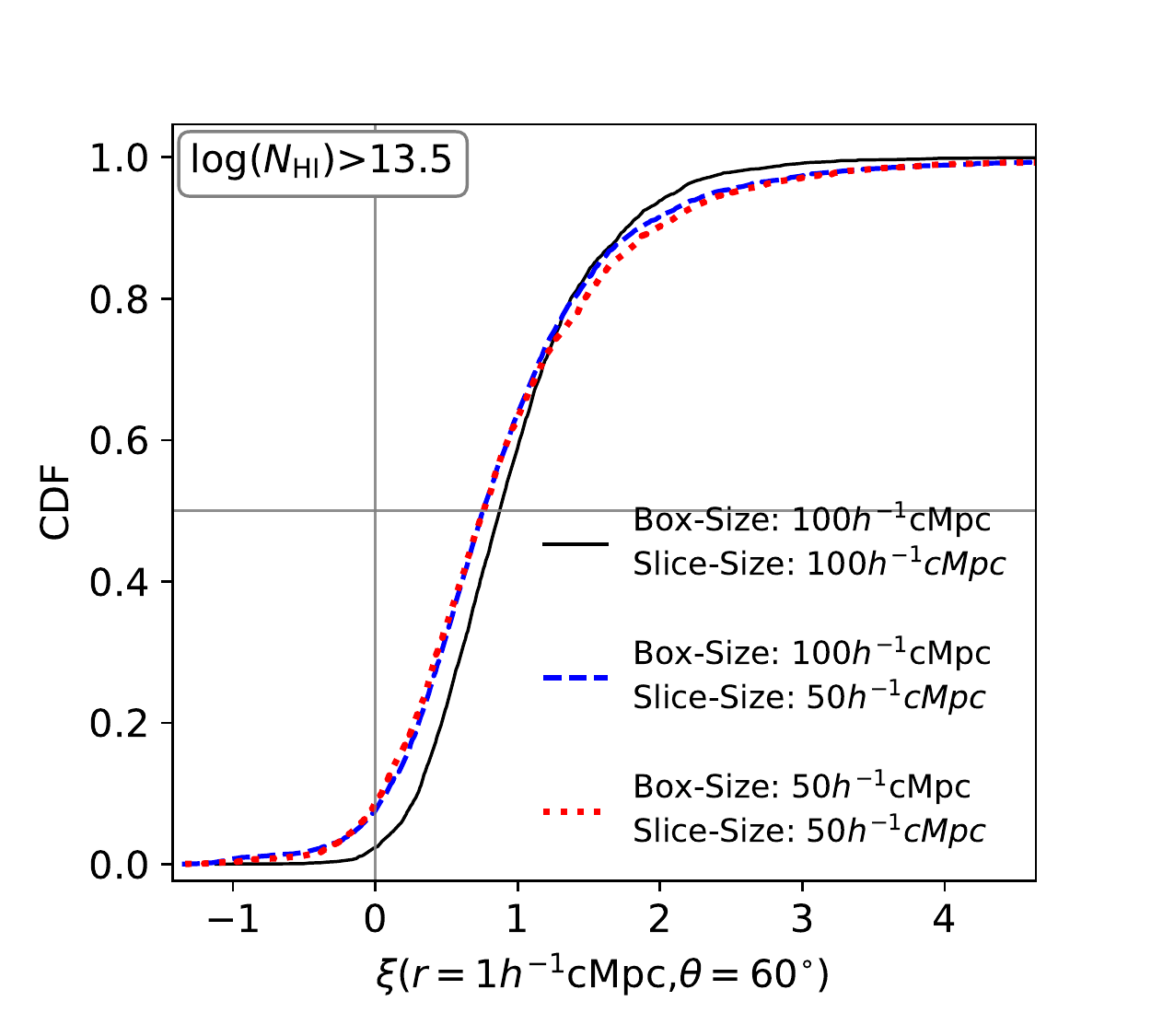}%
		\includegraphics[viewport=10 12 325 290,width=6cm, clip=true]{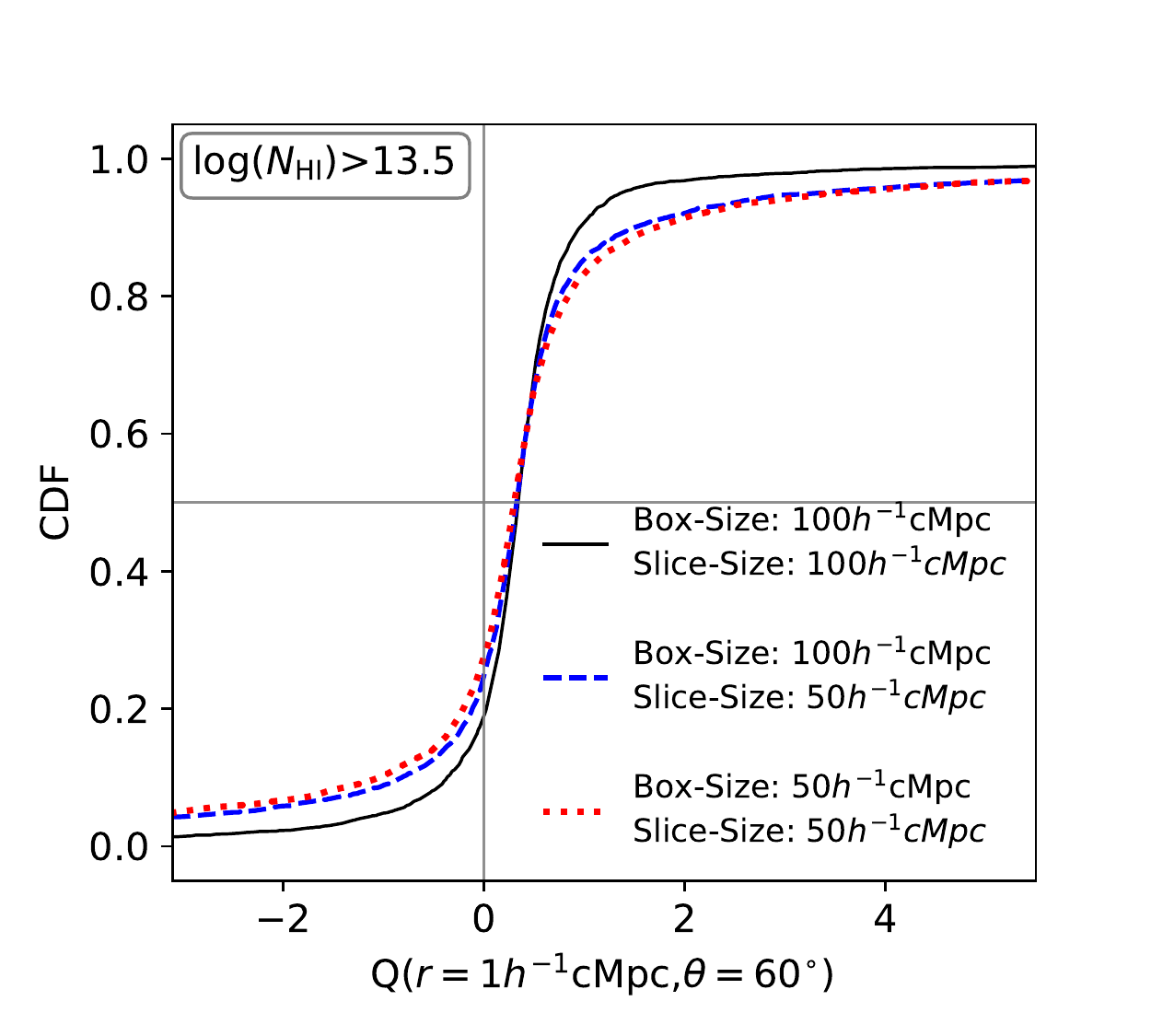}%
		\caption{Convergence test involving simulation box-size and longitudinal slice-size over which transverse three-point correlation is calculated for a single source configuration($r=1h^{-1}$cMpc, $\theta=60^{\circ}$) for clouds having $N_{\rm HI}>10^{13.5}$cm$^{-2}$. The cumulative distribution function of transverse three-point correlation is plotted for a box-size and slice-size of 100$h^{-1}$cMpc each (fiducial scheme; black curve),  box-size and slice-size of 100$h^{-1}$cMpc and 50$h^{-1}$cMpc respectively (blue-dashed curve) and  box-size and slice-size of 50$h^{-1}$cMpc each (red-dashed curve).}
		\label{Convergence}
	\end{figure*}

	We test the effect of changing the box size with the help of an additional 50$h^{-1}$cMpc simulation box run with our fiducial UVB with $\alpha=1.8$. We choose to investigate the effect of simulation box size on transverse three-point correlation at scale of $r=1h^{-1}$cMpc and $\theta=60^{\circ}$ for clouds having $N_{\rm HI}>10^{13.5}$cm$^{-2}$. Additionally, we also take our 100$h^{-1}$cMpc simulation box and calculate transverse three-point correlation for 50$h^{-1}$cMpc slices centered around the 100$h^{-1}$cMpc sightlines. So, we have transverse three-point correlations corresponding to:
	\begin{itemize}
		\item 100$h^{-1}$cMpc Simulation box size and 100$h^{-1}$cMpc sightline length (Slice size=100$h^{-1}$cMpc): 4000 triplet sightlines.
		\item 100$h^{-1}$cMpc Simulation box size and 50$h^{-1}$cMpc sightline length (Slice size=50$h^{-1}$cMpc): 4000 triplet sightlines.
		\item 50$h^{-1}$cMpc Simulation box size and 50$h^{-1}$cMpc sightline length (Slice size): 4000 triplet sightlines.
	\end{itemize}
	We also calculate transverse two-point and reduced three-point correlation for these cases for $r=1h^{-1}cMpc$ and with clouds having $N_{\rm HI}>10^{13.5}$cm$^{-2}$. The cumulative distribution function of transverse three-point ,two-point and reduced three-point correlations for all these cases are plotted in the left, middle and right panels of Fig.~\ref{Convergence}, respectively. 
	
	Comparing the CDF of transverse three-point and two-point correlation of the three cases, we see that the difference between the cases having similar slice size (50$h^{-1}$cMpc) and different box sizes is smaller compared to the cases having similar box size (100$h^{-1}$cMpc) and different slice sizes. The differences are smaller in case of two-point correlation. The median transverse three-point and two-point correlations (given by CDF=0.5) for cases having similar slice size (but different box size) is roughly the same. Same is not true for cases having similar box size (but different slice size). What this means is that the median transverse three-point and two-point correlation in clouds primarily depends on the redshift path length of the sightlines. We have reached convergence in box size at 50$h^{-1}$cMpc since the median correlation does not change with box size (50 and 100$h^{-1}$cMpc) for a fixed slice size (50$h^{-1}$cMpc). On the other hand, while the CDF for Q is different for the 3 different cases, the median Q remains similar. This means that while the median transverse three-point and two-point correlation in clouds depends on the redshift path length of the sightlines, median Q is independent of it.

	\subsection{Longitudinal binning for Transverse three-point correlation}\label{Binning}
	\begin{figure*}
		\centering
		\includegraphics[width=9cm]{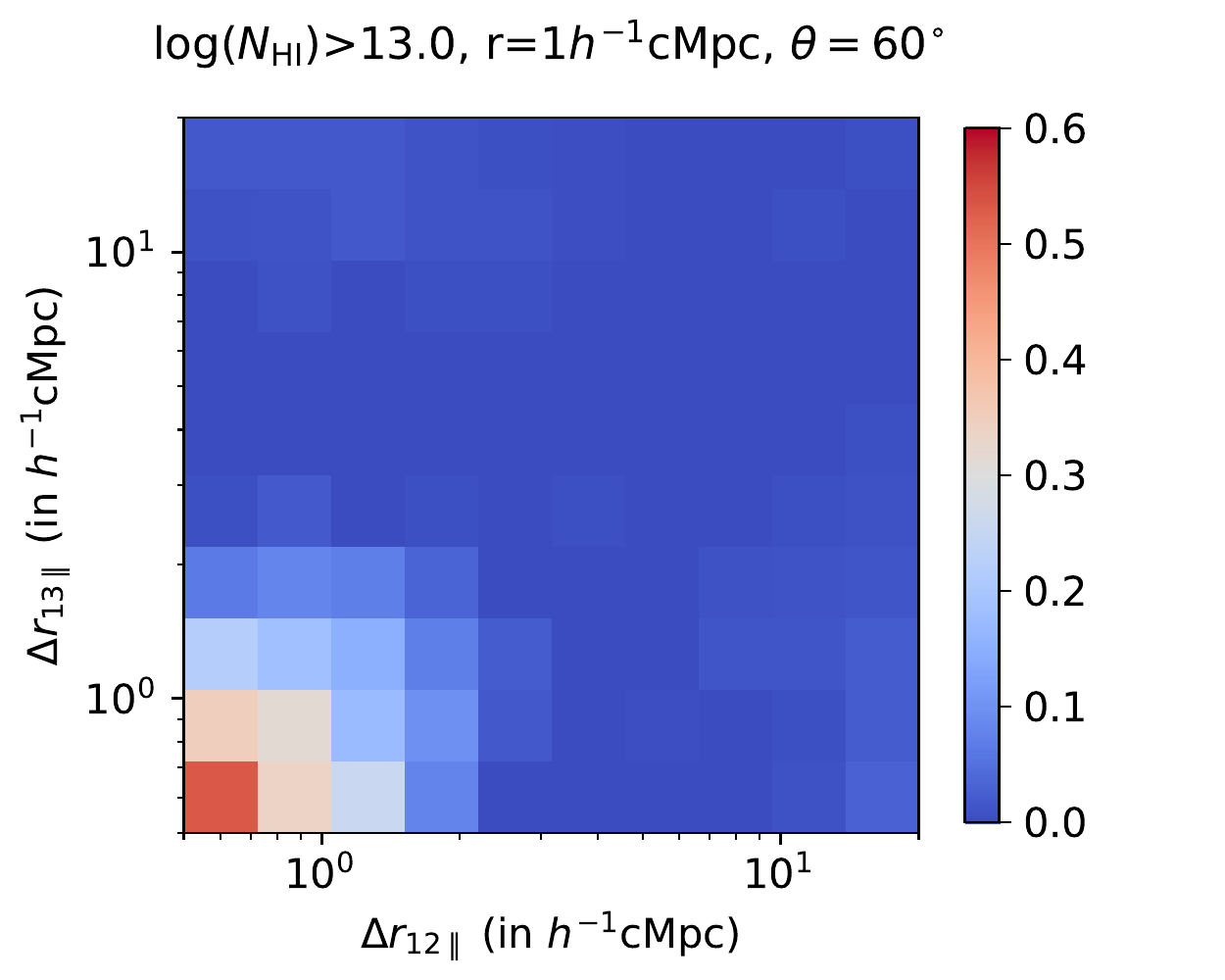}%
		\includegraphics[width=9cm]{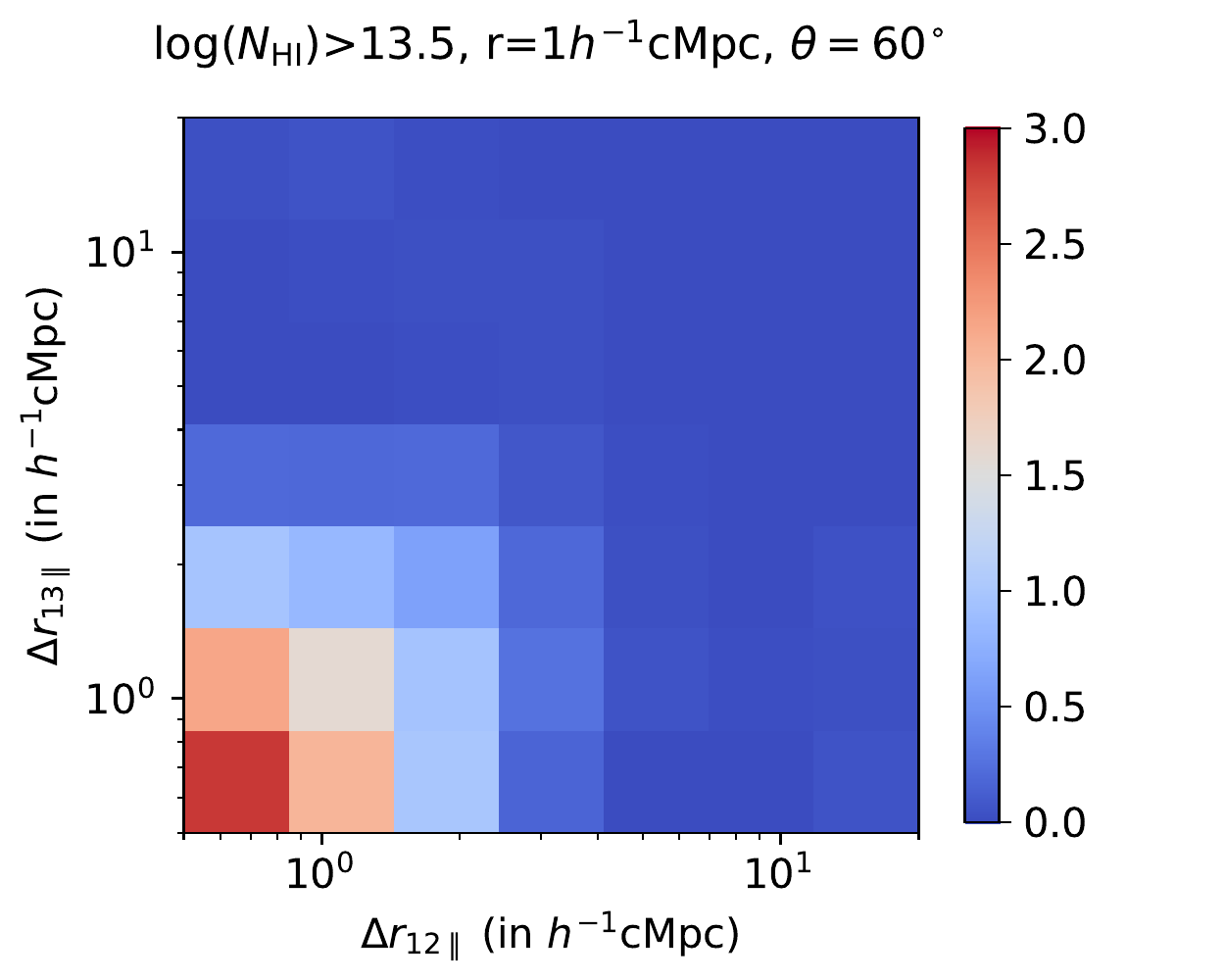}
		\caption{Three-point correlation as a function of redshift space separation of clouds along 2nd and 3rd sightlines with respect to the 1st sightline of the triplet. The colorbar represents the three-point correlation averaged for a bin.}
		\label{Corr_binning}
	\end{figure*}
	
\begin{figure*}
		\centering
		
		\includegraphics[viewport=5 10 300 300,width=4.5cm,clip=true]{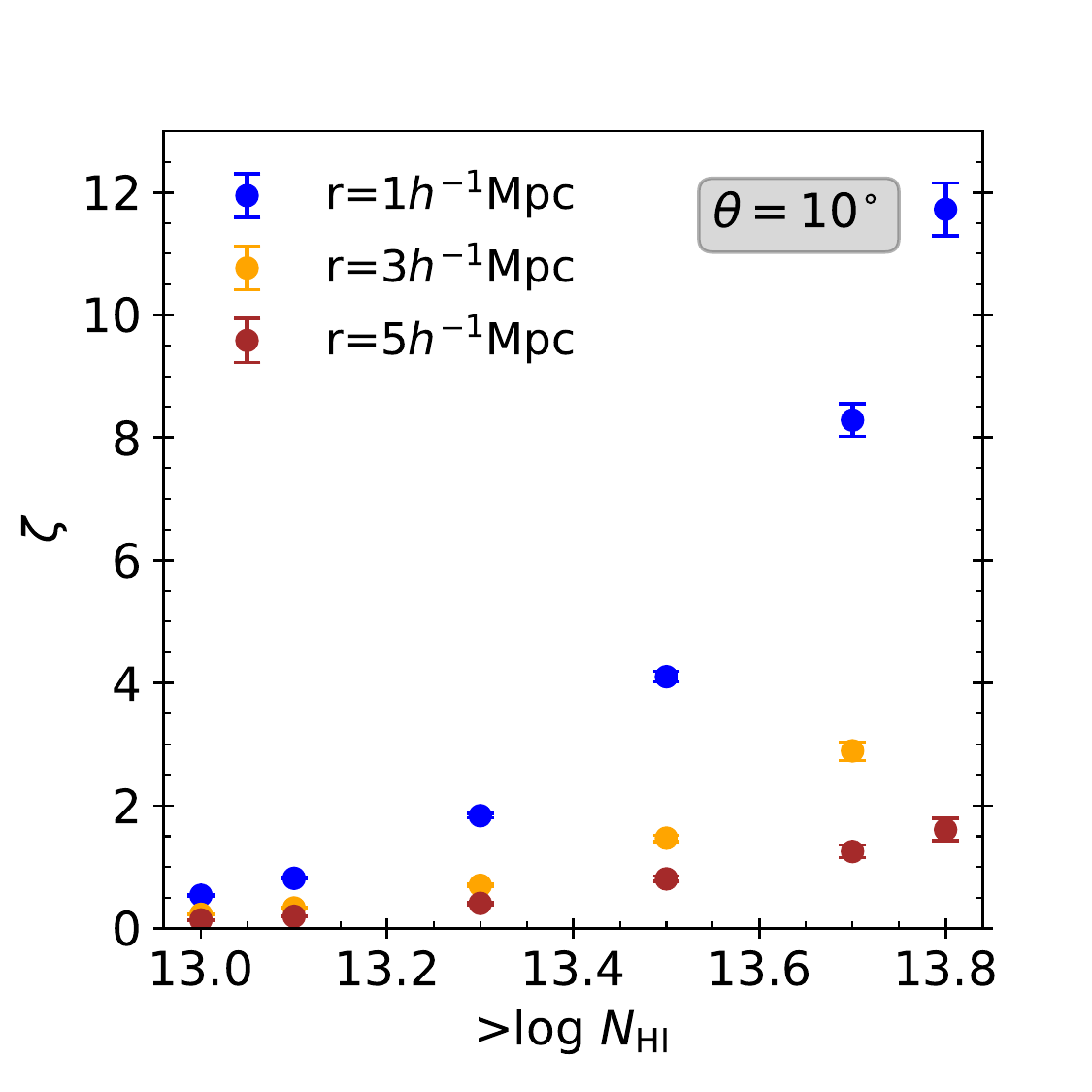}%
		\includegraphics[viewport=5 10 300 300,width=4.5cm,clip=true]{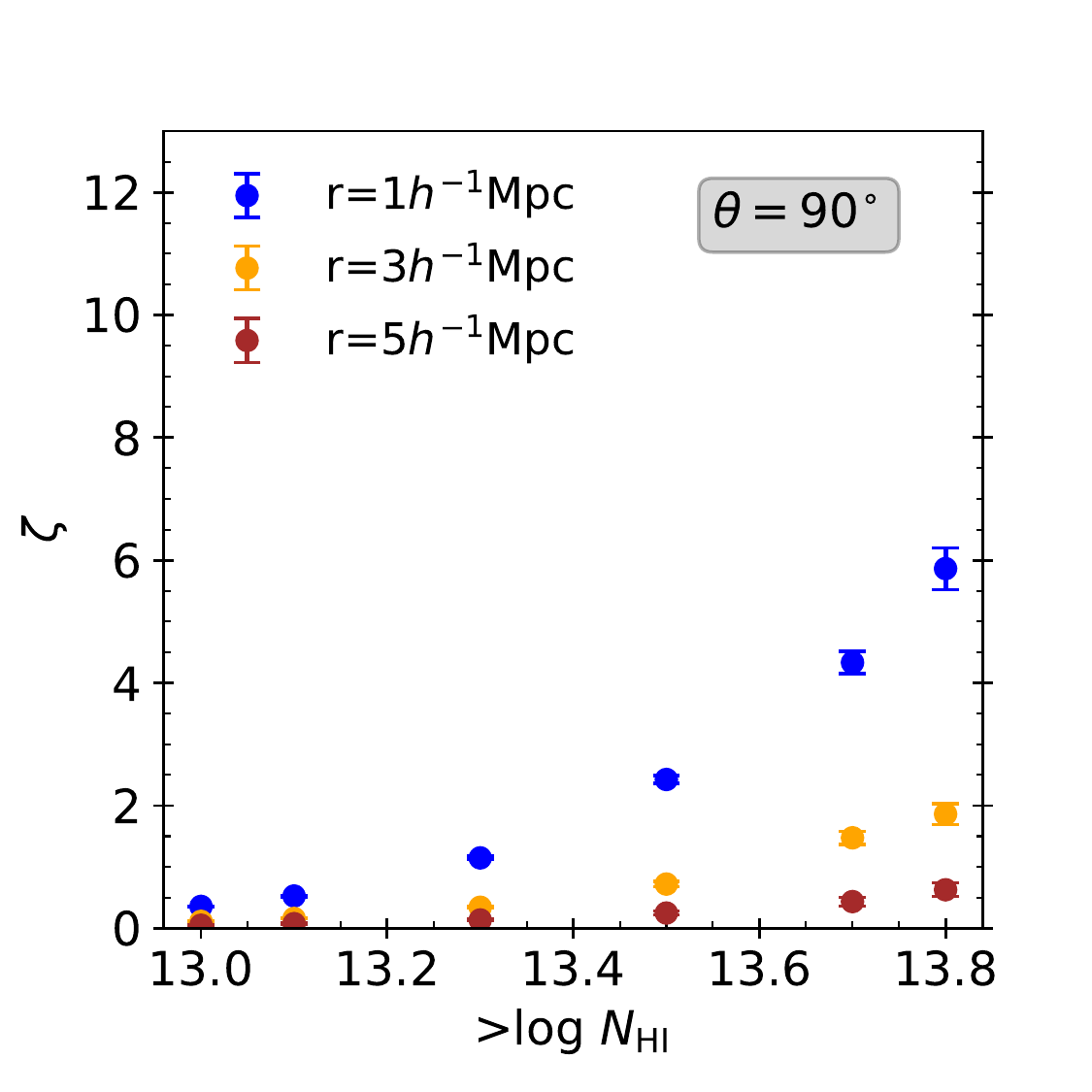}%
		\includegraphics[viewport=5 10 300 300,width=4.5cm,clip=true]{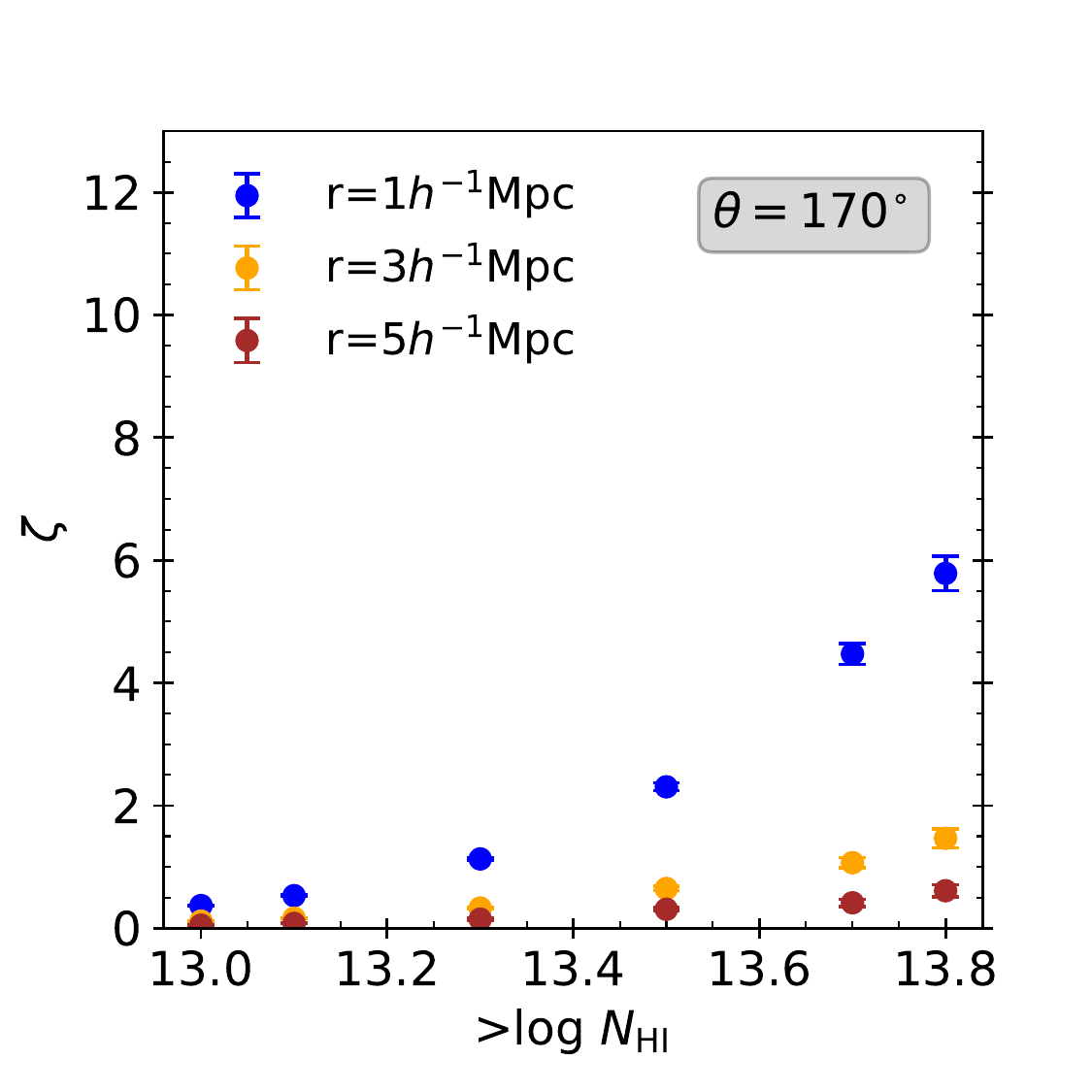}%
		\includegraphics[viewport=5 10 300 300,width=4.5cm,clip=true]{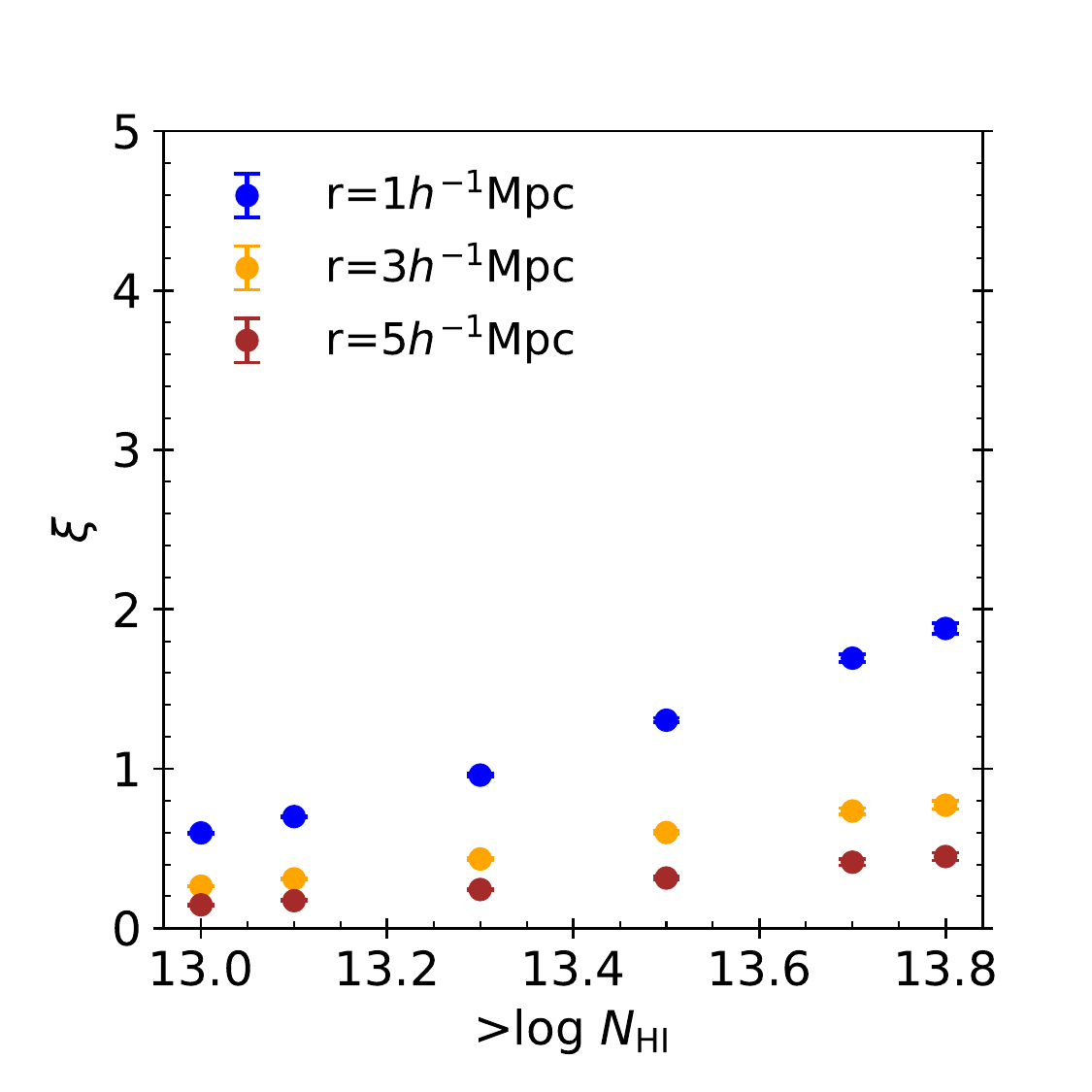}%
		
		\includegraphics[viewport=5 10 300 300,width=4.5cm,clip=true]{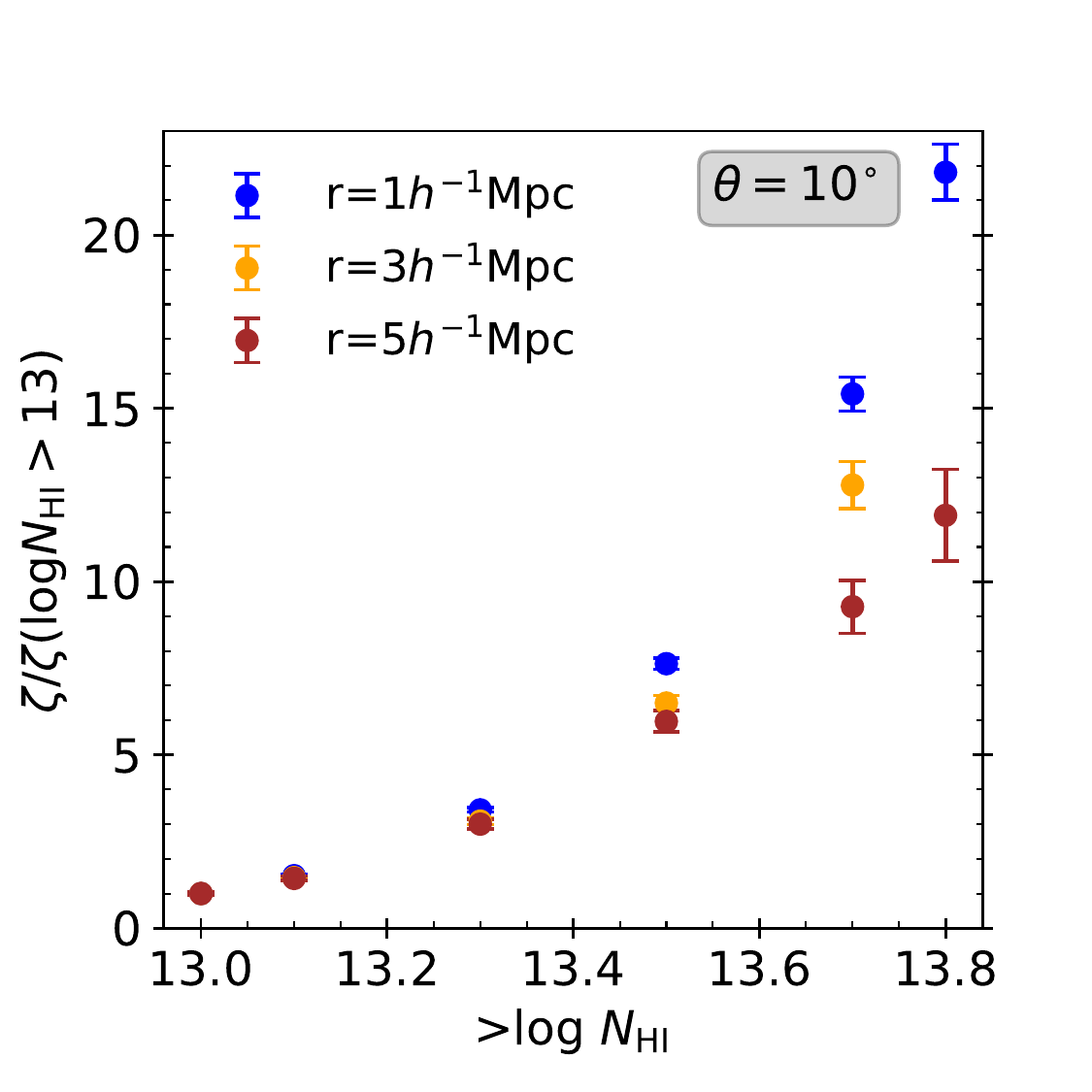}%
		\includegraphics[viewport=5 10 300 300,width=4.5cm,clip=true]{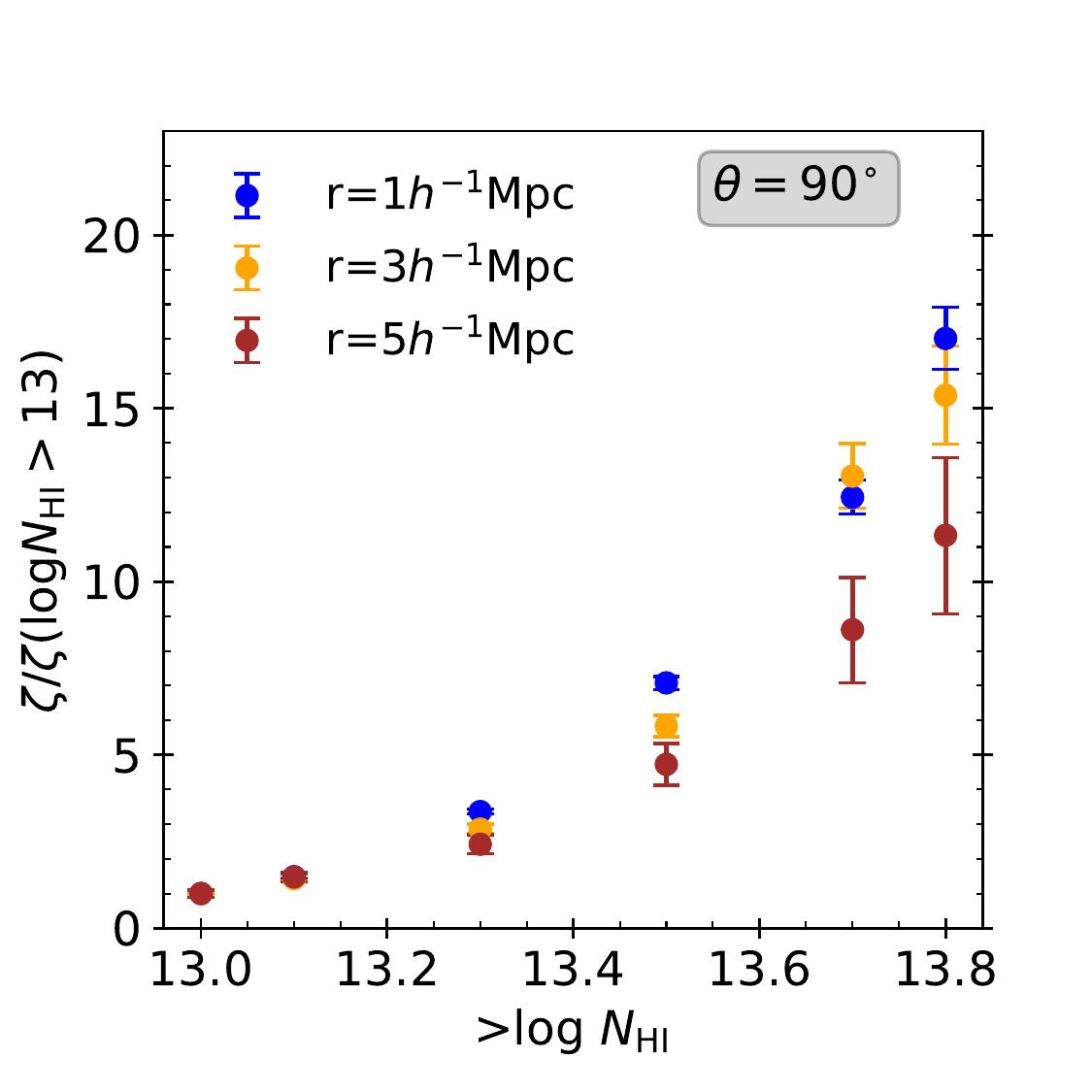}%
		\includegraphics[viewport=5 10 300 300,width=4.5cm,clip=true]{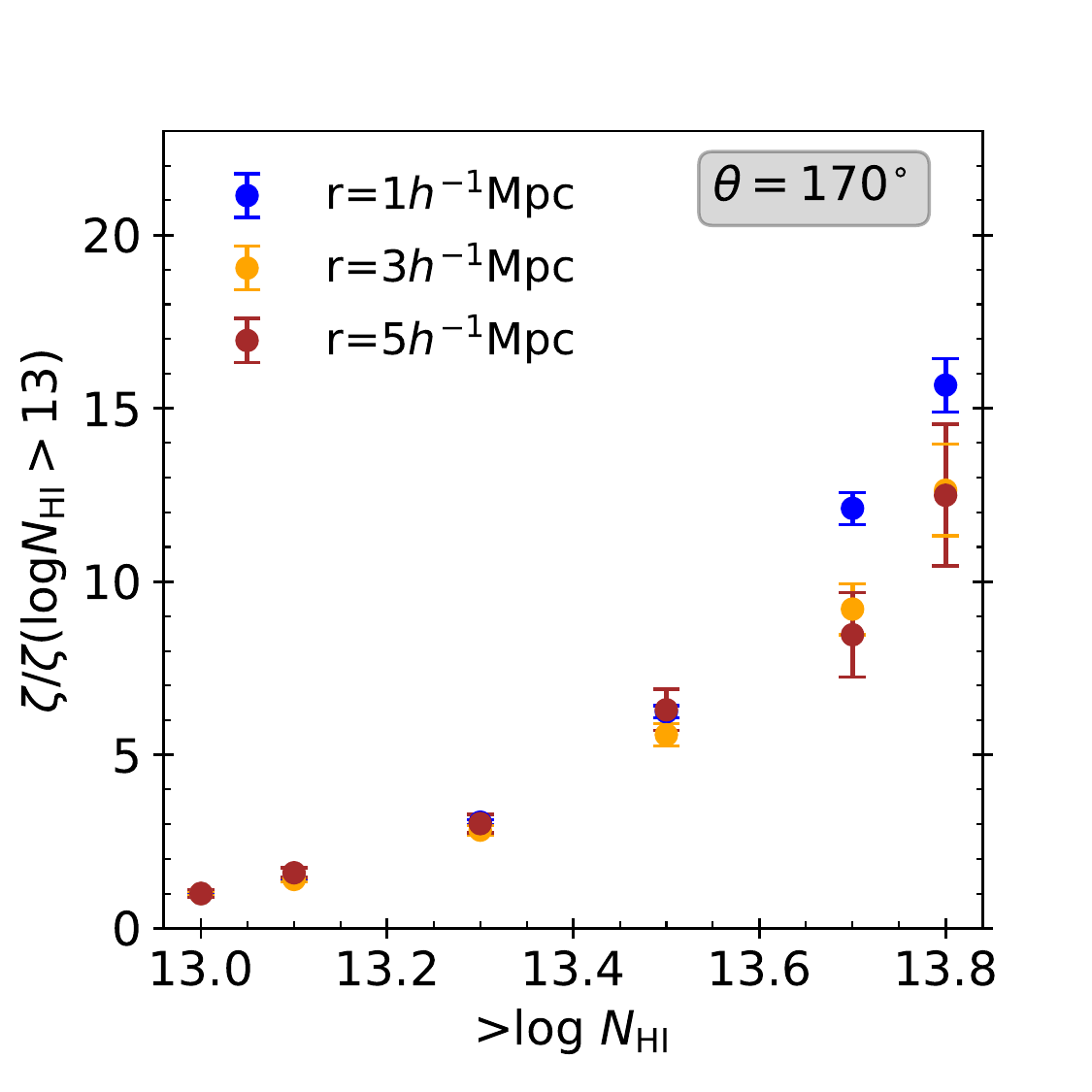}%
		\includegraphics[viewport=5 10 300 300,width=4.5cm,clip=true]{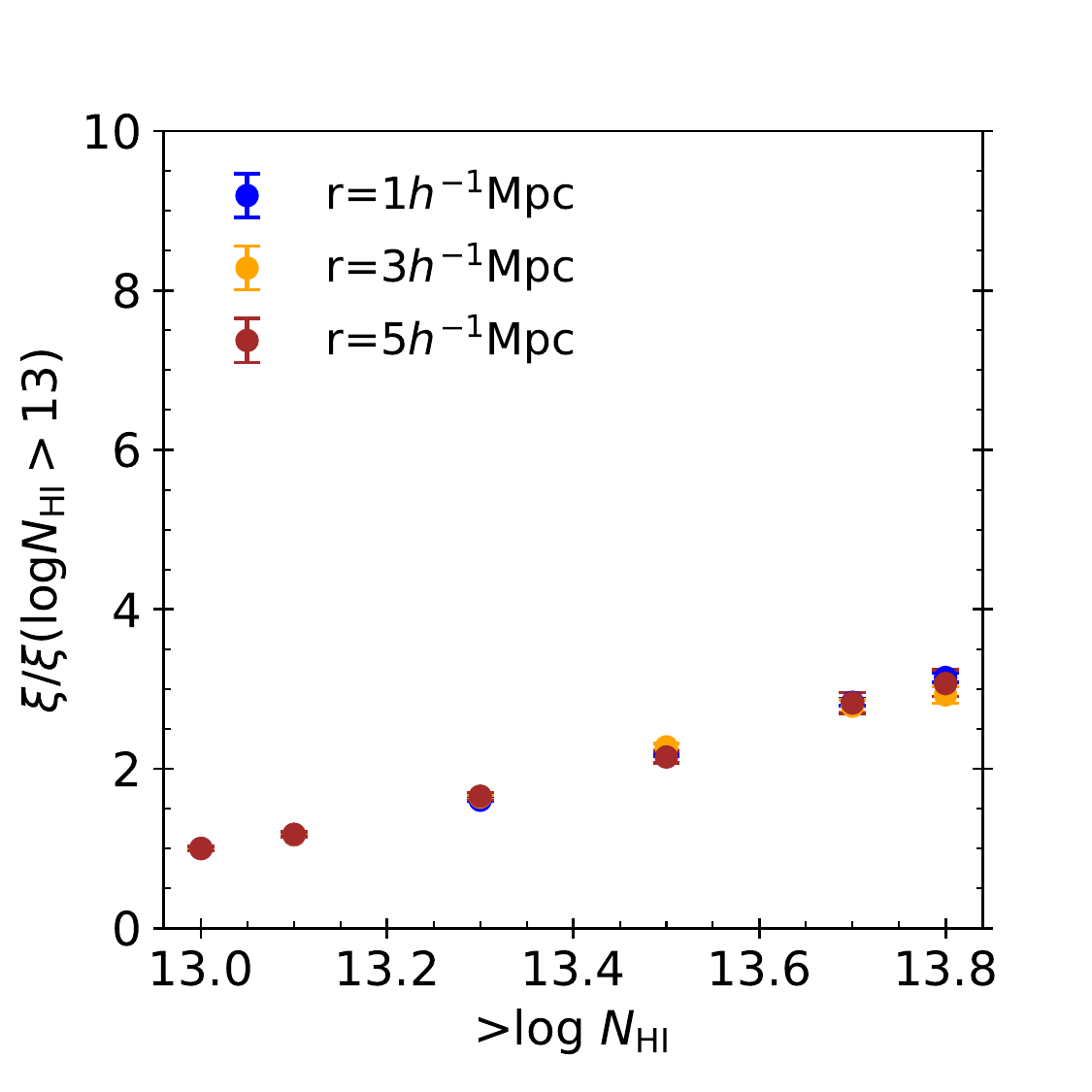}%

		\caption{Plots of transverse three-point correlation (Top left three panels) and transverse two-point correlation (Top rightmost panel) along with these normalized to 1 at initial point as a function of $N_{\rm HI}$ thresholds for different scales (Bottom panels). The longitudinal binning for transverse correlations have been taken as $\pm 1h^{-1}$cMpc. The angle of the configurations are taken as $\theta=10^{\circ}, 90^{\circ}$ and $170^{\circ}$. We plot this for $N_{\rm HI}$ threshold upto $10^{13.5}$cm$^{-2}$ only since higher $N_{\rm HI}$ threshold results in lesser number of absorbers and a shorter longitudinal binning gives noisy signal for small number of absorbers.}
		\label{Corr_cloud_NHI_1Mpc}
	\end{figure*}

\begin{figure*}
		\centering
		\includegraphics[viewport=40 30 1200 320,width=20cm, clip=true]{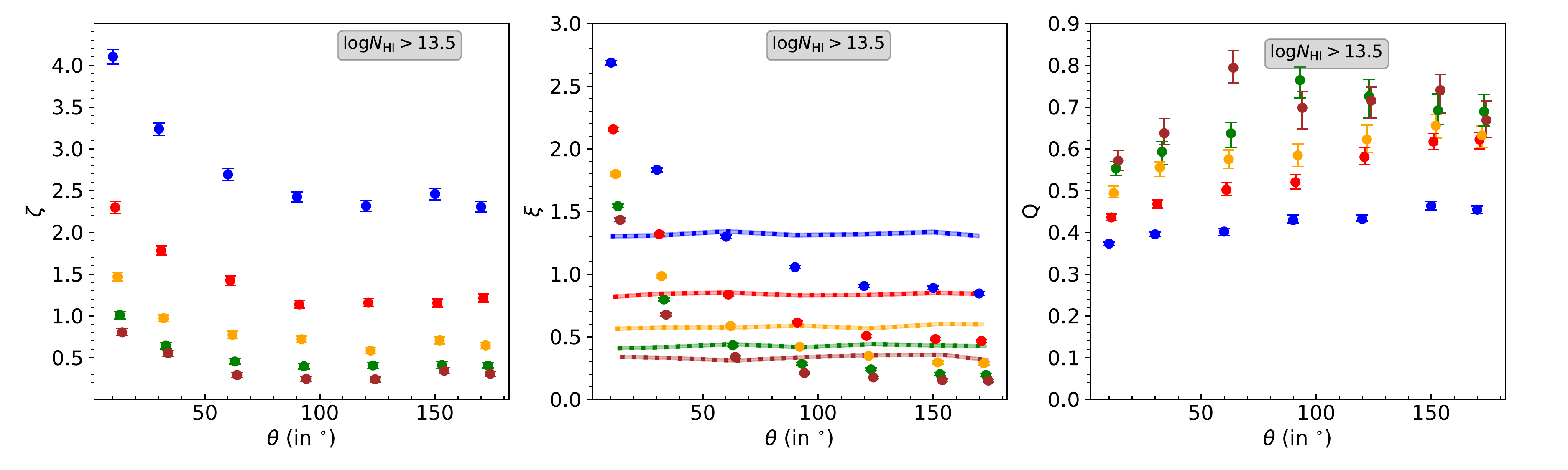}
		
		\includegraphics[viewport=40 10 1200 320,width=20cm, clip=true]{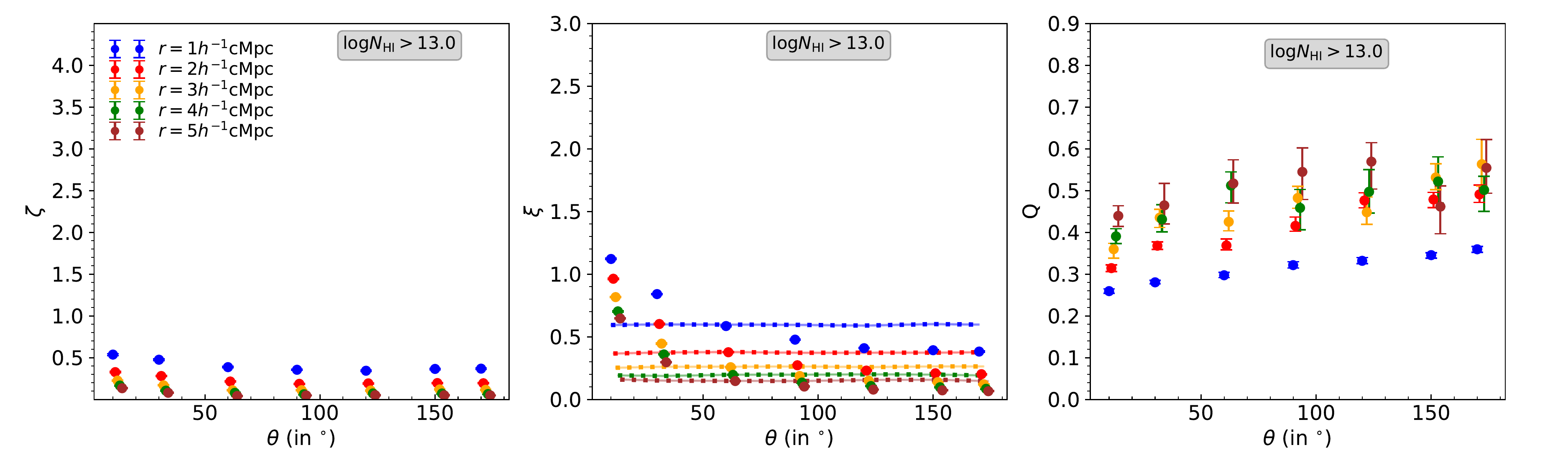}
		\caption{Angular dependence of transverse three-point correlation (Left panels), the associated two-point correlation (Middle panels) and the reduced three-point correlation (Right panels) functions for neutral hydrogen column density threshold $N_{\rm HI}>10^{13.5}$cm$^{-2}$ (Top panels) and $N_{\rm HI}>10^{13}$cm$^{-2}$ (Bottom panels) at scales 1-5$h^{-1}$cMpc.   The longitudinal binning for transverse correlations have been taken as $\pm 1h^{-1}$cMpc. The correlation statistics are calculated for UVB with spectral index $\alpha=1.8$.}
		\label{Corr_cloud_angle_1Mpc}
	\end{figure*}

	In this section, we discuss on our choice of longitudinal binning for transverse three-point or two-point correlation. We take a default bin size of $\pm 2h^{-1}$cMpc along the longitudinal direction. This bin size has been taken on the basis of the profile of three-point correlation as a function of redshift space separation of clouds along 2nd and 3rd sightline with respect to the 1st triplet sightline for $N_{\rm HI}>10^{13}$cm$^{-2}$ clouds. The profile has been shown in Fig.~\ref{Corr_binning} for $N_{\rm HI}>10^{13}$cm$^{-2}$ and $N_{\rm HI}>10^{13.5}$cm$^{-2}$. For  $N_{\rm HI}>10^{13}$cm$^{-2}$, we see that the correlation exists upto redshift space separation of $2h^{-1}$cMpc which is the basis for the choice of our longitudinal binning. For $N_{\rm HI}>10^{13.5}$cm$^{-2}$, the profile extends to a slightly larger distance. Hence, we can treat the longitudinal binning as a free-parameter.
	In Fig.~\ref{Corr_cloud_NHI_1Mpc} and Fig.~\ref{Corr_cloud_angle_1Mpc}, we show the $N_{\rm HI}$ and configuration dependence for a longitudinal bin size of $\pm 1h^{-1}$cMpc. We find that the amplitude of the observed correlations are larger compared to what one sees with longitudinal bin size of $\pm 2h^{-1}$cMpc.  However, the trends observed in transverse three-point and two-point correlations with $N_{\rm HI}$ thresholds and configuration of the triplet remains same irrespective of the choice of longitudinal bin size.

	%%%%%%%%%%%%%%%%%%%%%%%%%%%%%%%%%%%%%%%%%%%%%%%%%%

	% Don't change these lines
	\bsp	% typesetting comment
	\label{lastpage}

\end{document}